\documentclass[12pt]{ociamthesis} 
\usepackage{pstricks}
\usepackage{pst-all}
\usepackage{comment}
\usepackage{subeqnarray}
\usepackage{amssymb}
\usepackage{amsmath}
\usepackage{amsthm}
\usepackage{dsfont}
\usepackage{youngtab}
\usepackage{tensor}
\usepackage{mathptmx}       
\usepackage{helvet}         
\usepackage{setspace}
\usepackage{graphics}
\usepackage{fancyhdr}
\fancypagestyle{title}{
\fancyhf{}

\fancyfoot[R]{\thepage}
}
\fancypagestyle{plain}{
\fancyhf{}
\fancyhead[R]{\slshape\nouppercase{\leftmark}}
\fancyfoot[R]{\thepage}
}
\newcommand*{\I}{\mathrm{i}}
\newcommand*{\E}{\mathrm{e}}
\newcommand*{\D}{\mathrm{d}}

\newcommand{\bea}{\begin{eqnarray}}
\newcommand{\eea}{\end{eqnarray}}
\newcommand{\beq}{\begin{equation}}
\newcommand{\eeq}{\end{equation}}

\newcommand{\rishi}{\rangle\!\rangle}

\title{Random Matrices, \\[1ex]   
        Boundaries and Branes}  

\author{Benjamin Niedner}   
\college{Wolfson College} 
\degree{Doctor of Philosophy} 
\degreedate{Trinity 2015}     

\begin{document}
 
\newtheorem{thm}{Theorem}[section]
\theoremstyle{corollary}
\newtheorem{cor}[thm]{Corollary}
\theoremstyle{lemma}
\newtheorem{lem}[thm]{Lemma}
\theoremstyle{remark}
\newtheorem{rem}[thm]{Remark}
\theoremstyle{example}
\newtheorem{ex}[thm]{Example}
\theoremstyle{definition}
\newtheorem{defn}[thm]{Definition}
\theoremstyle{proposition}
\newtheorem{prop}[thm]{Proposition}

\setcounter{secnumdepth}{3}
\setcounter{tocdepth}{3}

\maketitle               
\begin{acknowledgements}
First and foremost, I would like to express my gratitude towards my supervisor, John Wheater, for his consistent support and guidance throughout the entire period of this work; numerous developments in this thesis would not have been possible without his shared insight and helpful suggestions. I am also immensely grateful for the pleasure and opportunity to collaborate with Max Atkin and Hirotaka Irie, whose ideas have been a significant driving force behind the projects in Chapters 3 and 5 of this thesis. Furthermore, I thank Stefan Zohren, Piero Nicolini as well as the Yukawa Institute for Theoretical Phyiscs, the mathematical physics group at Universit\'e catholique de Louvain, the Department of Physics at the Pontificial University Rio de Janeiro and the Frankfurt Institute for Advanced Studies for the kind hospitality during various research trips that facilitated this research. I also greatly acknowledge the essential financial support by the German National Academic Foundation, the Japan Society for the Promotion of Science and the STFC grant ST/J500641/1 that made my studies possible. Last but not least, I thank my friends and family for their continued love, support and understanding over the past years.
\end{acknowledgements}
   
\begin{abstractseparate}
This thesis is devoted to the application of random matrix theory to the study of random surfaces, both discrete and continuous; special emphasis is placed on surface boundaries and the associated boundary conditions in this formalism. In particular, using a multi-matrix integral with permutation symmetry, we are able to calculate the partition function of the Potts model on a random planar lattice with various boundary conditions imposed. We proceed to investigate the correspondence between the critical points in the phase diagram of this model and two-dimensional Liouville theory coupled to conformal field theories with global $\mathcal{W}$-symmetry. In this context, each boundary condition can be interpreted as the description of a brane in a family of bosonic string backgrounds. This investigation suggests that a spectrum of initially distinct boundary conditions of a given system may become degenerate when the latter is placed on a random surface of bounded genus, effectively leaving a smaller set of independent boundary conditions. This curious and much-debated feature is then further scrutinised by considering the double scaling limit of a two-matrix integral. For this model, we can show explicitly how this apparent degeneracy is in fact resolved by accounting for contributions invisible in string perturbation theory. Altogether, these developments provide novel descriptions of hitherto unexplored boundary conditions as well as new insights into the non-perturbative physics of boundaries and branes.
\end{abstractseparate}
          
\doublespacing
\begin{romanpages}          
\addcontentsline{toc}{chapter}{Table of Contents}
\tableofcontents            
\listoffigures             
\thispagestyle{title}
\addcontentsline{toc}{chapter}{List of Figures}

\end{romanpages}     

\chapter{Introduction}
\thispagestyle{title}

\noindent This thesis concerns the application of the theory of random matrices to the foundations of string theory and the quantum mechanical description of gravitation, or quantum gravity for short. A major motivation is that the tools of random matrix theory afford us a uniquely detailed window into the quantum physics of strings and gravity beyond the perturbative expansion in the string coupling. In particular, geometric objects such as boundaries and branes acquire a simple interpretation in terms of averages of characteristic polynomials of random matrices. Below we briefly review the history of work that has intertwined these subjects, with a view towards the open problems addressed in this work, and subsequently provide an outline of the following chapters.  
 
 \section{History of the subject}

 \noindent The application of the theory of random matrices to problems in physics was pioneered by Wigner in the fifities of the last century \cite{Wigner55}. Since its inception, this field has expanded tremendously and today, the applications of random matrix theory include areas as diverse as signal processing, number theory in mathematics, RNA folding in biology, and portfolio optimisation in finance \cite{akemann15}. In the subsequent decade, Tutte initiated the enumeration of \em planar maps \em \cite{Tutte62,Tutte63}, defined as graphs embeddable in the plane, modulo homeomorphisms. When endowed with a statistical lattice model defined thereon, the detailed knowledge of the asymptotic properties of maps with a large number of vertices allows a rigorous definition of the path integral for two-dimensional gravity coupled to matter or equivalently, bosonic string theory in a non-critical target space dimension. The classification of boundary conditions that can consistently be imposed on the boundary of the graph then provides important insights into the spectrum of the theory.
   
\begin{figure}[t]
\label{fig:intropic}
\centering
{\includegraphics[width=.4\linewidth]{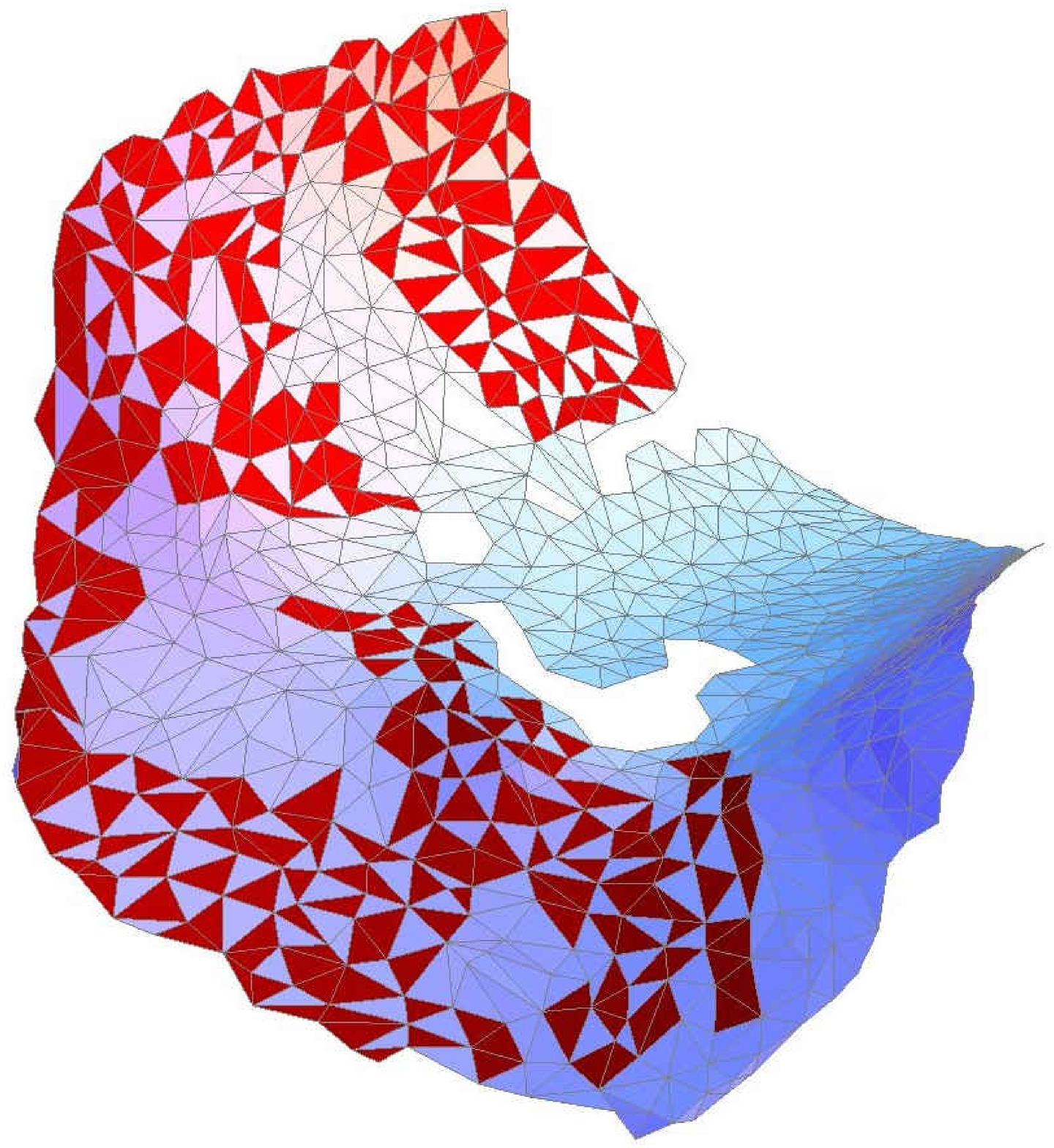}}
\hspace{.1\linewidth}
{\includegraphics[width=.4\linewidth]{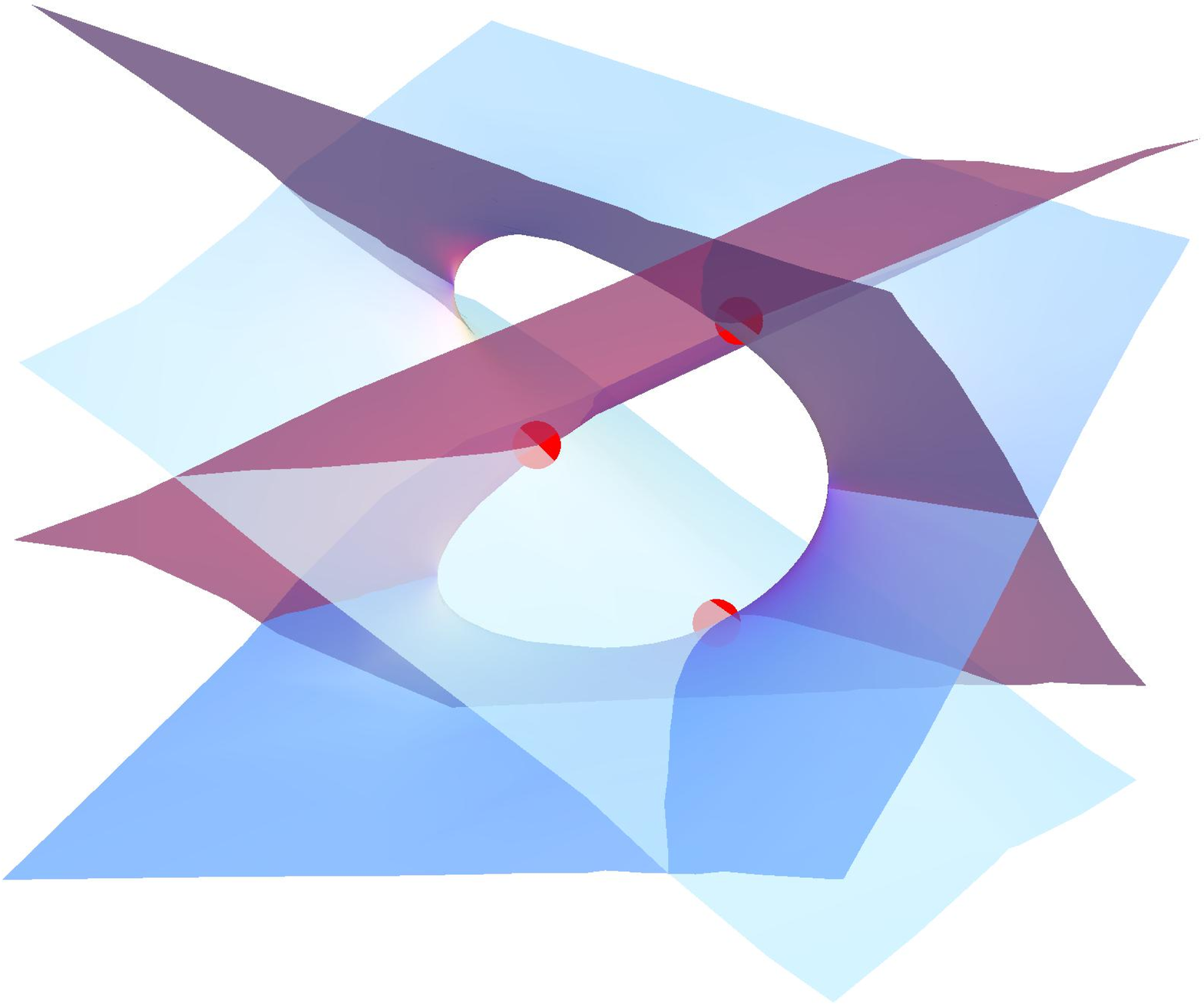}}
\caption{A configuration of the Ising model on a triangulated surface (left), and the Riemann surface arising from the analytic continuation at the critical point of the model (right).}
\end{figure}

   The intimate relationship between the above two subjects first emerged when `t Hooft observed that averages of infinite matrices admit an expansion in planar diagrams \cite{tHooft74}; the connection to the enumeration of planar maps was further fleshed out in the seminal works of Brezin et al. and Bessis et al. \cite{Brezin77,Bessis80}. In the eighties, David, Ambj{\o}rn et al. and Kazakov et al. exploited these insights to compute obervables in pure two-dimensional quantum gravity \cite{David85,Ambjorn85a,Ambjorn85b,Kazakov85} or equivalently, strings propagating in a zero-dimensional target space. These investigations revealed a powerful connection between the combinatorial problems at hand and algebraic geometry: for example, using random matrices, Boulatov and Kazakov discovered that the generating function for planar triangulations, weighted by the partition function of the Ising model defined thereon, can be obtained as the solution to a polynomial equation \cite{Boulatov86,Kazakov86}. As a result, upon analytic continuation, this generating function defines a Riemann surface called the \em spectral curve \em -- see Figure \ref{fig:intropic} for a cartoon of this correspondence.
      
     Shortly afterwards, Kazakov introduced a multi-matrix integral that describes the \em Potts model \em on a random lattice \cite{Kazakov88}, which is a generalisation of the Ising model dating back to \cite{Potts52}. In the same year, a breakthrough by Distler and Kawai allowed for the development of a complementary description of two-dimensional quantum gravity using Liouville conformal field theory \cite{Distler88}, leading to significant efforts to work out the correspondence to the random matrix description in the following decade: First, Lian and Zuckerman determined the physical Hilbert space of Liouville theory coupled to the so-called Virasoro minimal models \cite{LZ91}, of which the critical Ising model is a special case. Good agreement with the random matrix description was found following the solution of the so-called two-matrix model by Daul, Kazakov and Kostov \cite{Daul93}. These developments were preceeded by the discovery of the \em double scaling limit \em of matrix models by Douglas and Shenker \cite{Douglas89}, in which large maps of arbitrary genus contribute to the asymptoic behaviour. 
     
     Then in the mid-nineties, Daul found exact solutions for Kazakov's random matrix description of the Potts model on a random lattice \cite{Daul94}. Around the same time, Speicher reported on the connection between sums of \em independent \em random matrices and Voiculescu's free probability theory \cite{Speicher93,Speicher94}. Only a few years later, Carroll, Ortiz and Taylor calculated the partition function of the Ising model on the randomly triangulated disk for all independent boundary conditions by considering the average of the sum of two \em correlated \em random matrices \cite{Carroll96,Carroll97}, a task still to be completed for the Potts model on random planar maps. Indeed, for the Potts model on a \em fixed \em lattice, a complete set of boundary conditions was only described a year later by Affleck, Oshikawa and Saleur \cite{Affleck98}. With the advent of the new millenium, a classification of boundary conditions for Liouville theory was achieved by Fateev, the Zamolodchikov brothers and Teschner \cite{FZZ00,Teschner01,ZZ01}. Building on this, Seiberg and Shih subsequently developed the string theoretic interpretation of Liouville theory coupled to the Virasoro minimal model, in which the boundary conditions correspond to extended objects called \em branes \em \cite{Seiberg03,Seiberg04}. Curiously, a degeneracy in the states describing such boundary conditions was conjectured, implying that boundary conditions that are distinct for a matter system on a fixed background can be rendered indistinguishable when coupled to gravity. This conjecture was later challenged by matrix model calculations performed by Atkin, Wheater and Zohren \cite{Atkin11,Atkin12}.    
    
     Despite this being a fairly mature research field with a well-developed literature, a complete understanding of the allowed set of boundary conditions and their relationships has not yet been achieved even for simple models; the work described herein is an attempt to make progress towards filling these gaps. 

\section{Outline of the thesis}

 \noindent This thesis is organised as follows: In Chapter \ref{chap:intro}, we set the stage by introducing the concept of a matrix model and reviewing the description of boundaries and branes, employing the connections of random matrices to combinatorics, conformal field theory and string theory. After this prelude, the prerequisites are at hand to present the author's original contributions in three main chapters: 
 
 We begin by introducing Kazakov's multi-matrix model with permutation symmetry that describes the Potts model on a random lattice in Chapter \ref{chap:potts}. In this context, we generalise the work of Voiculescu and Speicher \cite{Voiculescu92,Speicher93} to the addition of \em correlated \em random matrices. Using Affleck et al.'s classification \cite{Affleck98} of boundary conditions for the Potts model as a guide, this enables us to compute the partition function of the model on the randomly triangulated disk for a whole family of boundary conditions exactly, thus extending the results \cite{Carroll96,Carroll97,Atkin11,Atkin12} for the Ising model obtained by Carroll and others. We deduce novel relationships between these boundary conditions and investigate the phase diagram to derive the scaling behaviour of the generating functions when the coupling constants approach a critical point. The results of this chapter have been reported in the publications \cite{ksm13,Atkin15}.
 
  In Chapter \ref{chap:cft}, we take advantage of the scale-invariance arising at the above-mentioned critical points to develop a description of the scaling functions using conformal field theory. The permutation symmetry of the matrix model and the conformal symmetry get enhanced to a larger continuous symmetry whose generators satisfy the so-called $\mathcal{W}$-algebra. We investigate the space of physical states for string worldsheets of both spherical and disk topology. On the sphere, our treatment extends the work of Lian, Zuckerman and Bouwknegt \cite{LZ91,Bouwknegt92} on Liouville theory coupled to the aforementioned minimal models, whose symmetries are captured by the smaller Virasoro algebra. Moreover, on the disk, the degeneracy of boundary conditions of the latter system observed by Seiberg and Shih \cite{Seiberg03} is found to persist in the more general case under study.
  
   In Chapter \ref{chap:wronskians}, employing the double scaling limit, we go beyond the conformal field theory description and study the description of branes non-perturbatively for specific cases. This will reveal novel and important differences that are not visible in the asymptotic expansion in the string coupling which follow from a careful counting of independent degrees of freedom. In particular, we find that the above-mentioned degeneracy is resolved upon inclusion of contributions from maps of unbounded genus and different boundary conditions capture truly independent degrees of freedom, thus potentially resolving the debate initiated in \cite{Atkin11,Atkin12}. The results of this chapter will be part of a forthcoming publication \cite{Wronskians}.
   
     Finally in Chapter \ref{chap:sum}, we summarise the key results that follow from the above investigations and comment on possible further applications and future developments.

\chapter{Review of the Hermitian Matrix Model}
\label{chap:intro}
\thispagestyle{title}

This chapter introduces the so-called Hermitian matrix model. This review will necessarily be incomplete and biased towards the applications in this thesis; more comprehensive reviews of these topics include \cite{Ginsparg93,DiFrancesco93,Ambjorn97}. We define the Hermitian matrix model as a probability measure for an $N\times N$ Hermitian matrix $X$,

\begin{equation}
\label{1-MM}
\D\mu(X)=\frac{1}{Z_N}\E^{-N\mathrm{tr}V(X)}\D X ,
\end{equation}

\noindent for a polynomial $V(x)=\sum_{m=2}^{k+2}t_m x^m/m$, where $\D X$ denotes the integration over independent components,

\begin{equation}
\mathrm{d}X=\prod_{1\leq i\leq j \leq N}\D \mathrm{Re}X\indices{^i_j}\prod_{1\leq i<j \leq N}\D\mathrm{Im}X\indices{^i_j}\ .
\end{equation}

\noindent Here, the \em partition function \em $Z_N$ normalises expectation values such that $\int\D\mu(X)=1$. The above measure is invariant under the adjoint action of the unitary group, in components

\begin{equation}
X\indices{^i_j}\longrightarrow \tilde{X}\indices{^i_j}=U\indices{^i_k}X\indices{^k_l}U\indices{_j^{l*}}\ ,\quad U\in U(N)\ .
\end{equation}

\noindent When $k=0$, $\mathrm{d}\mu(X)$ defines the so-called Gaussian unitary ensemble (GUE) -- one of Wigner's three ensembles\footnote{Besides the latter, these include the Gaussian orthogonal and symplectic ensembles, named analogously according to their respective symmetry groups.} \cite{Wigner55}. All results in this thesis pertain to the generalisation of this to the following probability measure on $q$ Hermitian matrices:

\begin{equation}
\label{pottsmeasure}
\D\mu(X_1,X_2,\dots X_q)=\frac{1}{Z_{N,q}}\prod_{\langle i j\rangle}\E^{N\mathrm{tr}X_i X_j}\times \prod_{i=1}^q\E^{-N\mathrm{tr}V_i(x)}\D X_i\ ,
\end{equation}

\noindent where $\langle ij \rangle$ denotes the product over distinct $i,j$. As the main Chapters \ref{chap:potts}, \ref{chap:cft} and \ref{chap:wronskians} will concern the so-called planar, scaling and double scaling limits of the model \eqref{pottsmeasure}, we introduce these limits in Sections \ref{int:planar}, \ref{int:scaling} and \ref{int:doublescaling} below for the simpler model \eqref{1-MM}. In doing so, we elucidate their connection to random planar maps, boundary conformal field theory and branes in string theory, respectively. Worked examples at the end of each section illustrate the relative ease with which results free of approximations can be obtained in these limits. 

\section{Planar Limit}
\label{int:planar}

\noindent In Chapter \ref{chap:potts}, we will be interested in the large-$N$ spectral density of sums of random matrices of the form $X_1+X_2+\dots X_p$, $1\leq p\leq q$, distributed according to \eqref{pottsmeasure}. To this end, we shall discuss the application of the saddle point method in the large $N$ limit in Subsection \ref{subsec:introspe}, first discussed in \cite{Brezin77}. To pave the way for the interpretation of the measure \eqref{pottsmeasure} as a description of the Potts model on a random lattice, we proceed to review the connection to statistical physics on planar surfaces in Subsection \ref{subsec:introstat}, which will reveal the origin of the term ``planar limit'' for $N\to\infty$. 

\subsection{Saddle point equations}
\label{subsec:introspe}
\noindent Given a Hermitian random matrix $X$, we woud like to compute  large-$N$ spectral density, defined as
\begin{equation}
\label{def-density}
\rho_X(x)=\lim_{N\to\infty}\frac{1}{N}\left\langle\sum_{i=1}^N\delta(x-x_i)\right\rangle\ ,
\end{equation}
\noindent where $\{x_i\}_{i=1}^N$ denote the eigenvalues of $X$. Introducing $U\in U(N)$ such that $UXU^{\dagger}$ is diagonal, we can integrate  out the off-diagonal components by performing the integration over the unitary group, allowing us to write the partition function as

\begin{equation}
\label{1MMdiagonalised}
Z_N=\frac{\mathrm{vol}\ U(N)}{\left(\mathrm{vol}\ U(1)\right)^N}\prod_{i=1}^N\int_\gamma\D x_i\ \E^{-NV(x_i)}\Delta^2(x)\ ,
\end{equation}

\noindent where $\Delta(x)=\det_{i,j}x_i^{j-1}$ is the Vandermonde determinant. To allow for odd values of $k$, we have generalised to normal matrices whose eigenvalues are supported on a one-dimensional cycle $\gamma\subset\mathbb{C}$. Note that when $k$ is even and $\mathrm{Re}\ t_{k+2}>0$, we can always let $X$ be Hermitian, i.e. choose $\gamma=\mathbb{R}$. The saddle points are then given by the eigenvalue configurations satisfying

 \begin{equation}
 \label{1-MM-SPE}
 V'(x_i)=\frac{2}{N}\sum_{j\neq i}\frac{1}{x_i-x_j}\ ,\quad 1\leq i\leq N\ .
  \end{equation}
 
  \noindent These are $N$ coupled algebraic equations, with a total of $N!\binom{N+k}{N}$ solutions, where the first factor arises from the invariance under permutations of eigenvalues, and the second from the dimensionality of the space of integration cycles. Since the left-hand side of \eqref{1-MM-SPE} is holomorphic, we can anlytically continue this equation for any $N$. At this stage it is useful to introduce the \em Stieltjes transform \em of $\rho_X$,
\begin{equation}
\label{stieltjes}
W_X(z)=\int_{\mathrm{supp}\ \rho_X}\D x\ \frac{\rho_X(x)}{z-x}\ ,\quad z\in\mathbb{C}\setminus\mathrm{supp}\ \rho_X\ .
\end{equation}
\noindent By construction, $W_X(z)$ computes the average of the trace of the resolvent $(z-X)^{-1}$ for large $N$:

\begin{equation}
\label{defn-discfct}
W_X(z)=\frac{1}{N}\left\langle\mathrm{tr}\frac{1}{z-X}\right\rangle+\mathcal{O}(1/N)\ .
\end{equation}

\noindent The normalised spectral density is obtained by inverting \eqref{stieltjes},
  
  \begin{equation}
  \label{stieltjesinv}
  \rho_X(x)=\frac{1}{\pi}\mathrm{Im}\ W_X(x)_+\ .
  \end{equation}

\noindent Here and in what follows we use the notation $f(z)_{\pm}=\lim_{\varepsilon\searrow 0}f(z\pm\I \varepsilon)$. We can consequently rewrite the saddle point equation \eqref{1-MM-SPE} as an equation for the Stieltjes transform of the spectral density,   
  
  \begin{equation}
  	\label{1MMspe}
  V'(z)=W_X(z)_++W_X(z)_-\ ,\quad z\in\mathbb{C}\ ,
  \end{equation}
  
  \noindent subject to the condition $\lim_{z\to\infty}zW_X(z)=1$. Writing
  
  \begin{equation}
  W_X(z)=\frac{1}{2}V'(z)-y(z)\ ,
  \end{equation}
  
  \noindent we are left to determine a function $y(z)$ holomorphic on $\mathbb{C}\setminus\mathrm{supp}\ \rho_X$. Throughout this thesis, we focus exclusively on those saddle points for which the spectral density has connected support\footnote{This poses an implicit restriction on the range of $t_m$.} as $N\to\infty$. Together with the requirement $y(z)=\frac{1}{2}V'(z)-z^{-1}+\mathcal{O}(z^{-2})$ for large $z$, this restriction fixes $y(z)$ entirely as the solution of a well-defined Riemann-Hilbert problem. 

\begin{ex} For the GUE ($k=0$), \eqref{1MMspe} implies
\begin{equation}
W_X(z)=\frac{t_2}{2}\left(z-\sqrt{z^2-4/t_2}\right)\ .
\end{equation}

\noindent From \eqref{stieltjesinv} it follows that the spectral density is given by the well-known semi-circle distribution
\begin{equation}
\label{semicircle}
\rho_X(x)=\begin{cases}
\frac{t_2}{2\pi }\sqrt{4/t_2-x^2}\ , &\quad |x|<\frac{2}{\sqrt{t_2}}\ ,\\
0\ ,&\quad |x|\geq \frac{2}{\sqrt{t_2}}\ .
\end{cases}
\end{equation}
\end{ex}

\begin{ex} When $V(z)$ is quartic ($k=2$) and even, we expect the support of the eigenvalue density to be symmetric under reflections of the real axis. Then $W_X(z)$ can be written as

\begin{equation}
\label{Wquartic}
W_X(z)=\frac{1}{2}\left(t_4z^3+t_2z-P(z)\sqrt{z^2-z_c^2}\right)\ ,
\end{equation}

\noindent where $P(z)$ is a quadratic polynomial and $z_c$ depends on $t_2$, $t_4$ only. Requiring $\lim_{z\to\infty}zW_X(z)=1$ fixes

\begin{equation}
P(z)=t_4z^2+\frac{1}{3}\left(2t_2+\sqrt{t_2^2+12t_4}\right)\ ,\quad z_c^2=\frac{2}{3t_4}\left(\sqrt{t_2^2+12t_4}-t_2\right)\ .
\end{equation}
\end{ex}

\subsection{Statistical physics on planar lattices}
\label{subsec:introstat}

\noindent The application of matrix integrals to the enumeration of random graphs was pioneered in \cite{tHooft74,Brezin77,Bessis80}. These developments have since been extended significantly and we refer the reader to \cite{DiFrancesco93,Ambjorn97} for a more comprehensive overview of these topics. In his seminal work \cite{tHooft74}, `t Hooft considered the new matrix integral obtained from $Z_N$ by expanding the exponential in the integrand and reversing the order of integration of summation:
\begin{equation}
\label{Zformal}
Z^{\mathrm{formal}}_N=\sum_{n=0}^\infty\frac{1}{n!}\int\D X\ \E^{-\frac{t_2}{2}N\mathrm{tr}X^2}\left(-N\mathrm{tr}V(X)\right)^n\ .
\end{equation}

\noindent Because each term is polynomial in $t_{m\geq 3}$, the above expression can be regarded as a formal power series in these parameters. The quantity \eqref{Zformal} is consequently referred to as a \em formal \em matrix integral \cite{Eynard11formal}, an a priori different quantity than the convergent expression \eqref{1MMdiagonalised}. Equally, throughout this section, we will regard averages $\langle \cdot\rangle$ as formal power series in the parameters $t_m$. An application of Wick's theorem tells us that \eqref{Zformal} can be evaluated by a sum over closed fatgraphs, in which a given graph $\mathcal{G}$ with $l$ internal lines and $n_m$ vertices of coordination number $m$ comes with a weight

\begin{equation}
\frac{N^{2-2h}}{|\mathrm{Aut}(\mathcal{G})|}t_2^{-l}\prod_{m=3}^{k+2}t_m^{n_m}\ ,
\end{equation}

\begin{figure}[t]
\centering
\begin{pspicture}(12,3.5)
\psset{unit=0.7cm}
\psset{arrowscale=1.5}

\psline[ArrowInside=->,ArrowInsidePos=0.7](0,2.1)(3,2.1)
\psline[ArrowInside=->,ArrowInsidePos=0.7](3,1.9)(0,1.9)
\rput(5,2){$=\frac{1}{Nt_2}\delta\indices{^i_l}\delta\indices{^k_j}$}
\rput(0,2.6){$i$}
\rput(3,2.6){$l$}
\rput(0,1.4){$j$}
\rput(3,1.4){$k$}

\psline[ArrowInside=->,ArrowInsidePos=0.7](9.9,4.24)(9.9,2.1)
\psline[ArrowInside=->,ArrowInsidePos=0.7](10.1,2.1)(10.1,4.24)
\psline[ArrowInside=->,ArrowInsidePos=0.7](9.9,2.1)(8,1.2)
\psline[ArrowInside=->,ArrowInsidePos=0.7](8.1,1)(10,1.9)
\psline[ArrowInside=->,ArrowInsidePos=0.7](10,1.9)(11.9,1)
\psline[ArrowInside=->,ArrowInsidePos=0.7](12,1.2)(10.1,2.1)
\rput(15.5,2){$=Nt_3\delta\indices{^j_k}\delta\indices{^l_m}\delta\indices{^n_i}$}
\rput(9.4,4.24){$j$}
\rput(10.6,4.24){$i$}
\rput(7.6,1.4){$k$}
\rput(8.4,0.6){$l$}
\rput(11.6,0.6){$m$}
\rput(12.4,1.4){$n$}

\end{pspicture}
\caption{Feynman rules for the matrix model with cubic potential; lines are oriented according to index positions.}
\label{fig:matrixrules}
\end{figure}

\noindent where $|\mathrm{Aut}(\mathcal{G})|$ is the order of the automorphism group of $\mathcal{G}$ and $h$ its genus. The dual graph is obtained by associating $m$-gons to $m$-valent vertices, with sides identified when connected by a propagator. In this way, the logarithm of \eqref{Zformal} enumerates maps, i.e. embeddings of connected graphs into surfaces; the parameters $\{t_m\}_{m=2}^{k+2}$ are the respective fugacities of $m$-gons in the map. Each series coefficient is polynomial in $N$ and at leading order in $1/N$, only planar graphs contribute to average\footnote{Note that this is inequivalent to the genuine $1/N$ expansion of convergent integrals as encountered in the next section.} -- hence the name ``planar limit'' for taking $N\to\infty$. Because the number of planar maps is exponentially bounded, averages will have a convergent power series expansion as $N\to\infty$ for $t_m$ of small enough modulus.

To see how the description of boundaries in the random graph can be achieved, note that applying Wick's theorem to the expansion of the first derivative of the planar free energy per degree of freedom

\begin{equation}
\label{momentsseries}
\frac{1}{N}\langle\mathrm{tr}X^j\rangle=\frac{1}{N^2}\frac{\partial}{\partial t_j}F_0+\mathcal{O}(1/N^2)\ ,\qquad F_0=-\lim_{N\to\infty}\ln Z_N^{\mathrm{formal}}\ ,
\end{equation}

\noindent gives a sum over all maps with $n_m$ $m$-gons and one marked $j$-gon called the \em root\em, whose links define the boundary of $\mathcal{G}$. Since the Stieltjes transform $W_X(z)$ of the spectral density is the generating function for the moments

\begin{equation}
\label{moments}
\lim_{N\to\infty}\frac{1}{N}\langle\mathrm{tr} X^j\rangle=\frac{1}{2\pi\I}\oint_C \D z\ z^j W_X(z)\ ,
\end{equation}

\noindent where the contour $C$ encloses $\mathrm{supp}\ \rho_X$ counter-clockwise, we see that $W_X(z)$ may be understood as the generating function for planar maps with connected boundary, i.e. maps of disk topology, where $z$ is the fugacity of a boundary link. For this reason, $W_X(z)$ is also referred to as the \em disk function\em. The partition function of the model on the disk is defined by dividing by the order of the automorphism group of the boundary -- which is simply the number of boundary links -- at each order in $z$. Equivalently, $W_X(z)$ is the first derivative of the disk partition function:

\begin{equation}
\label{Zdisk}
\frac{1}{N}\frac{\partial}{\partial z}\langle\mathrm{tr} \ln (z-X)\rangle=W_X(z)+\mathcal{O}(1/N)\ .
\end{equation}

\begin{ex} When $V$ is cubic ($k=1$), $Z_N^{\mathrm{formal}}$ is a power series in the single variable $t_3$,

\begin{equation}
\label{Zformalcubic}
Z^{\mathrm{formal}}_N=\sum_{n=0}^\infty\frac{(-1)^n}{n!}\left(\frac{Nt_3}{3}\right)^n \int\D X\E^{-\frac{t_2}{2}N\mathrm{tr} X^2}\left(\mathrm{tr}X^3\right)^n\ .
\end{equation}

\noindent The resulting Feynman rules are depicted in Figure \ref{fig:matrixrules}; an example of a term proportional to $t_3^4N^2$ arising from Wick's theorem applied to the above series in Figure \ref{fig:spheregraph}. 

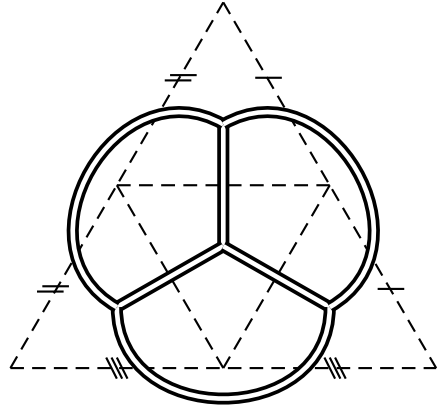
\begin{figure}[t]
\centering
\begin{pspicture}(6,6)
\psset{unit=0.7cm}

\psline[linestyle=dashed](1,1)(9,1)
\psline[linestyle=dashed](1,1)(5,7.928)
\psline[linestyle=dashed](5,7.928)(9,1)

\psline[linestyle=dashed](3,4.464)(7,4.464)
\psline[linestyle=dashed](3,4.464)(5,1)
\psline[linestyle=dashed](7,4.464)(5,1)

\psline(2.9,1.2)(3.1,0.8)
\psline(3,1.2)(3.2,0.8)
\psline(2.8,1.2)(3,0.8)

\psline(7.1,1.2)(7.3,0.8)
\psline(6.9,1.2)(7.1,0.8)
\psline(7,1.2)(7.2,0.8)

\psline(1.5,2.45)(2.0,2.45)
\psline(1.6,2.55)(2.1,2.55)

\psline(3.9,6.45)(4.4,6.45)
\psline(4,6.55)(4.5,6.55)

\psline(8.4,2.5)(7.9,2.5)
\psline(6.1,6.5)(5.6,6.5)

\psline[doubleline=true,linewidth=0.07](5,3.309)(5,5.618)
\psline[doubleline=true,linewidth=0.07](3,2.155)(5,3.309)
\psline[doubleline=true,linewidth=0.07](7,2.155)(5,3.309)

\psbezier[doubleline=true,linewidth=0.07](7,2.155)(7,-0.155)(3,-0.155)(3,2.155)
\psbezier[doubleline=true,linewidth=0.07](3,2.155)(1,3.309)(3,6.772)(5,5.618)
\psbezier[doubleline=true,linewidth=0.07](7,2.155)(9,3.309)(7,6.772)(5,5.618)
\end{pspicture}
\caption{Example of a Feynman graph dual to a triangulation of the sphere.}
\label{fig:spheregraph}
\end{figure}

\end{ex}

\begin{ex} When $V$ is quartic ($k=2$) and even, $Z_N^{\mathrm{formal}}$ is a power series in the single variable $t_4$. Using  \eqref{moments} for the explicit result \eqref{Wquartic} and setting $t_2=1$ without loss of generality, one recovers Tutte's famous result \cite{Tutte63} for the generating function of rooted planar quadrangulations:

\begin{equation}
\label{tuttequadrangulations}
\lim_{N\to\infty}\frac{1}{N}\langle\mathrm{tr}X^4\rangle=\sum_{n=0}^\infty (-t_4)^{n-1}\frac{2}{n!}\frac{3^n(2n)!}{(n+2)!}\ .
\end{equation}
\end{ex}

\section{Scaling limit}
\label{int:scaling}

\noindent In Chapter \ref{chap:potts}, we will derive scaling limit of averages computed with \eqref{pottsmeasure}; in Chapter \ref{chap:cft}, we will study the same system using conformal field theory. To set the stage, Subsections \ref{subsec:phasediagram} and \ref{subsec:cft} therefore introduce this limit and discuss its connection to conformal field theory, respectively.

\subsection{Phase diagram and critical points}
\label{subsec:phasediagram}
\noindent The description of continuous surfaces via the critical behaviour of large random matrices dates back to the seminal works of David, Ambj{\o}rn, Kazakov and collaborators \cite{David85,Ambjorn85a,Ambjorn85b,Kazakov85}. The planar limit may be understood as the thermodynamic limit of the eigenvalue statistics: at infinite $N$, averages are non-analytic in the fugacities $\{t_m\}_{m=2}^{k+2}$, which span a $k$-dimensional phase diagram. Consequently, power series like \eqref{momentsseries} generally have a finite radius of convergence. Tuning a single fugacity $t_j$ towards a critical value $t_{j,c}$ such that we approach a $(k-1)$-dimensional critical submanifold, the resulting universal behaviour can be characterised by the scaling exponent $\gamma_j$ of the second derivative of the planar free energy,

\begin{equation}
\label{susc}
\frac{\partial^2}{\partial t_j^2}F_0=\mathrm{const.}\times(t_j-t_{j,c})^{-\gamma_j}+\textrm{ terms analytic in}\ t_j\ .
\end{equation}

\noindent Generally, we see from  \eqref{momentsseries} that then

\begin{equation}
\frac{1}{N}\langle \mathrm{tr}X^j\rangle =\mathrm{const.}\times(t_j-t_{j,c})^{1-\gamma_j}+\textrm{ terms analytic in}\ t_j\ ,
\end{equation}

\noindent which implies that the number of rooted planar maps with $n_j$ $j$-gons is exponentially bounded by a constant multiple of $t_{j,c}^{-n_j}(n_j)^{\gamma_j-2}$ for large $n_j$. In particular, when $\gamma_j>0$, the terms analytic in $t_j$ are subleading and the average number of $j$-gons contributing to the average in \eqref{susc} diverges linearly with the distance from the critical point\footnote{When $\gamma_j<0$, the analogous conclusions follow after taking sufficiently many derivatives with respect to $t_j$, that is, for maps with a sufficient number of marked $j$-gons.}:

\begin{equation}
\langle n_j\rangle=\frac{\partial}{\partial t_j}\ln\left(\frac{\partial^2}{\partial t_j^2} F_0\right)\propto \frac{\gamma_j}{t_j-t_{j,c}}\quad \mathrm{as}\  t_j\to t_{j,c}\ .
\end{equation}

\noindent Assigning a fixed length $\varepsilon$ to each edge in the dual of $\mathcal{G}$ thus yields an expectation value of the dimensionful surface area proportional $\varepsilon^2\langle n_j\rangle$. Then the \em scaling limit \em is obtained by sending $t_j\to t_{j,c}$, $\varepsilon\to 0$, keeping $\mu=(t_j-t_{j,c})/\varepsilon^2$ and hence the dimensionful area fixed. For this reason, $\mu$ is also referred to as the renormalised \em cosmological constant\em. More generally, for higher-order critical behaviour near lower-dimensional critical submanifolds, the exact relationship between $\mu$ and the fugacities $\{t_m\}_{m=2}^{k+2}$ depends more intrically on the direction in which we approach the critical submanifold in question and was worked out by Moore et al. \cite{Moore91}, who coined the term \em conformal background \em for the scaling limit in which no other couplings besides $\mu$ are nonzero. A widely-used diagnostic discriminating between different universality classes is the scaling exponent $\gamma_s$ of the susceptibility

\begin{equation}
\label{susceptibilityscaling}
\frac{\partial^2}{\partial\mu^2}F_0=\mathrm{const}.\times \mu^{-\gamma_s}+\textrm{ terms analytic in}\ \mu \ .
\end{equation}

\noindent For a single random matrix with the measure \eqref{1-MM}, we can at most arrange for an algebraic singularity of the form $\gamma_s=1/2-k$; a more general algebraic singularity $\gamma_s=-(k+2)^{-1}$ can arise when we choose the measure \eqref{pottsmeasure} with $q=2$ \cite{Daul93}.

 For graphs with boundaries, we can similarly let the average number of boundary links diverge: as $N\to\infty$, the disk function $W_X(z)$ develops a branch cut located at the support of the eigenvalue density and the corresonding power series in $z$ displays a finite radius of convergence $z_c$. Recalling that $z$ is the fugacity of a boundary link, we introduce by analogy the \em boundary cosmological constant \em via $z=z_c(1-\varepsilon\mu_B)$. For $0>\gamma_s>-1$, the expansion of $W_X(z)$ near the critical point is then of the form 
 
\begin{equation}
\label{Wexpansion}
W_X(z)=W_X(z_c)+\varepsilon c_1\mu_B+\varepsilon^{1-\gamma_s} c_2\left.\frac{\partial D(\mu,\mu_B)}{\partial \mu_B}\right|_{\mu}+\dots ,
\end{equation} 

\noindent where the leading non-analytic term is the universal first derivative of the \em continuum \em disk partition function, and $c_1$ and $c_2$ are non-universal constants. It will turn out convenient to express the above in the dimensionless variables

\begin{equation}
\label{dimensionless}
\zeta=\frac{\mu_B}{\sqrt{\mu}}\ ,\quad Q=\mu^{(\gamma_s-1)/2}\left.\frac{\partial D(\mu,\mu_B)}{\partial \mu_B}\right|_{\mu}\ .
\end{equation}

\begin{ex} 
\label{ex-scaling}
Consider again the case $k=2$ with even $V$. Upon setting $t_2=1$ and applying Stirling's approximation to the coefficients in \eqref{tuttequadrangulations}, one finds the number of rooted planar quadrangulations grows asymptotically as $2/\sqrt{\pi}\times 12^nn^{-5/2}$, which implies $t_{4,c}=-1/12$ and $\gamma_s=-1/2$; this universality class describes a random planar surface which has been called the ``Brownian map'' in the mathematics literature \cite{legall13}. At the critical point, the spectral density \eqref{def-density} is proportional to $(8-x^2)^{3/2}$, so $z_c=\sqrt{8}$. Approaching this point via the parametrisation

\begin{equation}
t_4=t_{4,c}(1-\varepsilon^2\mu)\ ,\quad z=z_c(1+\varepsilon\mu_B/2)\ ,\quad \varepsilon\searrow 0\ ,
\end{equation}

 \noindent we find the following expansion of the disk function:

\begin{equation}
W_X(z)=\frac{\sqrt{2}}{3}-\varepsilon\frac{1}{\sqrt{2}}\mu_B+\varepsilon^{3/2}\frac{\sqrt{2}}{3}(2\mu_B-\sqrt{\mu})\sqrt{\mu_B+\sqrt{\mu}}+\mathcal{O}(\varepsilon^2)\ .
\end{equation}

\noindent In the dimensionless variables \eqref{dimensionless}, the above can be succinctly summarised as the solution to $T_3(\zeta)-T_2(Q)=0$, where $T_n(x)$ denotes the $n^{\mathrm{th}}$ Chebyshev polynomial of the first kind.
\end{ex}

\begin{rem} Had we considered cubic $V$ ($k=1$), $W_X(z)$ would have become the generating function for rooted planar triangulations. The scaling limit would have given different values of $z_c$, $c_1$ and $c_2$, but the same value of $\gamma_s$, and the same form $\zeta$ and $Q$ due to the universality of the Brownian map. 
\end{rem}

\subsection{Conformal field theory description}
\label{subsec:cft}

\noindent Generally, the existence of critical points suggests that the universal properties of the scaling limit can be captured by a scale-invariant field theory. To see how this expectation is borne out in Chapter \ref{chap:cft}, we will first need to recall some crucial results and widespread terminology. A general introduction to two-dimensional conformal field theory is \cite{DiFrancesco97}, the foundations of which will be assumed to be familiar to the reader. Generalities of BRST cohomology in the context of string theory are reviewed in \cite{Polchinski94}; for an overview on $\mathcal{W}$-algebras in conformal field theory, see \cite{Bouwknegt92b}.  

Distler and Kawai pioneered the definition of the measure on the space of physically distinct configurations of the continuum surface taking advantage of the fact that any two-dimensional metric $g$ may be written in the \em conformal gauge \em $g=f^*(\E^\varphi \hat{g})$, where the $f^*$ denotes the action of a diffeomorphism and the \em background metric \em $\hat{g}$ is specified by a unique point in \em moduli space \em -- the finite-dimensional compact space of two-dimensional metrics modulo diffeomorphisms and local Weyl transformations \cite{Distler88}. This change of variables contributes a Jacobian to the measure which is the product of a contribution from the non-invariance of the field measures under Weyl transformations, leading to the appearance of the Liouville action for the scalar field $\varphi$ and a determinant which as usual can be written as a functional integral over Grassmann-valued ``ghost'' fields $b$, $c$ of spin -1 and 2; the measure then displays a residual gauge invariance under the subset of diffeomorphisms which preserve $g$ up to a local Weyl transformation. In complex coordinates $z=x+\I y$, $\bar{z}=x-\I y$, the algebra of the latter is enlarged by the infinitely many additional generators $\ell_n=-z^{n+1}\partial_z$, $n\in\mathbb{Z}$ of two copies of the \em Witt algebra\em

\begin{equation}
\label{virasoro}
[\ell_m,\ell_n]=(m-n)\ell_{m+n}\ ,
\end{equation}

\noindent whose subalgebra with $|n|<2$ exponentiates to the \em conformal group \em $SL(2,\mathbb{C})/\mathbb{Z}_2$. One way to compute observables is via the \em conformal bootstrap\em, taking advantage of the fact that invariance under \eqref{virasoro} implies differential equations for correlation functions. Another route represents the algebra \eqref{virasoro} explicitly on a Fock space, which is the procedure we shall employ. On the latter, \eqref{virasoro} is represented only up to a phase, that is by operators $L_n$ generating the \em Virasoro algebra\em, which includes a central term

 \begin{equation}
 \label{virasoroc}
 \frac{c}{12}(m^3-m)\delta_{m+n,0}
 \end{equation}
 
  \noindent in addition to the right-hand side of \eqref{virasoro}. In particular, when the statistical system defined on the random surface approaches its critical point, the Liouville field $\varphi$ interacts with another conformal field theory, frequently referred to as the \em matter \em theory and the total \em central charge \em is given by $c=c_L+c_M-26$, where $c_L$ and $c_M$ denote the central charges of the two systems, offset by the negative contribution of the ghost fields. However, since here \eqref{virasoro} describes an algebra of residual gauge transformations arising from partial gauge fixing, it must be respected exactly so that $c$ must vanish, thus fixing $c_L$ given $c_M$. This allows for an interpretation of this theory as a description of bosonic strings propagating in a $c_M+1$-dimensional target space. For this reason, we will sometimes also refer to the random surface as the \em worldsheet \em of the string. As we shall see below, for $c_M\notin[1,25]$, the susceptibility exponent is determined by the celebrated Knizhnik-Polyakov-Zamolodchikov (KPZ) relation \cite{knizhnik88}:

\begin{equation}
\label{KPZ}
\gamma_s=\frac{1}{12}\left(c_M-1-\sqrt{(c_M-1)(c_M-25)}\right)\ .
\end{equation}

\noindent The subspace of physically distinct states -- i.e. those states in the Fock space that do not differ by just a gauge transformation -- is then obtained from the cohomology of the nilpotent Becchi-Rouet-Stora-Tyutin (BRST) operator, whose holomorphic component is expressed in terms of the respective stress-energy tensors and the anticommuting ghost field as\footnote{We omit a total derivative $3\partial^2c/2$ which has to be added to ensure $J$ transforms as a tensor.}

\begin{equation}
\label{brst}
\mathrm{d}=\oint\frac{\mathrm{d}z}{2\pi \I}J(z)\;,\qquad J(z)=\ :\left(T_M(z)+T_L(z)+\frac{1}{2}T_{gh}(z)\right)c(z):\;,
\end{equation} 

\noindent where we introduced standard notation for the normal ordering of free quantum fields \cite{DiFrancesco97},

\begin{equation}
:\mathcal{O}_1(z)\mathcal{O}_2(z):\ =\lim_{w\to z}\left(\mathcal{O}_1(z)\mathcal{O}_2(w)-\left\langle\mathcal{O}_1(z)\mathcal{O}_2(w)\right\rangle\right)\ .
\end{equation}

\noindent Finite representations of \eqref{virasoro} with $c_M\leq 1$ have been classified and are labelled by pairs of simply laced Dynkin diagrams \cite{Cappelli86}, constituing the Hilbert space of the so-called \em minimal models\em. In particular, the values of $\gamma_s$ mentioned in the previous section identify the universality classes describing highest critical point of a single random matrix with the $(A_1,A_{2k})$ minimal model with $c_M=10-6[k+2/(2k+1)]$ and those of \eqref{pottsmeasure} with $q=2$ with the $(A_{k+1},A_{k+2})$ minimal model with with $c_M=1-6/[(k+1)(k+2)]$, respectively coupled to Liouville theory; these will arise as special cases of the construction outlined below. For an explicit expression for the BRST operator and the space it acts on, let us fix some definitions for the matter, Liouville and ghost systems. Because in Chapter \ref{chap:cft} we are looking for a conformal field theory description of the scaling limit for \eqref{pottsmeasure}, we will consider the $\mathcal{W}_q$ minimal model as defining the matter theory, which reduces to the aforementioned minimal models for $q=1$ and $q=2$, respectively. For definiteness, we begin by discussing each system on the Riemann sphere, allowing us to choose the flat background metric $\hat{g}=\D z\D\bar{z}$ and drop the integration over moduli, before moving on to the inclusion of a single boundary. 
\\

\noindent \em Matter sector. \em The $\mathcal{W}_q$ minimal models are a family of rational conformal field theories dating back to 
\cite{FateevZamolodchikov87} that can be specified by a positive integer $q$ and a pair of coprime integers $(p,p')$; here we describe their free-field realisation paralleling the presentation in \cite{Caldeira03,Caldeira04}. Their action functional can be represented as

\begin{equation}
\begin{aligned}
\label{S_M}
S_M[\phi,\hat{g}]&=\frac{1}{8\pi}\int_{\mathbb{CP}^1}\mathrm{d}^2x\sqrt{\det \hat{g}}\ \left(\hat{g}^{\alpha\beta}\partial_{\alpha}\phi\cdot\partial_{\beta}\phi+2\I Q_0R[\hat{g}]\ \rho\cdot\phi \right)\;,
\end{aligned}
\end{equation}

\noindent where $\phi=(\phi^1,\dots \phi^{q-1})$ is a vector in the root space of $SU(q)$, $\rho$ is the Weyl vector, $R$ denotes the Ricci scalar and $Q_0=(p'-p)/\sqrt{pp'}$. Variation of \eqref{S_M} with respect to the background metric yields the stress-energy tensor, whose holomorphic component reads

\begin{equation}
\label{stressenergymatter}
T_M(z)=-\frac{1}{2}:\partial\phi\cdot\partial\phi:+\I Q_0\rho\cdot\partial^{2}\phi\;.
\end{equation} 

\noindent We shall group zero modes with the holomorphic field components, expanding the fields $\phi^i(z,\bar{z})=\phi^i(z)+\bar{\phi}^i(\bar{z})$ as

\begin{equation}
\phi^i(z)=\phi^i_0-\I a^i_0\ln z+\I \sum_{n\neq 0}\frac{a^i_n}{n} z^{-n}\ ,\quad \bar{\phi}^i(\bar{z})=-\I a^i_0\ln \bar{z}+\I \sum_{n\neq 0}\frac{\bar{a}^i_n}{n} \bar{z}^{-n}\ .
\end{equation} 

\noindent Adopting an orthogonal cartesian basis in root space, the commutation relations read

\begin{equation}
[\phi^i_0,a_n^j]=\I\delta_{n,0}\delta^{ij} \;, \qquad [a_n^i,a_m^j]=n\delta_{n+m,0}\delta^{ij}\;.
\end{equation} 

\noindent These imply that the generators

\begin{equation}
\label{Lnmatter}
L_n^M=\frac{1}{2}\sum_{m\in\mathbb{Z}}: a_m\cdot a_{n-m}:-(n+1)Q_0 \rho\cdot  a_n
\end{equation}

\noindent obey the Virasoro algebra\footnote{The Virasoro algebra generated by \eqref{Lnmatter} is merely a subalgebra of the larger, non-linear $\mathcal{W}_q$-algebra generated by the chiral spin-$s$ currents, $s=2\dots q$. These currents are primary with respect to Virasoro, but arise as descendants of the $\mathcal{W}_q$ vacuum module.} with central charge 

\begin{equation}
\label{cmatter}
c_M=(q-1)\left(1-\frac{(p'-p)^2}{pp'}q(q+1)\right)\;,
\end{equation}

\noindent where we used the Freudenthal-de-Vries `strange formula' $ \rho\cdot \rho=(q^3-q)/12$. The allowed highest-weight states are obtained by the action of the chiral vertex operators $V_{\alpha}^M(z)=\ :\exp(\I  \alpha\cdot \phi(z)):$ on the $SL(2,\mathbb{C})$-invariant vacuum,

\begin{equation}
|\lambda\rangle_M=\lim_{z\to0}\ V^M_{Q_0 \rho-\frac{1}{\sqrt{pp'}} \lambda}(z)|0\rangle_M\;,\qquad a_n^i|\lambda\rangle_M=0\ \forall\ n>0\;.
\end{equation}

\noindent In the above $ \lambda$ is short-hand for the $SU(q)$ representation weight $ \lambda = (p'r^i-ps^i) \omega_i$, where $\omega_i$ denotes the dual weight corresponding to the simple root $e_i$, satisfying

 \begin{equation}
  e_i\cdot  \omega_j =\delta_{ij}\ ,\quad  \omega_i\cdot \omega_j=\frac{i(q-j)}{q}\ ,\quad i \leq j\ .
 \end{equation}
 
   \noindent From the operator product expansion (OPE) with $T_M(z)$, we find that the primary field corresponding to $ \lambda$ has conformal weight

\begin{equation}
\label{matterweight}
L_0^M| \lambda\rangle_M=\left(\frac{ \lambda^2}{pp'}-Q_0^2\rho^2\right)| \lambda\rangle_M\;.
\end{equation}

\noindent Consistency of the fusion rules requires that $r^i$ and $s^i$ are positive integers satisfying
 
\begin{equation}
\label{Wprimaries}
\sum_{i=1}^{q-1} r^i<p\;, \qquad \sum_{i=1}^{q-1} s^i<p'\;.
\end{equation}

\noindent These restrictions identify the sets $[r^1,\dots,r^{q-1}]$ and $[s^1,\dots,s^{q-1}]$ as Dynkin labels for representations of $\widehat{\mathfrak{su}}(q)_k$ and $\widehat{\mathfrak{su}}(q)_{k+1}$, respectively, with the additional node of the affine diagram omitted and the level $k$ given by

\begin{equation}
\label{affinelevel}
k=\frac{p}{p'-p}-q\ .
\end{equation}

\noindent In particular, when $k$ is a positive integer and $q\geq 2$, all primary fields come with a positive weight and the model is unitary. Since the action of the Weyl group of $SU(q)$ leaves the root space inner product invariant, the weight \eqref{matterweight} is invariant under $\lambda\to w\lambda$, for $w$ an element of the symmetric group $S_q$ of order $q!$. Consequently the conditions \eqref{Wprimaries} still leave degeneracy in the spectrum: we need to restrict $\lambda$ further to a fundamental domain $\mathcal{B}^{(q)}_{p,p'}$ to avoid overcounting\footnote{See \cite{Caldeira04} for an explicit derivation of $\mathcal{B}^{(q)}_{p,p'}$.}. For each such $\lambda\in\mathcal{B}^{(q)}_{p,p'}$, we then define the Fock space

\begin{equation}
\label{Fockmatter}
\mathcal{F}_M(\lambda)=\mathrm{span}\left\{\prod_{i=1}^{q-1}\prod_{j=1}^{k_i}a^i_{-n_j^{(i)}}|\lambda\rangle_M\ |\ k_i\geq0,\ 0<n_1^{(i)}\leq\dots\leq n_{k_i}^{(i)}\right\}\;,
\end{equation}

\noindent which is a reducible $\mathcal{W}_q$-module. To obtain the irreducible $\mathcal{W}_q$-module $\mathcal{M}(\lambda)$ defining the subspace of physical states, we introduce the so-called Felder complex $(\mathcal{C}(\lambda),\D')$, where $\D'$ is a nilpotent operator assembled from integrals of products of field exponentials acting on

\begin{equation}
\label{felder}
\mathcal{C}(\lambda)=\bigoplus_{N^i\in\mathbb{Z}}\bigoplus_{w\in S_q}\mathcal{F}_M\left(\lambda^w-pp'N^i e_i\right)\;,
\end{equation}

\noindent with $\lambda^w=p'r^iw\omega_i-ps^i\omega_i$. The complex is called a \em resolution \em of $\mathcal{M}(\lambda)$; that is, the $n^{\mathrm{th}}$ cohomology cohomology group reads $H^{n}(\mathcal{C}_M(\lambda),\D')=\delta_{n0}\mathcal{M}(\lambda)$, where $n$ denotes the grading with respect to $\D'$ \cite{Bouwknegt92b}. The partition function thus obtained agrees with that obtained from other definitions of $\mathcal{M}(\lambda)$ \cite{Bouwknegt91}. This construction has first been described in detail for $q=2$ in \cite{Felder88} and for $q=3$ in \cite{Mizoguchi91}. The main feature of importance for Chapter \ref{chap:cft} in this construction is that the above-mentioned degeneracy in the definition of $\lambda$ implies that there exists more than one resolution for given $\mathcal{M}(\lambda)$.
\\

\noindent  \em Liouville sector. \em Liouville theory is governed by the action functional\footnote{Our normalisation of the Liouville field differs by a factor of $\sqrt{2}$ from \cite{Ginsparg93} and keeps our conventions close to the free-field treatments in \cite{DiFrancesco97,Bouwknegt92}.}

\begin{equation}
\begin{aligned}
\label{SLiouville}
S_L[\varphi,\hat g]&=\frac{1}{8\pi}\int_{\mathbb{CP}^1}\mathrm{d}^2x\sqrt{\det\ \hat g}\left(\hat{g}^{\alpha\beta}\partial_{\alpha}\varphi\partial_{\beta}\varphi+2Q_LR[\hat{g}]\varphi +8\pi\mu \E^{\sqrt{2}b\varphi}\right)\ .
\end{aligned}
\end{equation}

\noindent Requiring invariance under Weyl transformations fixes the background charge in terms of the Liouville coupling $b$ as $Q_L=(b+b^{-1})/\sqrt{2}$; the cosmological constant $\mu$ then corresponds to a marginal deformation. The relationship \eqref{KPZ} can now be inferred by inspecting the change of the action under translations in field space: according to the Gauss-Bonnet theorem, upon shifting $\varphi\to\varphi+\sigma$, the second term in \eqref{SLiouville} contributes $2Q_L\sigma$ to the change of the action. Moreover, choosing $\sigma=-\ln\mu/(\sqrt{2}b)$ renders the third term in \eqref{SLiouville} independent of $\mu$. Together with the invariance of the remaining contributions to the measure, this implies that the partition function on the sphere obeys $F_0(\mu)=\mu^{2-\gamma_s}F_0(1)$, where $\gamma_s=1-b^{-2}$; expressing $b$ as a function of $c_M$ then yields \eqref{KPZ}. Since the short-distance behaviour of fields is controlled by large negative values of $\varphi$, we do not expect the exponential interaction term to affect the expression for the central charge and conformal weights. This allows us to set $\mu=0$ throughout the remainder of this section. The holomorphic component of the stress-energy tensor is then

\begin{equation}
\label{stressenergyliouville}
T_L(z)=-\frac{1}{2}\ :\partial\varphi\partial\varphi:\ +Q_L\partial^{2}\varphi\ .
\end{equation}

\noindent Expanding the holomorphic component of $\varphi$ as

\begin{equation}
\varphi(z)=\varphi_0-\I\alpha_0\ln z+\I\sum_{n\neq 0}\frac{\alpha_n}{n} z^{-n}\;
\end{equation} 

\noindent results in the mode commutation relations 

\begin{equation}
\label{L-modes}
[\varphi_0,\alpha_n]=\I\delta_{n,0}\;, \qquad [\alpha_n,\alpha_m]=n\delta_{n+m,0}\;.
\end{equation} 

\noindent As a result, the generators

\begin{equation}
\label{LnLiouville}
L_n^L=\frac{1}{2}\sum_{m\in\mathbb{Z}}\ :\alpha_m\alpha_{n-m}:\ +\I Q_L(n+1)\alpha_n
\end{equation}

\noindent obey the Virasoro algebra with central charge

 \begin{equation}
 \label{cLiouville}
 c_L=1+12Q_L^2\ .
 \end{equation} 
 
 \noindent From the $SL(2,\mathbb{C})$-invariant vacuum we obtain a continuous family of highest-weight states $|P\rangle_L$ by acting with vertex operators $V^L_{\alpha}(z)=\ :\exp(\alpha\varphi(z)):$

\begin{equation}
\label{Liouvilleprimary}
|P\rangle_L=\lim_{z\to0}\ V^L_{Q_L+\I P}(z)|0\rangle_L\;,\qquad \alpha_n|P\rangle_L=0\ \forall\ n>0\;.
\end{equation}

\noindent The OPE with $T_L(z)$ reveals that the corresponding bulk fields are spinless primaries of weight

\begin{equation}
L_0^L|P\rangle_L=\frac{1}{2}(Q_L^2+P^2)|P\rangle_L\;,
\end{equation}

\noindent which is evidently invariant under the reflection $P\to -P$. The states \eqref{Liouvilleprimary} are normalisable iff $P\in\mathbb{R}$ - the corresponding operators then create macroscopic loops in the worldsheet as we will discuss below. On the other hand, operators with $\I P\in\mathbb{R}$ create non-normalisable states; on the latter, one must impose the Seiberg bound $\I P\geq0$ when $\mu> 0$ to avoid double-counting \cite{Seiberg90}. For given $P$, we define the Feigin-Fuchs module

\begin{equation}
\label{FockLiouville}
\mathcal{F}_L(P)=\mathrm{span}\left\{\prod_{i=1}^k \alpha_{-n_i}|P\rangle_L\ |\ k\geq 0,\  0<n_1\leq\dots\leq n_k\right\}\;.
\end{equation}
\\

\noindent \em Ghost sector. \em The action for the ghost system arising form the partial gauge-fixing of worldsheet diffeomorphisms is

\begin{equation}
\left.S_{gh}[b,c]\right|_{\hat{g}=\D z\D\bar{z}}=\frac{1}{2\pi}\int_{\mathbb{CP}^1}\D^2 z(b\bar{\partial}c+\bar{b}\partial\bar{c})\ ,
\end{equation}

\noindent with corresponding holomorphic stress-energy tensor 

\begin{equation}
\begin{aligned}
\label{stressenergyghost}
T_{gh}(z)&=\ :(2b\partial c +c\partial b):\;.
\end{aligned}
\end{equation}

\noindent The mode expansion of the holomorphic fields and commutation relations read

\begin{equation}
c(z)=\sum_{n\in\mathbb{Z}}c_n z^{-n+1}\;,\qquad b(z)=\sum_{n\in\mathbb{Z}}b_n z^{-n-2}\;,
\end{equation}

\begin{equation}
\{b_n,c_m\}=\delta_{n+m,0}\;.
\end{equation} 

\noindent so that the generators\footnote{The normal ordering constant in $L^{gh}_n$ is determined by $2L_0|0\rangle_{gh}=[L_1,L_{-1}]|0\rangle_{gh}=-2|0\rangle_{gh}$.}

\begin{equation}
\begin{aligned}
\label{Lnghost}
L^{gh}_n&=\sum_{m\in\mathbb{Z}}(m-n)\ :c_{-m}b_{m+n}:-\delta_{n,0}\;
\end{aligned}
\end{equation}

\noindent obey the Virasoro algebra with central charge $c_{gh}=-26$. Of the two possible ground states, we shall define the ghost vacuum by $b_{n-1}|0\rangle_{gh}=c_n|0\rangle_{gh}=0$ for $n>0$ and normalise the ghost number $\sum_{n\in\mathbb{Z}}:c_{-n}b_n:$ such that $|0\rangle_{gh}$ has ghost number zero. By repeated action of the creation operators on this state we generate the Fock space

\begin{equation}
\label{Fockghost}
\mathcal{F}_{gh}=\mathrm{span}\left\{\prod_{i=1}^kc_{-n_i}\prod_{j=1}^lb_{-m_j}|0\rangle_{gh}\ |\ k,l\geq 0\ ,\ 0<n_1<\dots<n_k,\ 0<m_1\dots<m_l\right\}\;.
\end{equation} 
\\

\noindent Let us now turn to the description of boundaries in the present context. Conformally invariant boundary conditions were determined in \cite{FZZ00,ZZ01,Teschner02} for Liouville theory and in \cite{Caldeira03,Caldeira04} for the $\mathcal{W}_q$ minimal model. Let us consider worldsheets with disk topology, and for concreteness we shall map the disk interior to the upper half plane $\{z\in\mathbb{CP}^1|\text{Im}\ z>0\}$ such that the boundary is located at $z=\bar{z}$.
\\

\noindent \em Matter sector. \em We briefly summarise the free field construction of the matter boundary states given in \cite{Caldeira04}. Let us define the coherent states

\begin{equation}
\label{coherentstate}
|B(\lambda)\rangle_\Lambda=\exp\left(\sum_{m>0}\frac{1}{m} a^T_{-m}\cdot\Lambda\cdot\bar{a}_{-m}\right)\lim_{z,\bar{z}\to 0}V^M_{Q_0\ \rho-\frac{1}{\sqrt{pp'}} \lambda}(z)\bar{V}^M_{Q_0 \rho+\frac{1}{\sqrt{pp'}} \lambda}(\bar{z})|0\rangle_{M}\;.
\end{equation}

\noindent The two possible forms of the $(q-1)\times(q-1)$ matrix $\Lambda$ compatible with conformal symmetry were determined in \cite{Caldeira03,Caldeira04}: either $\Lambda=-\mathbb{I}$ or $\Lambda=w_{\rho}$, where $w_{\rho}$ is the longest element of the Weyl group \footnote{Only the former choice of $\Lambda$ additionally preserves the higher-spin symmetries.}.

 The matter Ishibashi states corresponding to a bulk primary $\lambda$ are given by a sum of such coherent states over the Felder complex,

\begin{equation}
\label{matterishibashi}
|\lambda;\Lambda\rishi_M=\sum_{w\in S_q}\sum_{N^i\in\mathbb{Z}}\kappa^w_N\left|B(\lambda^w-pp'N^i e_i)\right\rangle_\Lambda\ ,
\end{equation}

\noindent with $ \lambda^w$ defined as in \eqref{felder} and $\kappa^w_N$ a pure phase which will be unimportant for our discussion. For each $\lambda\in\mathcal{B}^{(q)}_{p,p'}$, we obtain two physical boundary states that additionally obey the Cardy condition \cite{Cardy89}, one for each of the allowed choices of $\Lambda$:

\begin{equation}
\label{mattercardy}
\begin{aligned}
|\lambda\rangle_\mathrm{C}&=\sum_{\lambda^{\prime}\in\mathcal{B}}\Psi^*_{\lambda}(\lambda^{\prime})|\lambda^{\prime};-\mathbb{I}\rishi_M\;,\\
|\tilde{\lambda}\rangle_\mathrm{C}&=\sum_{\lambda^{\prime}\in\mathcal{B}}\Psi^*_{\lambda}(\lambda^{\prime})|\lambda^{\prime};w_\rho\rishi_M\;,
\end{aligned}
\end{equation}

\noindent where the one-point function the primary field $\lambda^{\prime}$ on the disk with boundary condition $\lambda$ is given in terms of the modular $S$-matrix 

\begin{equation}
\label{modsmatrix}
S_{\lambda\lambda^{\prime}}=\frac{Q_0^{q-1}}{\sqrt{\det A}}\sum_{w\in S_q}\sum_{w^{\prime}\in S_q}\det w\ \exp\left(2\pi \I Q_0^2\lambda^{\prime}\cdot w^{\prime}(p'r_0w-ps_0)\lambda\right)\ ,
\end{equation}

\noindent as $\Psi_{\lambda}(\lambda^{\prime})=S_{\lambda\lambda^{\prime}}/\sqrt{S_{\rho\lambda^{\prime}}}$, with $\lambda=\rho$ the identity field and $A$ the Cartan matrix of $SU(q)$. In the above, $r_0$ and $s_0$ denote the unique\footnote{The uniqueness of $r_0$ and $s_0$ is a consequence of B\'ezout's idenity.} pair of integers integers satisfying $1\leq r_0\leq p-1$, $1\leq s_0\leq p'-1$ and $p'r_0-ps_0=1$. Note that states with $\Lambda=w_\rho$ may decouple so that in general we do not obtain two boundary states per primary field as \eqref{mattercardy} might suggest. For example, for $(q,k)=(3,1)$, there are 6 primary fields, but only 8 independent boundary states \cite{Caldeira03}.
\\

\noindent \em Liouville sector. \em In presence of a boundary, the boundary cosmological constant $\mu_B$ arises as an additional marginal coupling as the Liouville action has to be augmented by a boundary term, which in our coordinates is simply

\begin{equation}
\begin{aligned}
\label{liouvillebdry}
\mu_B\int_{\mathbb{R}}\mathrm{d}s\ \E^{b\varphi}\;.
\end{aligned}
\end{equation}

\noindent The parameter $\mu_B$ labels a family of Neumann boundary conditions on $\varphi$ along the real axis,

\begin{equation}
\label{LiouvilleNeumann}
\I(\partial-\bar{\partial})\varphi=4\pi\mu_B\ \E^{b\varphi}\ .
\end{equation}

\noindent At sufficiently strong coupling $b$, we need to account for the presence of the semiclassically invisible, marginal counterterms \cite{Teschner01}

\begin{equation}
\label{liouvillecounter}
\tilde{\mu}\int_{\text{Re}\ z,\bar{z}>0}\mathrm{d}^2z\ \E^{2\varphi/b}\ ,\qquad \tilde{\mu}_B\int_{\mathbb{R}}\mathrm{d}s\ \E^{\varphi/b}\ ,
\end{equation}

\noindent rendering the action \eqref{SLiouville} invariant under the \em strong-weak duality \em transformation $(b,\mu,\mu_B)\to(b^{-1},\tilde{\mu},\tilde{\mu_B})$. The boundary conditions thus defined by the dimensionless ratios

\begin{equation}
\label{bdrycc2}
\zeta^2=\frac{\mu_B^2}{\mu}\sin(\pi b^2)\ , \quad \eta^2=\frac{\tilde{\mu}_B^2}{\tilde{\mu}}\sin(\pi /b^2)\ ,
\end{equation}

\noindent can be parametrised by a single variable $\sigma$, which we define as 

\begin{equation}
\label{bdrycc}
\zeta=\cosh(\pi b\sigma)\ ,\quad \eta=\cosh(\pi\sigma/b)\ .
\end{equation}

\noindent The Liouville Ishibashi states are given by

\begin{equation}
|P\rishi_L=\ \exp\left(-\sum_{k>0}\frac{1}{k}\alpha_{-k}\bar{\alpha}_{-k}\right)\lim_{z,\bar{z}\to 0}V^L_{Q_L+\I P}(z)\bar{V}^L_{Q_L-\I P}(\bar{z})|0\rangle_L\;,
\end{equation}

\noindent where now $P$ is real. The physical boundary state corresponding to the boundary condition \eqref{bdrycc} was worked out in \cite{FZZ00,Teschner02}
\begin{equation}
\label{BLiouville}
|\sigma\rangle_{\text{FZZT}}=\int_{0}^\infty\mathrm{d}P\ \Psi^*_{\sigma}(P)|P\rishi_L\ ,
\end{equation}

\noindent where the disk one-point function with boundary condition $\sigma$ was found to be

\begin{equation}
\Psi_{\sigma}(P)=\mu^{-\I P/b}\frac{\cos(2\pi\sigma P)}{\I P} \Gamma(1+2\I P/b)\Gamma(1+2\I Pb)\ ,
\end{equation}

\noindent which is manifestly invariant under the strong-weak duality. In the context of string theory, this defines the so-called Fateev-Zamolodchikov-Zamolodchikov-Teschner (FZZT) brane. An infinite discrete family of Dirichlet boundary conditions has been found in \cite{ZZ01} -- the Zamolodchikov-Zamolodchikov (ZZ) brane -- but will not be of concern in this thesis. 
\\

\noindent \em Ghost sector. \em For completeness, we finally spell out the conformally invariant boundary state for the ghost system \cite[p.226]{Polchinski98}

\begin{equation}
\label{Bghost}
|B\rangle_{gh}=(c_0+\bar{c}_0)\exp\left(-\sum_{k>0}(b_{-k}\bar{c}_{-k}+\bar{b}_{-k}c_{-k})\right)|0\rangle_{gh}\ .
\end{equation}

\section{Double scaling limit}
\label{int:doublescaling}

\noindent In Chapter \ref{chap:wronskians}, we study averages of products of characteristic polynomials in the double-scaling limit, employing the operator approach developed by Douglas \cite{Douglas89} and applied to the Hermitian two-matrix model by Daul, Kazakov and Kostov \cite{Daul93}. Below we shall define the double scaling limit in the context of this formalism. The application of the theory of monodromy preserving deformations of linear ordinary differential equations \cite{JMU81a,JMU81b,JMU81c} to the present context was developed by Moore \cite{Moore90a,Moore90b} -- see also \cite{Chan:2013wya} for a more recent discussion. Further details on this formalism can be found in the reviews \cite{Ginsparg93,DiFrancesco93}; the connection to the theory of integrable systems is reviewed in \cite{Morozov94}. 

In the previous section, we saw that planar maps are an exponentially bounded family, yielding convergent expressions for series expansions in $\{t_m\}_{m=3}^{k+2}$ for generic values of the latter. Resting on the fact that generating functions for maps of fixed positive genus display the same radius of convergence, the topological recursion algorithm \cite{Eynard07} yields finite answers for the free energy $F_h$ for maps of arbitrary genus $h$ from the initial data at $h=0$. For large $N$, the free energy to all orders can then be estimated by an asymptotic series\footnote{Though finite at each order in $1/N$, this estimate is formal because this series, neglecting exponentially small corrections, has vanishing radius of convergence.}, in the following denoted by `$\simeq$':

\begin{equation}
\label{Zasympt}
Z_N\simeq\exp\left(\sum_{h=0}^\infty N^{2-2h}F_h\right)\ ,\quad N\to \infty\ .
\end{equation} 

\noindent At the boundary of the domain of analyticity, the scaling relation \eqref{susceptibilityscaling} generalises for the genus-$h$ free energy as

\begin{equation}
F_h=\varepsilon^{(2-\gamma_s)(2-2h)}\mathcal{F}_h(\mu)\ ,
\end{equation}

\noindent where $\mu$ is the renormalised cosmological constant as defined in the previous section. Introducing the \em string coupling \em $g_s=\varepsilon^{\gamma_s-2}/N$, we recast \eqref{Zasympt} as an asymptotic expansion in $g_s$,

\begin{equation}
\label{asymptDSL}
Z_N\simeq\exp\left(\sum_{h=0}^\infty g_s^{2h-2}\mathcal{F}_h(\mu)\right)\ ,\quad g_s\to 0\ .
\end{equation}

\noindent We can thus retain significant contributions from all topologies even at large $N$ by taking the double scaling limit $N\to\infty$, $\varepsilon\to 0$, keeping $g_s$ fixed.

To study this limit effectively, we need a handle on contributions from all worldsheet topologies that does not rely on the asymptotic expansion \eqref{asymptDSL} about zero string coupling. One way to achieve this is to employ the operator approach  \cite{Douglas89} for the measure \eqref{pottsmeasure} with $q=2$ and $\mathrm{deg}\ V_1=p$, $\mathrm{deg}\ V_2=p'$ for coprime integers $p$ and $p'$, whose scaling limit about the highest critical point is described by the $(A_{p-1},A_{p'-1})$ minimal model coupled to Liouville theory \cite{Daul93}. In this context, it is useful to study the exponentiation of the operator $N^{-1}\mathrm{tr}\ln(x-X)$ encountered in equation \eqref{Zdisk} of Subection \ref{subsec:introstat}: denoting the contribution from connected surfaces by $\langle \cdot \rangle_c$, the latter can be written as

\begin{equation}
\begin{aligned}
\langle \det(x-X)\rangle&=\exp\left(\left\langle \E^{\mathrm{tr}\ln (x-X)}-1\right\rangle_c\right)\\
&\simeq \exp\left(\sum_{n=1}^{\infty}\frac{1}{n!}\left\langle\left(\mathrm{tr}\ln(x-X)\right)^n\right\rangle_c\right)\ .
\end{aligned}
\end{equation}

\noindent The large-$N$ expansion of the above can be represented graphically as

\begin{equation}
\begin{aligned}
\label{determinantgraphical}
 \langle\det(x-X)\rangle\simeq\exp\left[N\vcenter{\hbox{
\begin{pspicture*}(0,0)(.55,.5)
\psset{unit=.5cm}
\psellipse(.5,.5)(.3,.5)
\pscurve(.5,0)(1,.25)(1,.75)(.5,1)
\end{pspicture*}
}}+\vcenter{\hbox{
\begin{pspicture*}(0,0)(.9,1.25)
\psset{unit=.5cm}
\psellipse(.5,.5)(.3,.5)
\psellipse(.5,2)(.3,.5)
\pscurve(.5,1)(.75,1.25)(.5,1.5)
\pscurve(.5,0)(1.5,.5)(1.5,2)(.5,2.5)
\end{pspicture*}

}}+
 \frac{1}{N}\left(\vcenter{\hbox{
\begin{pspicture*}(0,-.1)(1.5,1.1)
\psset{unit=.5cm}
\psellipse(.5,1)(.3,.5)
\pscurve(.5,.5)(1,.5)(2,0)(2.6,.5)(2.85,1)(2.6,1.5)(2,2)(1,1.5)(.5,1.5)
\pscurve(1.75,1.1)(2,0.85)(2.25,1.1)
\pscurve(1.85,1)(2,1.1)(2.15,1)
\end{pspicture*}
}}+\frac{1}{3!}\vcenter{\hbox{\begin{pspicture*}(0,0)(.9,2)
\psset{unit=.5cm}
\psellipse(.5,.5)(.3,.5)
\psellipse(.5,2)(.3,.5)
\psellipse(.5,3.5)(.3,.5)
\pscurve(.5,1)(.75,1.25)(.5,1.5)
\pscurve(.5,2.5)(.75,2.75)(.5,3)
\pscurve(.5,0)(1.5,.75)(1.5,3.25)(.5,4)
\end{pspicture*}
}}\right)
 +\mathcal{O}\left(\frac{1}{N^2}\right)\right]\ ,
\end{aligned}
\end{equation}

\noindent thus accounting for an arbitrary number of boundaries with the same boundary condition labelled by $x$. In the context of string theory, this expectation value is therefore interpreted as the partition function of a brane at target space position $x$. More generally, the principal objects of interest are averages of characteristic polynomials labelled by $1\leq n\leq N$,

\begin{equation}
\begin{aligned}
\label{charpol}
\alpha_n(x)&=\left\langle\det(x-X_1)\right\rangle_{n\times n}\ , \qquad \beta_n(y)&=\left\langle\det(y-X_2)\right\rangle_{n\times n}\ ,
\end{aligned}
\end{equation}

\noindent where $\langle \cdot \rangle_{n\times n}$ denotes the average with respect to the measure on $n\times n$ minors,

\begin{equation}
\begin{aligned}
\label{measure-n}
\D\mu_{n\times n}(X_1,X_2)&=\frac{1}{Z_n}\E^{-N\mathrm{tr}[V_1(X_1)+V_2(X_2)-X_1X_2]}\D^{n^2} X_1\ \D^{n^2} X_2\ ,\\
\D^{n^2}X&=\prod_{1\leq i\leq j \leq n}\D \mathrm{Re}X\indices{^i_j}\prod_{1\leq i<j \leq n}\D\mathrm{Im}X\indices{^i_j}\ .
\end{aligned}
\end{equation}

\noindent Then the \em Baker-Akhiezer functions\em

\begin{equation}
\begin{aligned}
\label{wavefct}
\psi_n(x)=\frac{\E^{-NV_1(x)}}{\sqrt{h_n}}\alpha_n(x)\ ,\quad \chi_n(y)=\frac{\E^{-NV_2(y)}}{\sqrt{h_n}}\beta_n(y)\ ,
\end{aligned}
\end{equation}

\noindent are bi-orthonormal, i.e. $\int\mathrm{d}x\ \mathrm{d}y\ \psi_n(x)\chi_m(y)\E^{xy}=\delta_{mn}$ for a suitable choice of the normalisation constant $h_n$, and obey recursion relations of the form

\begin{equation}
\begin{aligned}
\label{recursionrel}
& x \psi_n(x)=P_n(z)\psi_n(x)\ , \qquad &\frac{1}{N}\frac{\partial}{\partial x}\psi_n(x)&=-Q_n(z)^T\psi_n(x)\ ,\\ & y \chi_n(y)=Q_n(z)\chi_n(y)\ , \qquad &\frac{1}{N}\frac{\partial}{\partial y}\chi_n(y)&=-P_n(z)^T\chi_n(y)\ ,\\
\end{aligned}
\end{equation}

\noindent where the difference operators $P_n$ and $Q_n$ have an expansion in powers of $z=\exp(-\partial_n)$ and satsify $[P_n,Q_n]=1/N$. Here, the transpose is defined by $(f(n)\E^{s\partial_n})^T=\E^{-s\partial_n}f(n)$. Introducing the scaling variables

\begin{equation}
\label{def-dsl}
g_s=\varepsilon^{-\frac{p+p'}{p}}/N\ , \quad t=\varepsilon^{-\frac{p+p'-1}{p}}(N-n)/N\ ,
\end{equation}

\noindent so that $\varepsilon^{-1/p}\partial_n=-g_s\partial_t$ and taking the double scaling limit $N\to\infty$, $\varepsilon\to 0$, keeping $g_s$ and $t$ finite, the difference operators become

\begin{equation}
\begin{aligned}
P_n(z)=x_c+\varepsilon\mathbb{P}(t;\partial_t)\ , \quad Q_n(z)^T=y_c-\varepsilon^{p'/p}\mathbb{Q}(t;\partial_t)\ ,\quad \partial_t\equiv g_s\frac{\partial}{\partial t}
\end{aligned}
\end{equation}

\noindent where the $p^{\mathrm{th}}$ and $(p')^{\mathrm{th}}$ order differential operators can be brought to the form

\begin{subequations}
\begin{align}
\mathbb{P}(t;\partial_t)&=2^{p-1}\partial_t^p+\sum_{n=2}^pu_n(t)\partial_t^{p-n}\ ,\\
\mathbb{Q}(t;\partial_t)&=\beta_{p,p'}\left(2^{p'-1}\partial_t^{p'}+\sum_{n=2}^{p'}v_n(t)\partial_t^{p'-n}\right)\ ,
\end{align}
\end{subequations}

\noindent where $\beta_{p,p'}=(-1)^{p+p'}\beta_{p',p}$ is a real constant and the coefficients $\{u_n(t)\}_{n=2}^p$, $\{v_n(t)\}_{n=2}^{p'}$ solve the \em string equation \em \cite{Douglas89}

\begin{equation}
\label{stringeqn}
[\mathbb{P},\mathbb{Q}]=g_s\ .
\end{equation}

\noindent Upon a suitable rescaling of $\mu_B=\varepsilon^{-1}(x-x_c)$, $\widetilde{\mu}_B=\varepsilon^{-p'/p}(-1)^{p+1}(y-y_c)$ and introducing the dimensionless variables $\zeta=\mu_B/\sqrt{\mu}$, $\eta=\beta_{p,p'}^{-1}\widetilde{\mu}_B/\sqrt{\widetilde{\mu}}$, the functions \eqref{wavefct} then satisfy the overdetermined couples of differential equations

\begin{subequations}
\label{psi-system}
\begin{align}
\label{Ppsi}
\zeta \psi(t;\zeta)&=\mathbb{P}(t;\partial_t)\psi(t;\zeta)\ ,\\
\label{Qpsi}
\partial_\zeta\psi(t;\zeta)&=\mathbb{Q}(t;\partial_t)\psi(t;\zeta)\ ,\quad \partial_\zeta\equiv g_s\frac{\partial}{\partial \zeta}
\end{align}
\end{subequations}

\noindent and similarly, with $(f(t)\partial^n)^T=(-\partial)^nf(t)$,

\begin{subequations}
\label{chi-system}
\begin{align}
\eta\chi(t;\eta)&=\frac{(-1)^{p'}}{\beta_{p,p'}}\mathbb{Q}^T(t;\partial_t)\chi(t;\eta)\ ,\\
\partial_\eta \chi(t;\eta)&=(-1)^{p'}\beta_{p,p'}\mathbb{P}^T(t;\partial_t)\chi(t;\eta)\ ,\quad \partial_\eta\equiv g_s\frac{\partial}{\partial \eta}
\end{align}
\end{subequations}

\noindent whose compatibility is expressed by \eqref{stringeqn}. In the theory of integrable systems, the operators $(\mathbb{Q},\mathbb{P})$ are often called the \em Lax pair \em and the set of differential equations for $u_n(t)$ and $v_n(t)$ summarised by \eqref{stringeqn} is referred to as the \em $p^{\mathrm{th}}$ reduction of the Kadomtsev-Petviashvili (KP) hierarchy\em. They are invariant under 
\begin{enumerate}
\item the $SL(2,\mathbb{C})$-family of linear canonical transformations 
\begin{equation}
\label{linearcanonical}
(\mathbb{P},\mathbb{Q})\longmapsto(a\mathbb{P}-c\mathbb{Q},d\mathbb{Q}-b\mathbb{P})\ ,\quad \det\begin{pmatrix}a & b\\ c & d\end{pmatrix}=1\ ,
\end{equation}
\item the ``charge conjugation'' 
\begin{equation}
\label{chargeconj}
(\mathbb{P},\mathbb{Q})\longmapsto(\mathbb{P}^T,-\mathbb{Q}^T)\ .
\end{equation}
\end{enumerate}

\noindent Each $\gamma\in SL(2,\mathbb{C})$ with $b\neq 0$ can be represented by an integral transform acting on functions of $\zeta$,
\begin{equation}
\gamma[f](\eta)=\E^{-\frac{d}{2b}\eta^2}\int\mathrm{d}\zeta\ f(\zeta) \E^{-\frac{1}{b}(a\zeta^2/2-\eta\zeta)} \ .
\end{equation}
\noindent The particular case $a=d=0$, $c=-b$ yields the Laplace transform, for which we reserve the notation
\begin{equation}
\label{laplace}
\mathcal{L}_b [f](\eta)=\int\mathrm{d}\zeta \E^{\eta\zeta/b}f(\zeta)\ ,
\end{equation}

\noindent and drop the subscript when $b=-1$. A symmetry of particular importance in Chapter \ref{chap:wronskians} is the \em duality \em transformation $(p,p')\to(p',p)$ as first introduced in this context in \cite{FKN92}. From \eqref{measure-n} we see this amounts to the interchange of the matrices $X_1$ and $X_2$ -- a definition that extends to finite $N$ and has been considered in \cite{BEH01a,BEH01b,BEH02,BEH04} under the name of \em spectral duality\em. We observe that the effect of this transformation on the differential equations \eqref{psi-system} and \eqref{chi-system} is, up to a coefficient, the composition of the charge conjugation \eqref{chargeconj} and a Laplace transform \eqref{laplace}, namely

\begin{equation}
\label{duality}
(\mathbb{P},\mathbb{Q})\longmapsto\left((-1)^{p'}\beta_{p,p'}^{-1}\mathbb{Q}^T,(-1)^{p'}\beta_{p,p'}\mathbb{P}^T\right)\ ,
\end{equation}  

\noindent and thus preserves the string equation \eqref{stringeqn}. 

We close this section by introducing a few concepts originating from the application of the inverse monodromy problem to \eqref{psi-system} and \eqref{chi-system}, which will be useful for the definition of the spectral curve and the analysis of the semiclassical limit $g_s\to 0$. In brief, defining the $p$- and $p'$-vectors

\begin{subequations}
\begin{align}
\vec \psi (t;\zeta)&=\left(\psi(t;\zeta),\partial_\zeta\psi(t;\zeta),\dots,\partial^{p-1}_\zeta\psi(t;\zeta)\right)^T\ ,\\
\vec \chi (t;\eta)&=\left(\chi(t;\eta),\partial_\eta\chi(t;\eta),\dots,\partial^{p'-1}_\eta\chi(t;\eta)\right)^T\ ,
\end{align}
\end{subequations}

\noindent the relations \eqref{psi-system} can be expressed as linear differential systems with $p\times p$ resp. $p'\times p'$ matrix-valued coefficients that are rational functions\footnote{These relations are also frequently expressed in other variables such as $\lambda=(2\zeta)^{1/p}/2$ \cite{Chan:2013wya}.},

\begin{subequations}
\begin{align}
\label{isomonodromysystem}
& \partial_\zeta\vec \psi(t;\zeta)=\mathcal{Q}(t;\zeta)\vec \psi(t;\zeta)\ ,\quad & \partial_t\vec \psi(t;\zeta)=\mathcal{B}(t;\zeta)\vec \psi(t;\zeta)\ ;\\
& \partial_\eta\vec \chi(t;\eta)=\widetilde{\mathcal{Q}}(t;\eta)\vec \chi(t;\eta)\ ,\quad & \partial_t\vec \chi(t;\eta)=\widetilde{\mathcal{B}}(t;\eta)\vec \chi(t;\eta)\ .
\end{align}
\end{subequations}

\noindent Accordingly, the spectrum of the Lax operators $\mathbb{P}$, $\mathbb{Q}$ and their dual images under \eqref{duality} can conveniently be encoded in the zero locus of the polynomials

\begin{subequations}
\begin{align}
\label{spectralcurvedsl}
G(t;\zeta,Q)&=\det(Q\mathbb{I}_{p\times p}-\mathcal{Q}(t;\zeta))\ ,\\
\widetilde{G}(t;\eta,P)&=\det(P\mathbb{I}_{p'\times p'}-\widetilde{\mathcal{Q}}(t;\eta))\ ,
\end{align}
\end{subequations}

\noindent which defines a Riemann surface $\mathcal{C}_{p,p'}(t)=\{(\zeta,Q)\in\mathbb{C}^2|G(t;\zeta,Q)=0\}$, called the \em spectral curve\em\footnote{This is to be contrasted with the proposal to define a `quantum' spectral curve by generalising the notion of a Riemann surface by allowing non-commuting coordinates in accordance with \eqref{stringeqn}, see e.g. \cite{Moore90a,Moore90b,Ambjorn05}.}. In the semiclassical limit $g_s\to 0$, the coefficients $v_n(t)$ and $u_n(t)$ are approximately constant. To compare this limit to the results of conformal field theory, we need to choose the conformal background \cite{Moore91} described in Subsection \ref{subsec:phasediagram}. This limit was first computed in \cite{Daul93}, with the result\footnote{Here and often in what follows, we suppressed the unimportant real constant $\beta_{p,p'}$.}
 
 \begin{equation}
 \label{spectral-conf-backgr}
 \lim_{g_s\to 0}G(t;\zeta,Q)=\frac{1}{2^{p-1}}\left(T_p(Q)-T_{p'}(\zeta)\right)\ ,
 \end{equation}
 
\noindent where $T_n(x)$ denotes the $n^{\mathrm{th}}$ Chebyshev polynomial. This corresponds to the algebraic equation satisfied by the dimensionless disk amplitude \eqref{dimensionless}; as will become evident in later chapters, the duality transformation $(p,p')\to(p',p)$ then reduces to the strong-weak duality $b\to b^{-1}$ of the Liouville part of the conformal field theory.

\begin{ex} $(p,p')=(2,1)$. This describes the double scaling limit of the GUE. The recursion relations \eqref{recursionrel} are solved by the $n^{\mathrm{th}}$ Hermite polynomial $H_n$,

\begin{equation}
\langle\det(x-X)\rangle_{n\times n}=\left(\frac{1}{2t_2}\right)^{n/2}H_n\left(x\sqrt{\frac{t_2}{2}}\right)\ .
\end{equation}

\noindent Using the integral representation of the latter \cite{Gradshteyn07} to write

 \begin{equation}
\langle\det(x-X)\rangle_{n\times n}=\sqrt{\frac{t_2}{2\pi}}\int_\mathbb{R}(x+\I z)^n\E^{-t_2z^2/2}\D z
 \end{equation}
 
  \noindent and taking the double scaling limit with $z=t_2^{-1/2}(\I-\varepsilon^{1/2}g_s^{1/3}s)$ and $g_s$, $t$ as in \eqref{def-dsl}, this becomes the so-called Airy function, 

\begin{equation}
\psi(t;\zeta)=\mathrm{Ai}\left(g_s^{-2/3}(\zeta+t)\right)\ ,\quad \mathrm{Ai}(x)=\frac{1}{2\pi}\int_\mathbb{R}\E^{\I(s^3/3+xs)}\D s\ .
\end{equation}

\noindent Upon rescaling $g_s\to g_s/\sqrt{2}$, this indeed solves \eqref{psi-system} with
\begin{equation}
\mathbb{P}=2\partial_t^2+u_2(t)\ ,\quad \mathbb{Q}=\partial_t\ ,
\end{equation}

\noindent where the string equation \eqref{stringeqn} demands $\dot{u}_2(t)=-1$. The spectral curve is then given by the zeroes of

\begin{equation}
G(t;\zeta,Q)=Q^2-\frac{1}{2}(\zeta+t)\ .
\end{equation}
\end{ex}

\begin{ex} 
\label{intex:puregrav}
$(p,p')=(3,2)$. The universality class of the critical cubic matrix model is controlled by \eqref{psi-system} with
\begin{equation}
\mathbb{P}=4\partial_t^3+u_2(t)\partial_t+u_3(t)\ ,\quad \mathbb{Q}=\beta_{3,2}\left(2\partial_t^2+v_2(t)\right)\ .
\end{equation}

\noindent The string equation \eqref{stringeqn} requires $u_2(t)=3v_2(t)$ and $u_3(t)=3\dot{v}_2(t)/2$, where $v_2(t)$ solves the first Painlev\'e equation: $\ddot{v}_2(t)=6v_2(t)+t$. The spectral curve is given by the zeroes of 
\begin{equation}
G(t;\zeta,Q)=Q^3-\frac{\zeta^2}{2}-Q\left(\frac{3v_2^2}{4}+\frac{\ddot{v}_2}{2}\right)-\frac{v_2^3}{4}-\frac{v_2\ddot{v}_2}{2}+\frac{\dot{v}_2^2}{2}\ ,
\end{equation}

\noindent see also Example \ref{ex-puregrav} in Subsection \ref{subsec:diff} of Chapter \ref{chap:wronskians}. In the limit $g_s\searrow 0$, $v_2(t)=-1+\mathcal{O}(g_s)$ so that to leading order, the above becomes $\left(T_3(Q)-T_2(\zeta)\right)/4$. After a duality transformation $(p,p')\to(p',p)$, this describes the universality classs encountered in Example \ref{ex-scaling} of the previous section.
\end{ex}

\chapter{Sums of Random Matrices and the Potts Model on Planar Maps}
\label{chap:potts}
\thispagestyle{title}
\section{Overview}
\label{sec:overview}

\noindent We compute the partition function $W_{(p)}(z)$ of the $q$-states Potts model \cite{Potts52} on a random planar lattice with $1\leq p\leq q$ allowed, equally weighted colors on a connected boundary, where $z$ denotes the fugacity of a boundary link. In the particular cases $q=2$ and $q=3$, all of these correspond to boundary conditions of the Ising and Potts lattice models that were found to be integrable on a fixed lattice by Behrend and Pearce \cite{Behrend00}. To this end, we employ its matrix model formulation \eqref{pottsmeasure} as proposed long ago by Kazakov \cite{Kazakov88}, who used it to solve the limits $q\to 0$ and $q\to 1$.

 In the particular case of random \em triangulations\em, $W_{(1)}(z)$ was first found by Daul \cite{Daul94} and later Zinn-Justin \cite{ZJ99} in the saddle point approximation for integer $0\leq q\leq 4$, and by Bonnet and Eynard \cite{Bonnet99,Eynard99} using the method of loop equations, who found an algebraic equation for $W_{(1)}(z)$ when $\arccos((q-2)/2)/\pi$ is rational; see also \cite{Guionnet10,Bernardi11} for related results. More recently, the authors of \cite{Borot12} considered a combinatorial approach using the so-called ``loop-gas'' representation of the Potts model on planar maps without reference to a matrix integral, from which a pair of coupled functional equations for $W_{(1)}(z)$ and a function related to $W_{(q)}(z)$ was obtained and solved. For $q=2$, the relationship between $W_{(1)}(z)$ and $W_{(2)}(z)$ has been expressed succinctly from the perspective of the boundary renormalisation group \cite{Carroll96,Carroll97}, a picture which later was extended to non-planar geometries \cite{Atkin11} and arbitrary face degrees \cite{Atkin12}. Indeed, these investigations revealed that different boundary conditions yield inequivalent algebraic equations satisfied by the corresponding generating functions. However, a systematic understanding of the relationship between different boundary conditions for more general values of $q$ and $p$ appears to be lacking and herein we report on some progress on this matter. 
 
 As will be discussed in Section \ref{sec:bc}, $W_{(p)}(z)$ is given by the Stieltjes transform of the spectral density of the sum $p$ Hermitian random matrices of infinite size. A more mathematically inclined characterisation of the problem solved in this chapter thus goes as follows: given a set of Hermitian random matrices $\{X_i\}_{i=1}^q$ distributed according to \eqref{pottsmeasure} and a positive integer $p\leq q$, what is the spectral density of the sum $X_1+X_2+\dots X_p$ as $N\to\infty$? For the simpler case of uncorrelated matrices, the answer has been neatly summarised in the context of free probability  \cite{Speicher93,Speicher94}, going back to Voiculescu's observation of the asymptotic freeness of Gaussian independent random matrices \cite{Voiculescu92}: given a spectral density $\rho_X(z)$, define the \em R-transform \em via the functional inverse of its Stieltjes transform $W_X(z)$

\begin{equation}
\label{rtrafo} 
R^X(z)=W_X^{-1}(z)-\frac{1}{z}\ .
\end{equation}

\noindent Now assume $Y$ is freely independent from $X$. Then the \em free (additive) convolution \em $\rho_X\boxplus\rho_Y$ is defined by $\rho_{X+Y}$. The results of free probability theory \cite{Voiculescu92} state that the latter is obtained from $\rho_X$ and $\rho_Y$ by

\begin{enumerate}
\item computing $R^{X+Y}$ by adding the respective R-transforms,
\begin{equation}
R^{X+Y}(z)=R^X(z)+R^Y(z)\ ,
\end{equation}
\item inverting the relationship \eqref{rtrafo}, 
\begin{equation}
\begin{aligned}
\label{Voicu}
W_{X+Y}^{-1}(z)&=R^{X+Y}(z)+\frac{1}{z}\ .
\end{aligned}
\end{equation}
\end{enumerate}

\noindent The spectral density for the sum $X+Y$ can then be read off from the imaginary part of the inverse function,
\begin{equation}
\begin{aligned}
\rho_{X+Y}(x)&=\frac{1}{\pi}\mathrm{Im}\ W_{X+Y}(x)_+\ .
\end{aligned}
\end{equation}

\noindent We follow \cite{ZJ99-2} in referring to the key relationship \eqref{Voicu} as Voiculescu's formula. Clearly the matrices $\{X_i\}_{i=1}^q$ distributed according to \eqref{pottsmeasure} are \em not \em freely independent -- their correlations prevent us from applying Voiculescu's formula to compute the spectral densities for sums like $X_1+X_2+\dots X_p$. Our strategy to obtain the disk partition function of the Potts model involves a suitable generalisation of the R-transform and using it to evaluate the spectral density and hence $W_{(p)}(z)$. 

This chapter is organised as follows: Section \ref{sec:bc} reviews the matrix model formulation and defines the observables of interest. In Section \ref{sec:planar}, we state the main results in Propositions \ref{prop-curve} and \ref{prop-ycurve} and discuss how our results reduce to Voiculescu's formula when the interactions of the Potts model are turned off. In Section \ref{sec:cases}, we study hard dimers, the Ising model and the $3$-states Potts model on planar triangulations as simple examples in greater detail. We derive explicit expressions for the spectral curve for given $p$ and compare our results to the literature where available. In Section \ref{sec:crit}, we proceed to investigate the phase diagram of the model when $0<q<4$ and comment on the conformal field theory description of the scaling behaviour associated with the critical points. Finally, we discuss the implications of our results in Section \ref{sec:pottsdiscussion}.

\section{Definition of the model}
\label{sec:bc}

\noindent Following \cite{Kazakov88,Daul94,Bonnet99,Eynard99}, we use the measure \eqref{pottsmeasure} to compute observables of the $q$-states Potts model on a random planar lattice. A distinguishing feature are the exponentials of $\mathrm{tr}X_i X_j$ in \eqref{pottsmeasure}, breaking the overall $U(N)\times O(q)$-invariance of the remaining factors. Here we confine our study to the case $V_i(z)=U(z)+z^2/2$ $\forall i$ for a fixed polynomial $U(z)=\sum_{m=2}^{k+2}t_mz^m/m$, rendering the $q$ states of the statistical system indistinguishable. In this case, the measure \eqref{pottsmeasure} remains invariant under the overall symmetries

\begin{equation}
\label{sym}
	X_i\to U^{\dagger}X_iU\ ,\quad U\in U(N)\ ,\quad\mathrm{and}\quad X_i\to X_{\sigma(i)}\ ,\quad \sigma\in S_q\ ,
\end{equation}

\noindent where $S_q$ dentoes the symmetric group of order $q!$. This is to be contrasted with the ``multi-matrix chain'' studied for example in \cite{ZJ98,Eynard03}, for which $\mathbb{Z}_2$ is preserved in place of $S_q$. Our definition includes a subset of the statistical RSOS models on a random lattice, which are indexed by simply laced Dynkin diagrams \cite{Pasquier86} and have been described using matrix integrals by Kostov \cite{Kostov95}. In particular, for $(q,k)=(2,1)$, \eqref{pottsmeasure} describes the $A_3$ model and for $(q,k)=(3,1)$ the $D_4$ model on random triangulations, respectively.

  The desired quantities $W_{(p)}(z)$ can now be defined along the lines of our discussion in Subection \ref{subsec:introstat} of the previous chapter: Given $\sigma\in S_q/(S_p\times S_{q-p})$, we define the partition function of the model on a random lattice with $p$ allowed, equally weighted colors on a single connected boundary containing a marked point as\footnote{Note that the $X_{(p|\sigma)}$ inherit a covariant transformation behaviour under $S_q\subset O(q)$.}

\begin{equation}
\label{discfct}
\begin{aligned}
	W_{(p|\sigma)}(z)=\frac{1}{N}\left\langle\mathrm{tr}\frac{1}{z-X_{(p|\sigma)}}\right\rangle\ ,\quad X_{(p|\sigma)}=\sum_{i=1}^pX_{\sigma(i)}\ ,\quad 1\leq p \leq q\ ,
\end{aligned}
\end{equation}

\noindent where here and in what follows $\langle\cdot\rangle$ denotes the average with respect to \eqref{pottsmeasure} and $z$ denotes the fugacity of a boundary link. As a result of the permutation symmetry, for given $p$, all $|S_q/(S_p\times S_{q-p})|=\binom{q}{p}$ partition functions $W_{(p|\sigma)}(z)$ are described by the same function, so that we henceforth abbreviate $W_{(p)}(z):=W_{(p|\sigma_0)}(z)$ for a representative $\sigma_0$ and denote the spectral density of the sum $X_{(p|\sigma)}$ by $\rho_{(p)}(z)$. Note that for $p=1$, our definition of $W_{(p)}(z)$ reduces to the one studied in \cite{Kazakov88,Daul94,ZJ99,Bonnet99,Eynard99}. 

We conclude this section with a helpful lemma which expresses the partition function as single integral over effective matrix variables $X_0$ and $P_\pm$ by a series of integral transformations\footnote{Thinking of $\{X_i\}_{i=1}^q$ as coordinates on configuration space, these simply correspond to linear canonical transformations on the corresponding phase space.}. This circumvents a notorious difficulty presented by the first factor in \eqref{pottsmeasure}, which leads to a complicated integral over the unitary group when the latter is written as a function of the eigenvalues of the matrices $X_i$ with $i>0$ \cite{Kazakov88,Daul94}. 

\begin{lem} 
\label{lem-Z} Let $h>0$ and abbreviate the integral transformations

\begin{subequations}
\begin{align}
\gamma_{\pm}(X)&=\int_{\mathbb{R}}\D P_\pm f(P)\E^{-\frac{N}{2}\mathrm{tr}P_\pm^2}\E^{N\mathrm{tr}P_\pm X/\sqrt{\E^{\pm 2h}-1}}\ ,\\
\gamma'_{\pm}(P)&=\int_\Gamma\D X f(X)\E^{N\mathrm{tr}PX\sqrt{1-\E^{\mp 2h}}}\ ,
\end{align}
\end{subequations}

\noindent where the subscripts below the integrals indicate the integration cycle for the corresponding eigenvalues. Then up to an overall constant, the partition function in \eqref{pottsmeasure} can be written as

\begin{subequations}
\begin{align}
	\label{Z1}
Z_{N,q}
	&=\int_{\mathbb{R}}\mathrm{d}P_+\ \E^{-\frac{N}{2}(1-\E^{-2h})\mathrm{tr}P_+^2}\left(\gamma'_+\left[\E^{-N\mathrm{tr}U}\right](P_+)\right)^q\\
	\label{Z2}
	&=\int_\mathbb{R}\mathrm{d}X_0\ \gamma_+\left[(\gamma'_+[\E^{-N\mathrm{tr}U}])^p\right](X_0)\ \gamma_-\left[(\gamma'_-[\E^{-N\mathrm{tr}U}])^{q-p}\right](X_0)\ \\
	\label{Z3}
	&=\int_\mathbb{R}\mathrm{d}X_0\left(\prod_{i=1}^q\int_\Gamma\mathrm{d}X_i\ \E^{-N\mathrm{tr}U(X_i)}\right)\nonumber\\&\quad\times\gamma_+[1]\left(X_0+2\sinh(h)\sum_{i=1}^pX_i\right)\ \gamma_-[1]\left(X_0-2\sinh(h)\sum_{i=p+1}^q X_i\right)\ .
\end{align}
\end{subequations}
\end{lem}

\begin{proof}
We begin by showing equality of \eqref{Z2} and \eqref{Z3}, and then equality to $Z_{N,q}$. Subsequently showing equality to \eqref{Z1} completes the proof. By definition, we can write

\begin{equation}
\begin{aligned}
\label{integralexp}
\gamma_{\pm}\left[\left(\gamma'_\pm[\E^{-N\mathrm{tr}U}]\right)^n\right](X_0)&=\int_\mathbb{R}\mathrm{d}P_\pm\ \E^{-N\mathrm{tr}P^2/2}\ \E^{N\mathrm{tr}X_0 P_\pm/\sqrt{\E^{\pm 2h}-1}}\\&\quad\times\left(\prod_{i=1}^n\int_\Gamma\mathrm{d}X_i\ \E^{-N\mathrm{tr}U(X_i)}\ \E^{N\mathrm{tr}P_\pm X_i\sqrt{1-\E^{\mp 2h}}}\right)\ .
\end{aligned}
\end{equation}

\noindent In general, there are $\mathrm{deg}\ U'=k+1$ independent cycles $\Gamma$ that render this iterated integral absolutely convergent for finite $N$. Hence we can apply the Fubini-Tonelli-theorem, that is, exchange the order of integration:

\begin{equation}
\begin{aligned}
\gamma_{\pm}\left[\left(\gamma'_\pm[\E^{-N\mathrm{tr}U}]\right)^n\right](X_0)
	&= \left(\prod_{i=1}^n\int_\Gamma\mathrm{d}X_i\ \E^{-N\mathrm{tr}U(X_i)}\right)\gamma_\pm[1]\left(X_0\pm2\sinh(h)\sum_{i=1}^n X_i\right)\ .
\end{aligned}
\end{equation}

\noindent Inserting this result into \eqref{Z2} proves equality to \eqref{Z3}. To obtain equality to $Z_{N,q}$, note that up to an overall multiplicative constant,

\begin{equation}
\begin{aligned}
&\int_\mathbb{R}\mathrm{d}X_0\ \gamma_+[1]\left(X_0+2\sinh(h)\sum_{i=1}^p X_i\right)\gamma_-[1]\left(X_0-2\sinh(h)\sum_{i=p+1}^q X_i\right)=\E^{N\mathrm{tr}\left(\sum_{i=1}^qX_i\right)^2/2}\ .
\end{aligned}
\end{equation}

\noindent Inserting this result into \eqref{Z3} and interchanging the order of integration between $X_0$ and $X_i$ by the same argument proves equality to $Z_{N,q}$. It remains to show equivalence to \eqref{Z1}. Starting from \eqref{Z3}, we may use \eqref{integralexp} to write the action of $\gamma_+$ on $(\gamma'_+[\E^{-N\mathrm{tr}U}])^p$ and of $\gamma_-$ on $(\gamma'_-[\E^{-N\mathrm{tr}U}])^{q-p}$ as Gaussian integrals over two matrices $P_+$, $P_-$, respectively. Performing the integration over $P_-$ and subsequently $X_0$, we are left with \eqref{Z1}. As a cross-check, it is straightforward to confirm that \eqref{Z1} equals our initial definition in \eqref{pottsmeasure} of $Z_{N,q}$ by writing out the $q^{\mathrm{th}}$ power of $\gamma'_+[\E^{-N\mathrm{tr}U}]$ as a product of integrals over $X_i$, $i=1\dots q$ and reversing the order of integration with $P_+$.
\end{proof}

\section{Planar limit}
\label{sec:planar}

\noindent This section is concerned with the explicit evaluation of $W_{(p)}(z)$ in the planar limit and is organised as follows: Subsection \ref{subsec:saddle} expresses $W_{(p)}(z)$ via the $p$-independent spectrum of the matrix $Y\equiv \sqrt{1-\E^{-2h}}P_+$ in \eqref{Z1}. The latter is then provided explicitly in Subsection \ref{subsec:solution} -- a problem first solved in \cite{Daul94,ZJ98} and rederived here for arbitrary $q\neq 4$. Finally, in Subsection \ref{subsec:voiculescu}, we discuss how Voiculescu's formula \eqref{Voicu} arises as a special case of our results in the limit of vanishing interaction strength of the Potts model. To streamline the presentation, for any pair of $N\times N$ matrices $(X,P)$, we define the averages

\begin{equation}
\label{def-g}
\begin{aligned}
	G^X_P(z)&=\frac{1}{N}\frac{\partial}{\partial z}\ln\left\langle\det_{1\leq k, l\leq N}\E^{Nx_kp_l}\right\rangle_{p_N=z}\ ,\quad z\notin \mathrm{supp}\ \rho_P\ ,\\
	G^P_X(z)&=\frac{1}{N}\frac{\partial}{\partial z}\ln\left\langle\det_{1\leq k, l\leq N}\E^{Nx_kp_l}\right\rangle_{x_N=z}\ ,\quad z\notin \mathrm{supp}\ \rho_X\ .
\end{aligned}
\end{equation} 

\noindent The key property of the above functions is that $G^X_P(G_X^P(z))=z\left(1+\mathcal{O}(1/N)\right)$ for large $N$ \cite{Matytsin93,ZJ98}.

\subsection{Saddle point equations}
\label{subsec:saddle}

\noindent We begin by stating the main result of this section. This rests on Lemmas \ref{lem-loop} and \ref{lem-SPE}, which we derive from the $q+1$-matrix integral \eqref{Z3} and the pair of $1$-matrix integrals \eqref{Z1}, \eqref{Z2} at large $N$, respectively. After presenting the latter, we conclude this section with the proof of the main statement.

\begin{prop} 
\label{prop-curve}
Let the matrix $P_+$ be defined as in Lemma \ref{lem-Z}, and set $Y=\sqrt{1-\E^{-2h}}P_+$. Then for $N\to\infty$, the spectral density of the sum of $p$ matrices distributed according to \eqref{pottsmeasure} is given by

\begin{equation}
\label{curve1}
\rho_{(p)}(z)=\frac{1}{2\pi\I}\left[G^Y_{(p)}(z)_+-G^Y_{(p)}(z)_-\right]\ ,
\end{equation} 
\noindent where $G^Y_{(p)}(z)$ is the functional inverse of 
\begin{equation}
\label{curve2}
G_Y^{(p)}(z)=\frac{p}{q}(z-W_Y(z)_-)+\frac{q-p}{q}W_Y(z)_+\ .
\end{equation}
\end{prop}

\begin{rem} Generally, $G^Y_{(p)}(z)$ is a multi-valued function so that we need to specify the sheet on which \eqref{curve1} is evaluated. This ambiguity is fixed by the condition that $\lim_{z\to\infty}z\ W_{(p)}(z)=1$. 
\end{rem}

\begin{cor} 
\label{curve3}
When $G_{(p)}^Y(z)$ satisfies an algebraic equation of the form $F_{(p)}(z,G_{(p)}^Y(z))=0$, then $G_{(q-p)}^Y(z)$ follows from

\begin{equation}
F_{(p)}\left(G_{(q-p)}^Y(z)-z,G_{(q-p)}^Y(z)\right)=0\ .
\end{equation}
\end{cor}

\noindent As announced, we proceed to formulate the main lemmas involved in the proof of the above results:

\begin{lem} 
\label{lem-loop} In the limit $N\to\infty$, the matrix $M=\E^{-h}\sum_{i=1}^pX_i+\E^{h}\sum_{i=p+1}^qX_i$ satisfies
\begin{equation}
W_{X_0}(z)=W_M(z-W_{X_0}(z))\ .
\end{equation}
\end{lem}

\begin{proof}
This result follows from the translation invariance of the measure \eqref{pottsmeasure}: Setting $\sigma=\mathrm{id}$ in \eqref{Z3} without loss of generality, consider the shift by a small Hermitian matrix\footnote{See also \cite{Eynard99} for an earlier application of this method of ``loop equations'' to the Potts matrix model.}

\begin{equation}
X_0\longrightarrow X_0'=X_0+\varepsilon\left(\frac{1}{z-X_0}\frac{1}{z'-M}+\mathrm{h.c.}\right)\ ,\quad \varepsilon\ll 1\ ,
\end{equation}

\noindent as a formal power series in $z$, $z'$. When $M=\E^{-h}\sum_{i=1}^pX_i+\E^{h}\sum_{i=p+1}^qX_i$, the variation of the product of the two Gaussian integrals 

\begin{equation}
I\left(\{X_i\}_{i=0}^q\right)\equiv\gamma_+[1]\left(X_0+2\sinh(h)\sum_{i=1}^pX_{i}\right)\ \gamma_-[1]\left(X_0-2\sinh(h)\sum_{i=p+1}^q X_{i}\right)
\end{equation}

\noindent and the measure $\mathrm{d}X_0$ is respectively given to leading order by

\begin{subequations}
\begin{align}
I\left(\{X_i\}_{i=0}^q\right)& \longrightarrow I(\{X_i\}_{i=0}^q)+\varepsilon\mathrm{tr}\frac{1}{z-X_0}\frac{1}{z'-M}(M-X_0)+\mathcal{O}(\varepsilon^2)\ ,\\
\mathrm{d}X_0 & \longrightarrow \mathrm{d}X_0\left(1+\varepsilon\mathrm{tr}\frac{1}{z-X_0}\mathrm{tr}\frac{1}{z-X_0}\frac{1}{z'-M}+\mathcal{O}(\varepsilon^2)\right)\ , 
\end{align}
\end{subequations}

\noindent Demanding invariance of $Z_{N,q}$ to first order in $\varepsilon$ and approximating $\langle\mathrm{tr}A\mathrm{tr}B\rangle=\langle\mathrm{tr}A\rangle\langle\mathrm{tr}B\rangle+\mathcal{O}(1/N)$ yields

\begin{equation}
W_M(z')-W_{X_0}(z)=\frac{1}{N}\left\langle\mathrm{tr}\frac{1}{z-X_0}\frac{1}{z'-M}\right\rangle\left(W_{X_0}(z)-z+z'\right)+\mathcal{O}(1/N^2)\ .
\end{equation}

\noindent Evaluating the above at $z'=z-W_{X_0}(z)$ proves the lemma.
\end{proof}

\begin{lem} 
\label{lem-SPE}
Let the matrices $P_+$, $X_0$ be defined as in Lemma \ref{lem-Z}, and set $P_+=Y/\sqrt{1-\E^{-2h}}$ and $X_0=2\sinh(h)\bar{X}_0$. Then for $N\to\infty$,

\begin{equation}
\label{SPE}
\mathrm{Re}\ G_Y^{\bar{X}_0}(z)=\left(\frac{2p}{q}-1\right)\mathrm{Re}\ W_Y(z)+\left(\frac{1}{1-\E^{-2h}}-\frac{p}{q}\right)z\ ,\quad z\in\mathrm{supp}\ \rho_Y\ .
\end{equation}
\end{lem}

\begin{proof} 
This result follows from the saddle point approximation to the integrals \eqref{Z1} and \eqref{Z2} in Lemma \ref{lem-Z}: Setting 

\begin{equation}
\label{diagYX0}
Y=U\mathrm{diag}(\{y_n\}_{n=1}^N)U^{\dagger}\ , \quad \frac{X_0}{2\sinh(h)}=V\mathrm{diag}(\{x_n\}_{n=1}^N)V^{\dagger}\ ,
\end{equation}

\noindent with $U, V \in U(N)$, we can perform the integration over $U^{\dagger}V$ using the well-known result \cite{Harishchandra56,Itzykson80} 

\begin{equation}
\label{IZint}
	\int_{U(N)}\mathrm{d}U\ \E^{\lambda N\mathrm{tr}[YUXU^{\dagger}]}=\mathrm{const.}\times \frac{\det_{1\leq k,l\leq N}\E^{\lambda Ny_kx_l}}{\Delta(x)\Delta(y)}\quad\forall\lambda\in\mathbb{C}\ ,
\end{equation}

\noindent where $\mathrm{d}U$ is the normalised Haar measure. It follows that for the exponent of the integrand in \eqref{Z2} to have an extremum, the eigenvalues in \eqref{diagYX0} must satisfy

\begin{subequations}
\begin{align}
\label{SPE1}
0&=\frac{1}{N}\left[\frac{\partial}{\partial y_n}\ln\det_{k,l}\E^{Ny_kx_l}+\sum_{k\neq n}\frac{1}{y_n-y_k}+p\frac{\partial}{\partial y_n}\ln\gamma'_+[\E^{-N\mathrm{tr}U}]\left(\frac{Y}{\sqrt{1-\E^{-2h}}}\right)\right]-\frac{y_n}{1-\E^{-2h}}\ .
\end{align}
\end{subequations}

\noindent On the other hand, from \eqref{Z1} we find, that when \eqref{SPE1} holds, then also

\begin{equation}
\label{SPE3}
0=\frac{2}{N}\sum_{k\neq n}\frac{1}{y_n-y_k}+\frac{q}{N}\frac{\partial}{\partial y_n}\ln\gamma'_+[\E^{-N\mathrm{tr}U}]\left(\frac{Y}{\sqrt{1-\E^{-2h}}}\right)-y_n\ ,
\end{equation}

\noindent which allows us to eliminate $\gamma'_+[\E^{-N\mathrm{tr}U}](Y/\sqrt{1-\E^{-2h}})$ between the above and \eqref{SPE1}\footnote{It is the analysis of this quantity that leads us to the exact solution for $W_Y(z)$ in the next subsection.}. Taking $N\to\infty$ and using the definition \eqref{def-g} yields \eqref{SPE} as advertised. 
\end{proof}

\begin{proof}[Proof of Proposition \ref{prop-curve} and Corollary \ref{curve3}] 
We begin by deriving the form of \eqref{curve2}, then \eqref{curve1}. Firstly, the form of \eqref{curve2} follows from Lemma \ref{lem-SPE} after analytic continuation: Following an argument in \cite{ZJ99}, we note that the derivative w.r.t. $x_N$ of the logarithm of \eqref{IZint} is an entire function of $x_N$, which implies that as $N\to\infty$, $G_X^Y(z)$ and $W_X(z)$ have the same discontinuity across the real axis. Applying this to our situation, we conclude that when $z\in\mathrm{supp}\ \rho_Y$,

\begin{subequations}
\begin{align}
\label{discont1}
G_Y^{\bar{X}_0}(z)_{\pm}&=\mathrm{Re}\ G_Y^{\bar{X}_0}(z)\pm \I\pi\rho_Y(z)\ ,\\
\label{discont2}
G^Y_{\bar{X}_0}(z)_{\pm}&=\mathrm{Re}\ G^Y_{\bar{X}_0}(z)\pm \I\pi\rho_{\bar{X}_0}(z)\ .
\end{align}
\end{subequations}

\noindent For $h>0$, it follows from \eqref{Z1} that $Y$ is Hermitian, so $W_Y(z)$ has no singularities in the complex plane away from the real axis. Hence we can analytically continue \eqref{SPE} to $z\in\mathbb{C}\setminus\mathrm{supp}\ \rho_Y$ using

\begin{equation}
G_Y^{\bar{X}_0}(z)_+-G_Y^{\bar{X}_0}(z)_-=W_Y(z)_+-W_Y(z)_-\ ,
\end{equation}

\noindent which results in

\begin{equation}
\label{spe-complex}
G_Y^{\bar{X}_0}(z)=\frac{p}{q}\left(W_Y(z)_+-z\right)-\frac{q-p}{q}W_Y(z)_-+\frac{z}{1-\E^{-2h}}\ .
\end{equation}

\noindent Secondly, to obtain \eqref{curve1}, note first that from Lemma \ref{lem-loop}, we find

\begin{equation}
\begin{aligned}
W_{\bar{X}_0}(z)=W_{M/(2\sinh(h))}\left(z-\frac{1}{4\sinh(h)^2}W_{\bar{X}_0}(z)\right)\ ,
\end{aligned}
\end{equation}

\noindent where we used the property $W_X(z)=\lambda W_{\lambda X}(\lambda x)$ for real $\lambda$. In the limit $h\to\infty$, $M/(2\sinh(h))\to\sum_{i=p+1}^qX_i$ and consequently, from the above,

\begin{equation}
W_{\bar{X}_0}(z)\to W_{(q-p)}\left(z+\mathcal{O}(\E^{-2h})\right)\quad \mathrm{as}\ h\to\infty\ .
\end{equation}

\noindent We infer that in this limit, $\rho_{\bar{X}_0}(z)\to \rho_{(q-p)}(z)$, which in conjunction with \eqref{discont2} yields

\begin{equation}
\rho_{(q-p)}(z)=\lim_{h\to\infty}\frac{1}{2\pi\I}\left[G^Y_{\bar{X}_0}(z)_+-G^Y_{\bar{X}_0}(z)_-\right]\ .
\end{equation}

\noindent We thus obtain the desired expressions \eqref{curve1} and \eqref{curve2} from the above and \eqref{spe-complex} by identifying  $G^Y_{\bar{X}_0}(z)=G_{(q-p)}^Y(z)$, and noting that according to \eqref{spe-complex}, the analytic continuation of $G_Y^{(q-p)}(z)$ through $\mathrm{supp}\ \rho_Y$ is given by $z-G_Y^{(p)}(z)$,

\begin{equation}
\label{continuationGp}
G_Y^{(q-p)}(z)_\pm=z-G_Y^{(p)}(z)_\mp\ .
\end{equation}
 
\noindent The functional inversion relation then follows from $G^Y_{\bar{X}_0}\circ G_Y^{\bar{X}_0}=\mathrm{id}$. Finally, to show Corollary \ref{curve3}, observe that according to \eqref{continuationGp}, for an algebraic function $F$ in two variables,

\begin{equation}
F_{(p)}\left(z,G_{(p)}^Y(z)\right)=0\quad \mathrm{implies}\quad F_{(p)}\left(z'-G_Y^{(q-p)}(z'),z'\right)=0\ ,
\end{equation}

\noindent since the analytic continuation merely takes us from one solution to the above equation to another. Evaluating at $z'=G^Y_{(q-p)}(z)$ proves the corollary.
\end{proof}

\subsection{General solution}
\label{subsec:solution}

\noindent The main result in Proposition \ref{prop-curve} is expressed via the functional inverse of the quantity \eqref{curve2}. Generally, this functional inversion is most easily achieved by means of an explicit parametric form of $W_Y(z)$; Proposition \ref{prop-ycurve} below provides just that for general $q\neq 4$ when $U(z)$ is cubic, i.e. $k=1$. 

\begin{prop} 
\label{prop-ycurve}
Let $k=1$, $\nu=\arccos((q-2)/2)/\pi$ and assume $\mathrm{supp}\ \rho_Y=[z_-,z_+]\subset\mathbb{R}$ as $N\to\infty$. Then $W_Y(z)=W_Y^{\mathrm{reg.}}(z)+W_Y^{\mathrm{sing.}}(z)$ with\footnote{Our conventions for elliptic functions are those of Gradshtein and Ryzhik \cite{Gradshteyn07} and are spelled out in Appendix \ref{app:auxprob}.}
	
\begin{equation}
	\begin{aligned}
	z(\sigma)&=\delta_U+\sqrt{(z_+-\delta_U)(z_--\delta_U)}\left(\frac{\vartheta_2(\pi\sigma|\tau)}{\vartheta_3(\pi\sigma|\tau)}\right)^2\ ,\\
	W_Y^{\mathrm{reg.}}\left(z(\sigma)\right)&=\frac{1}{4-q}\left(\frac{qt_2}{t_3}+2z(\sigma)\right)\ ,\\
		W_Y^{\mathrm{sing.}}\left(z(\sigma)\right)&=\sum_{n\geq 0}\frac{f_n}{n!}\frac{
		\partial^n}{\partial\sigma^n}\left(\E^{\I\pi\nu\sigma}\frac{\vartheta_3(\pi\sigma+\pi\tau\nu/2|\tau)}{\vartheta_3(\pi\sigma|\tau)}+\E^{-\I\pi\nu\sigma}\frac{\vartheta_3(\pi\sigma-\pi\tau\nu/2|\tau)}{\vartheta_3(\pi\sigma|\tau)}\right)\ ,
	\end{aligned}
\end{equation}

\noindent where $\tau$ and $\delta_U$ are implicit functions of $t_2$, $t_3$, and the coefficients $f_n$ are determined by the requirement $\lim_{z\to\infty}zW_Y(z)$.
\end{prop}

\begin{proof} 
We can determine the spectrum of $Y$ from the saddle point approximation \eqref{SPE3} to the integral \eqref{Z3}, which is precisely the problem first considered in \cite{Daul94,ZJ99}. To our knowledge, the first large $N$ analysis of $\gamma'_+[\E^{-N\mathrm{tr}U}]$ appearing in \eqref{SPE3} for cubic $U$ was done, if in a slightly different context, by Gross and Newman in \cite{Gross91}. Using \cite[eqns. (2.10), (2.11)]{Gross91}, equation \eqref{SPE3} can be expressed as 

\begin{equation}
	\label{spe1}
	\begin{aligned}
		z=2\mathrm{Re}\ W_Y(z)&+\frac{q}{2}\int_{z_-}^{z^+}\frac{\mathrm{d}z'}{\sqrt{z'-\delta_U}}\frac{\rho_Y(z')}{\sqrt{z-\delta_U}+\sqrt{z'-\delta_U}}\\ &+q\frac{\sqrt{z-\delta_U}}{\sqrt{t_3}}-q\frac{t_2}{2t_3}\ ,\qquad z\in[z_-,z_+]\ ,
	\end{aligned}
\end{equation}

\noindent where $\delta_U$ solves the implicit equation 

\begin{equation}
\frac{t_2}{4t_3}+\delta_U=\frac{\sqrt{t_3}}{t_2^{3/2}}\int_{z_-}^{z^+}\mathrm{d}z\ \frac{\rho_Y(z)}{\sqrt{z-\delta_U}}\ .
\end{equation}

\noindent Let us resolve the branch point at $\delta_U$ by the change of variables $w(z)=\sqrt{z-\delta_U}$, and denote $w(z_{\pm})=w_{\pm}$. Introducing the auxiliary function

\begin{equation}
	\label{auxfct}
	f(w)=\int_{w_-}^{w_+}\mathrm{d}w'\frac{\rho_Y(\delta_U+w^{\prime 2})}{w-w'}\ ,
\end{equation}

\noindent we derive the two identities

\begin{equation}
	\begin{aligned}
		\mathrm{Re}\ W_Y(z)&=\mathrm{Re}\ f(w(z))+f(-w(z))\;,\qquad z\in[z_-,z_+]\;,\\
		f(-w(z))&=-\frac{1}{2}\int_{z_-}^{z_+}\frac{\mathrm{d}z'}{\sqrt{z'-\delta_U}}\frac{\rho_Y(z')}{\sqrt{z-\delta_U}+\sqrt{z'-\delta_U}}\;.
	\end{aligned}
\end{equation}

\noindent We can then rewrite \eqref{spe1} in the equivalent form
	
\begin{equation}
	2\mathrm{Re}\ f(w)+(2-q)f(-w)=\delta_U+w^2-q\frac{w}{\sqrt{t_3}}+q\frac{t_2}{2t_3}\ , \quad w\in[w_-,w_+]\ .
\end{equation}

\noindent A particular polynomial solution to the above equation when $q\neq 4$ is

\begin{equation}
	f_{\mathrm{reg.}}(w)=\frac{qt_2}{2t_3(4-q)}-\frac{w}{\sqrt{t_3}}+\frac{\delta_U+w^2}{4-q}\ .
\end{equation}

\noindent The general solution will therefore differ from the above by a function $f_{\mathrm{sing.}}(w)=f_{\mathrm{reg.}}(w)-f(w)$ holomorphic on $\mathbb{C}\setminus[w_-,w_+]$ satsifying the homogenous equation

\begin{equation}
	\label{hom1}
	2\mathrm{Re}\ f_{\mathrm{sing.}}(w)+(2-q)f_{\mathrm{sing.}}(-w)=0\ , \quad w\in[w_-,w_+]\ .
\end{equation}

\noindent We recover $\rho_Y(z)$ by inverting the relationship \eqref{auxfct}, which, using the fact that $f_{\mathrm{reg.}}(w)$ is analytic, becomes
	
\begin{equation}
	\label{lambdadens}
	\begin{aligned}
		\rho_Y(z)&=\frac{1}{2\pi\I}\left[f_{\mathrm{sing.}}\left(\sqrt{z-\delta_U}\right)_--f_{\mathrm{sing.}}\left(\sqrt{z-\delta_U}\right)_+\right]\ ,\quad z\in[z_-,z_+]\ .\\
	\end{aligned}
\end{equation}
	
\noindent From the above expressions it then follows that for $z\notin[z_-,z_+]$, $W_Y(z)$ is given by

\begin{equation}
	\label{lambdares}
	\begin{aligned}		W_Y(z)&=2\int_{w_-}^{w^+}\zeta'\mathrm{d}w'\frac{\rho_Y(\delta_U+w^{\prime 2})}{z-\delta_U-w^{\prime 2}}\\
		&=\frac{1}{4-q}\left(\frac{qt_2}{t_3}+2z\right)-f_{\mathrm{sing.}}\left(\sqrt{z-\delta_U}\right)-f_{\mathrm{sing.}}\left(-\sqrt{z-\delta_U}\right)\ ;
	\end{aligned}
\end{equation}

\noindent The general solution to the homogeneous equation \eqref{hom1} was first derived in \cite{Eynard95} in the context of the $O(n)$ model and is presented in more detail in Appendix \ref{app:auxprob}. There we recall how $f_{\mathrm{sing.}}(w)$ can be parametrised in terms of elliptic functions as\footnote{By abuse of notation, we distinguish the functions $w(\sigma)$ and $w(z)=\sqrt{z-\delta_U}$ solely by their arguments.}
	
\begin{equation}
	\label{ellipticsol}
	\begin{aligned}
		f_{\mathrm{sing.}}(w(\sigma))&=\sum_{n\geq 0}\frac{f_n}{n!}\frac{
		\partial^n}{\partial\sigma^n}\left(\E^{\I\pi\nu(\sigma-1)}\frac{\vartheta_3(\pi\sigma+\pi\tau\nu/2|\tau)}{\vartheta_3(\pi\sigma|\tau)}+\E^{-\I\pi\nu(\sigma-1)}\frac{\vartheta_3(\pi\sigma-\pi\tau\nu/2|\tau)}{\vartheta_3(\pi\sigma|\tau)}\right)\ ,\\
w(\sigma)&=\sqrt{w_-w_+}\frac{\vartheta_2(\pi\sigma|\tau)}{\vartheta_3(\pi\sigma|\tau)}\ ,
	\end{aligned}
\end{equation}
	
\noindent where $\nu=\arccos((2-q)/2)/\pi$ and $\tau=\I K'/K$, with $K$ and $K'$ respectively given by the complete elliptic integral of the first and second kind (cf. eqn. \eqref{ellipticK}); the coefficients $\{f_n\}$ are entirely determined by the condition that $\lim_{z\to\infty} z\ W_Y(z)=1$. Inserting the above parametrisation into \eqref{lambdares} completes the proof.
\end{proof}

\subsection{Derivation of Vouiculescu's formula for free convolution}
\label{subsec:voiculescu}

\noindent Here we show how our results imply a non-trivial generalisation of Voiculescu's formula for free convolution of probability distributions to a non-free situation. This is essentially an adaption of the derivation in \cite{ZJ99-2} to the case where the ``external'' matrix follows a Gaussian distribution\footnote{Equivalent results were previously obtained by Zee in \cite{Zee96}.}; gradually turning off the $O(q)$-symmetry breaking interactions of the Potts model, our formulae should reduce to Voiculescu's for free random variables. To confirm this is the case, it is convenient to consider the slight generalisation of \eqref{pottsmeasure},

\begin{equation}
\label{pottsmeasure2}
\D\mu(X_1,X_2,\dots X_q)=\frac{1}{Z_{N,q}^{\lambda}}\prod_{\langle i j\rangle}\E^{\lambda N\mathrm{tr}X_i X_j}\times \prod_{i=1}^q\E^{-N\mathrm{tr}V_i(x)}\D X_i\ ,\quad \lambda \geq 0\ ,
\end{equation}

\noindent which reduces to \eqref{pottsmeasure} for $\lambda\to1$ and should yield Voiculescu's formula for $\lambda\to 0$\footnote{Of course, the parameter $\lambda$ is redundant in that we may equivalently obtain \eqref{pottsmeasure} by a suitable resacling of $X_i$ and $\{t_m\}_{m=2}^{k+2}$; we are thus not departing from the initial parameter space of the model.}. Then the following holds for averages with respect to \eqref{pottsmeasure2}:

\begin{prop} 
Take $N\to\infty$. Then as $\lambda\to 0$, 
\begin{subequations}
\begin{align}
\label{freelimit1}
G^M_{\lambda Y}(z)-W_Y(z)&\to R^M(z)\ ,\\
\label{freelimit2}
W_{(q)}^{-1}(z)&\to \sum_{i=1}^qW_{X_i}^{-1}(z)-\frac{q-1}{z}\ ,
\end{align}
\end{subequations}

\noindent where $R^M(z)$ denotes the R-transform \eqref{rtrafo} of $W_M(z)$. 
\end{prop}

\begin{proof} 
According to Lemma \ref{lem-Z}, we can write the partition function in \eqref{pottsmeasure2} via the fiducial matrix $Y=\sqrt{1-\E^{-2h}}P_+$ as

\begin{subequations}
\begin{align}
\label{Zalpha1}
Z_{N,q}^{\lambda}&=\int_\mathbb{R}\mathrm{d}Y\ \E^{-N\mathrm{tr}Y^2/2}\prod_{i=1}^q\int_\Gamma\mathrm{d}X_i\ \E^{-N\mathrm{tr}[V_i(X_i)+X_i^2/2-\lambda X_iY]}\\
\label{Zalpha2}
&\equiv \int_\mathbb{R}\mathrm{d}Y\ \E^{-N\mathrm{tr}Y^2/2}\left(\prod_{i=1}^q\int_\Gamma\mathrm{d}X_i\ \E^{-N\mathrm{tr}[V_i(X_i)+X_i^2/2]}\right)\E^{\lambda N\mathrm{tr}Y(X_1+X_2+\dots X_q)}\ .
\end{align}
\end{subequations}

\noindent Diagonalising the matrices and integrating over the unitary group, we can write, taking the limit $N\to\infty$,

\begin{equation}
\frac{1}{N}\frac{\partial}{\partial z}\ln\left.\int_\Gamma\mathrm{d}X_i\ \E^{-N\mathrm{tr}[V_i(X_i)+X_i^2/2-\lambda X_iY]}\right|_{y_N=z}=G_{\lambda Y}^{X_i}(z)-W_{\lambda Y}(z)\ ,\quad i=1\dots q\ ,
\end{equation}

\noindent where we used the definition \eqref{def-g}; comparing \eqref{Zalpha1} and \eqref{Zalpha2}, this implies

\begin{equation}
\label{Voicu-gen}
G_{\lambda Y}^{X_1+X_2+\dots X_q}(z)-W_{\lambda Y}(z)=\sum_{i=1}^q\left(G_{\lambda Y}^{X_i}(z)-W_{\lambda Y}(z)\right)\ .
\end{equation}

\noindent Now consider the limit $\lambda\to 0$. On the one hand,

\begin{equation}
\lim_{\lambda\to 0}\frac{\E^{\lambda Nx_i y_j}}{\Delta(x)\Delta(y)}=1\ ,
\end{equation} 

\noindent from which it follows that 

\begin{subequations}
\begin{align}
\lim_{\lambda\to 0}\left(G_{X_i}^{\lambda Y}(z)-W_{X_i}(z)\right)&=0\ ,\\
\lim_{\lambda\to 0}\left(G_{X_1+X_2+\dots X_q}^{\lambda Y}(z)-W_{(q)}(z)\right)&=0\ .
\end{align}
\end{subequations} 

\noindent On the other hand, as can be seen from \eqref{SPE3} in this limit, the matrix $Y$ will follow a Gaussian distribution, so its spectral density approaches the semi-circle \eqref{semicircle}. As a result, the spectral density of the rescaled matrix $\lambda Y$ approaches a delta function so that $W_{\lambda Y}(z)\to 1/z$. Together with the relation $G_X^Y\circ G_Y^X=\mathrm{id}$, this means that indeed 

\begin{equation}
G^M_{\lambda Y}(z)-W_{\lambda Y}(z)\to R^M(z)\quad\mathrm{as}\ \lambda\to 0\ ,
\end{equation}

\noindent from comparison with the definition \eqref{rtrafo}. Lastly, inserting the above into \eqref{Voicu-gen} yields \eqref{freelimit2}.
\end{proof}

\noindent For $q=2$, \eqref{freelimit2} indeed gives Voiculescu's formula \eqref{Voicu}. It is in this sense that the function $G^M_{\lambda Y}(z)-W_{\lambda Y}(z)$ lifts the notion of the R-transform, so that \eqref{Voicu-gen} represents a nontrivial extension of Voiculescu's formula to the addition of \em correlated \em random matrices, distributed according to \eqref{pottsmeasure2}. It is instructive to compare \eqref{Voicu-gen} for $\lambda=1$ to the expressions in Proposition \ref{prop-curve} of the previous section more explicitly. Since from \eqref{Zalpha1} and \eqref{Zalpha2}

\begin{equation}
G_Y^{X_i}(z)=\frac{q-1}{q}W_Y(z)_++\frac{1}{q}\left(z-W_Y(z)_-\right)\ ,\quad G_Y^{X_1+\dots X_q}(z)=z-W_Y(z)_-\ ,
\end{equation}

\noindent we observe upon comparison to \eqref{curve2} that indeed

\begin{equation}
\label{identificationG}
G_Y^{(1)}(z)=G_Y^{X_i}(z)\ ,\quad G_Y^{(q)}(z)=G_Y^{X_1+\dots X_q}(z)\ .
\end{equation}

\noindent Hence, for the $S_q$-invariant case\footnote{The above expressions indicate that the generalisation of Proposition \ref{prop-curve} to $\lambda\neq 1$ and $U_i\neq U_j$ for $i\neq j$ is straightforward.} $V_i(z)\equiv U(z)-z^2/2\ \forall i$, our main result in Proposition \ref{prop-curve} \em further \em generalises this result to the sum of $p\leq q$ matrices: the function $G^{(p)}_Y(z)-W_Y(z)$ generalises the R-transform of $W_{(p)}(z)$, and \eqref{curve2} generalises Voiculescu's formula. 

\section{Case studies}
\label{sec:cases}

\noindent Here we consider the cases $(q,k)=(1,2)$, $(2,1)$, and $(3,1)$, which describe hard dimers, the $A_3$ and the $D_4$ model on planar triangulations, respectively. For the first two models, the functions $W_{(p)}(z)$ have been known for a while \cite{Staudacher89,Carroll96,Carroll97} -- the fact that our general formula in Proposition \eqref{prop-curve} reproduces these results lends crecedence to our extension to the $D_4$-model. Unlike models with irrational values of $\arccos((q-2)/2)/\pi$, all of these share the simplification that they can be described by polynomial equations: We derive explicit expressions for the polynomials $F_{(p)}(x,y)$ satisfying 

\begin{equation}
\label{pol}
F_{(p)}\left(z,G_{(p)}^Y(z)\right)=0\ ,\quad 1\leq p\leq q\ ,
\end{equation}

\noindent which define a family of algebraic curves $\mathcal{C}_{(p)}=\{(x,y)\in\mathbb{C}^2|F_{(p)}(x,y)=0\}$. In Appendix \ref{app:asympt}, we describe the resulting analytic structure of $G_Y^{(p)}(z)$ and $G_{(p)}^Y(z)$. The coefficients in \eqref{pol} may be fixed as follows: As stated in the introduction, herein we restrict ourselves to solutions for which the spectral densities have connected support, translating into a single cut in the complex $z$-plane for the Stieltjes transform $W_{(p)}(z)$. To ensure this property, the condition that the curve $\mathcal{C}_{(p)}$ be of genus zero is sufficient, though not in general necessary, as is clear from the geometry of Riemann surfaces\footnote{See also \cite{Eynard02} for a discussion of the relationship between the connectedness of the spectral density and a vanishing genus of the spectral curve.}. Nonetheless, we remark that this slightly stronger condition on the solution set guarantees the existence of a (non-unique) rational parametrisation of the curve. Requiring consistency of the latter with the deduced asymptotic behaviour for large $z$ in turn determines the constants $c^{(p)}_{i,j}$ that appear in the expressions for $F_{(p)}(x,y)$ entirely as functions of $\{t_m\}_{m=2}^{k+2}$. 

\subsection{\texorpdfstring{$(q,k)=(1,2)$}{(q,k)=(1,2)} -- Hard dimers}
\label{subsec:q=1}

\noindent This model describes hard dimers on planar triangulations and was first solved on the sphere by Staudacher \cite{Staudacher89}. According to \eqref{pottsmeasure} and \eqref{Z1}, the partition function can be written as both a one- and two-matrix integral,

\begin{subequations}
\begin{align}
Z_{N,1}&=\int\mathrm{d}X\ \E^{-N\mathrm{tr}[U(X)-X^2/2]}\\
\label{Z2-2}
	&=\int\mathrm{d}Y\ \E^{-N\mathrm{tr}Y^2/2}\int\mathrm{d}X\ \E^{-N\mathrm{tr}[U(X)-XY]}\ .
\end{align}
\end{subequations}

\noindent Using the defintion \eqref{def-g} in the planar limit, the above expressions imply the following relations: 

\begin{subequations}
\begin{align}
\label{case1a}
z&=W_Y(z)_-+G_Y^X(z)_+\ ,\\
\label{case1b}
U'(z)&=W_{(1)}(z)_-+G_X^Y(z)_+\ ,\\
\label{case1c}
U'(z)&=W_{(1)}(z)_-+W_{(1)}(z)_++z\ ,
\end{align}
\end{subequations}

\noindent The first line \eqref{case1a} is indeed consistent with \eqref{curve2} in Proposition \ref{prop-curve} and the relations \eqref{identificationG} in Subsection \ref{subsec:voiculescu}. Via the relation $G_Y^X\circ G_X^Y=\mathrm{id}$, equations \eqref{case1a} and \eqref{case1b} dictate the analytic structure and asymptotic behaviour of $G_Y^X(z)$, the result of which is spelled out in Appendix \ref{app:asympt}, Example \ref{ex-dimers}. This allows us to compute the spectral curve using \eqref{curve2},

\begin{equation}
\begin{aligned}
\label{polynomial0}
F_{(1)}(x,y)&=x^4-x^3y+\frac{t_3}{t_4}x^3+\frac{y^2}{t_4}-\frac{t_3}{t_4}x^2y+\frac{t_2+t_4}{t_4}x^2\\&\quad-\frac{t_2+1}{t_4}xy-c^{(1)}_{0,0}x+ c^{(1)}_{1,1}y+c^{(1)}_{1,0}\ .
\end{aligned}
\end{equation}

\noindent According to Corollary \ref{curve3}, the functions $G_{(p)}^Y(z)$ then satisfy

\begin{equation}
F_{(1)}\left(G_{(0)}^Y(z)-z,G_{(0)}^Y(z)\right)=0\ ,\quad F_{(1)}\left(z,G_{(1)}^Y(z)\right)=0\ ,
\end{equation} 

\noindent which in turn determines their analytic structure and asymptotic behaviour on all sheets -- see Appendix \ref{app:asympt}. Finally, comparing to \eqref{case1c}, we conclude that 

\begin{equation} 
\label{G-1}
G_{(1)}^Y(z)_+=z+W_{(1)}(z)\ ,\quad G_{(1)}^Y(z)_-=t_4z^3+t_3z^2+t_2z-W_{(1)}(z)\ .
\end{equation}

\subsection{\texorpdfstring{$(q,k)=(2,1)$}{(q,k)=(2,1)} -- Ising model}
\label{subsec:q=2}

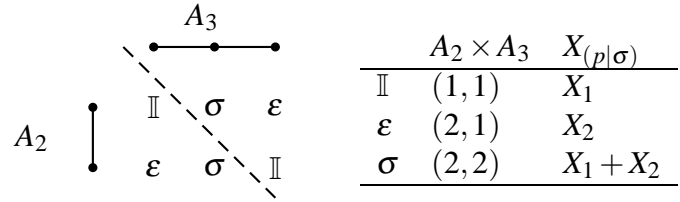
\begin{figure}[h]
\centering
\begin{pspicture}(9,3.5)
\psset{unit=.8cm}

\rput(2.8,4.5){$A_3$}
\psdot[dotsize=.15](2,4)
\psdot[dotsize=.15](3,4)
\psdot[dotsize=.15](4,4)
\psline(2,4)(3,4)
\psline(4,4)(3,4)

\rput(0,2.5){$A_2$}
\psdot[dotsize=.15](1,3)
\psdot[dotsize=.15](1,2)
\psline(1,3)(1,2)

\rput(2,3){$\mathbb{I}$}
\rput(3,3){$\sigma$}
\rput(4,3){$\varepsilon$}
\rput(2,2){$\varepsilon$}
\rput(3,2){$\sigma$}
\rput(4,2){$\mathbb{I}$}


\psline[linestyle=dashed](1.5,4)(4,1.5)

\rput(8,3){$
\begin{tabular}[b]{lll}
		&$A_2\times A_3$&$X_{(p|\sigma)}$\\
		\hline
		$\mathbb{I}$&$(1,1)$&$X_1$\\
		$\varepsilon$&$(2,1)$&$X_2$\\
		$\sigma$&$(2,2)$&$X_1+X_2$\\
		\hline
\end{tabular}$}
\end{pspicture}
\vspace{-1cm}
	\caption{Integrable boundary conditions for the Ising model ($q=2$) on a fixed lattice are labelled by the nodes of the graph $A_2\times A_3$; the dashed line separates two equivalent choices of a fundamental domain.}
	\label{fig:statesising}
\end{figure}

\noindent This corresponds to the Ising model on planar triangulations, which is the $A_3$ model in the classification of \cite{Pasquier86} and was first solved on the sphere by Kazakov and Boulatov \cite{Boulatov86,Kazakov86} using the much-studied  $\mathbb{Z}_2$-symmetric Hermitian two-matrix model. The 3 integrable boundary conditions of this model are captured by the linear combinations of $X_i$ shown in Figure \ref{fig:statesising} \cite{Cardy89}: $W_{(1)}(z)$ captures the $S_2\simeq\mathbb{Z}_2$-doublet $\{\mathbb{I},\varepsilon\}$, $W_{(2)}(z)$ the $\mathbb{Z}_2$-singlet $\{\sigma\}$. From \eqref{Z1}, we see that in this case the partition function can be written as

\begin{subequations}
\begin{align}
\label{Z2-1}
Z_{N,2}&=\int\mathrm{d}X_1\mathrm{d}X_2\ \E^{-N\mathrm{tr}[U(X_1)+U(X_2)]}\E^{N\mathrm{tr}(X_1+X_2)^2/2}\\
\label{Z2-2}
	&=\int\mathrm{d}Y\ \E^{-N\mathrm{tr}Y^2/2}\left(\int\mathrm{d}X\ \E^{-N\mathrm{tr}[U(X)-XY]}\right)^2\ .
\end{align}
\end{subequations}

\noindent On the other hand, changing variables to $X_\pm=X_1\pm X_2+t_2/t_3$ and integrating out $X_-$, we obtain the equivalent one-matrix representation going back to \cite{Eynard92},

\begin{equation}
\begin{aligned}
Z_{N,2}&=\mathrm{const.}\times\int\frac{\mathrm{d}X_+\ \E^{-N\mathrm{tr}U_+(X_+)}}{\sqrt{\mathrm{Det}(X_+\otimes\mathbb{I}+\mathbb{I}\otimes X_+)}}\ ,
\end{aligned}
\end{equation}

\noindent where $U'_+(z)=t_3z^2/4-z-t_2(4-t_2)/(4t_3)$ and capital $\mathrm{Det}$ denotes the determinant on $N^2\times N^2$ matrices. Using the defintion \eqref{def-g} in the planar limit, the above expressions respectively imply the following set of equations:

\begin{subequations}
\begin{align}
\label{case2a}
z&=W_Y(z)_--W_Y(z)_++2G_Y^X(z)_+\ ,\\
\label{case2b}
U'(z)&=W_{(1)}(z)_-+G_X^Y(z)_+\ ,\\
\label{case2c}
U'(z)&=W_{(1)}(z)_-+G_{X_1}^{X_2}(z)_++z\ ,\\
\label{case2d}
U'_+\left(z+t_2/t_3\right)&=W_{(2)}(z)_-+W_{(2)}(z)_++W_{(2)}(-z)\ .
\end{align}
\end{subequations}

\noindent Again, the first line is consistent with \eqref{curve2} in Proposition \ref{prop-curve} and the relations \eqref{identificationG} in Subsection \ref{subsec:voiculescu}. Equations \eqref{case2a} and \eqref{case2b} dictate the analytic structure and asymptotic behaviour of $G_Y^X(z)$, cf. Appendix \ref{app:asympt}, Example \ref{ex-ising}. As before, this allows us to compute the spectral curve using \eqref{curve2},

\begin{subequations}
\begin{align}
\label{polynomial1}
F_{(1)}(x,y)&=x^4-2x^3y-\frac{1}{t_3}y^3+\frac{1-t_2}{t_3^2}y^2+x^2y^2-\frac{t_2+2}{t_3}x^2y+\frac{t_3^2-t_2^2}{t_3^2}x^2\nonumber\\
&\quad+\frac{t_2+2}{t_3}xy^2+\frac{t_2^2-t_3^2}{t_3^2}xy+c^{(1)}_{1,1}x+c^{(1)}_{1,0}\ ,\\
F_{(2)}(x,y)&=x^4+\frac{4t_2}{t_3}x^3+\frac{4}{t_3}y^3-x^2y^2-\frac{4+2t_2}{t_3}x^2y-\frac{2 t_2}{t_3}xy^2+\frac{4t_2^2+2t_3^2}{t_3^2}x^2+\frac{8t_2}{t_3^2}y^2\nonumber\\&\quad-\frac{4t_2(2+t_2)}{t_3^2}xy-c^{(2)}_{0,0}x+c^{(2)}_{1,1}y+c^{(2)}_{1,0}\ .
\end{align}
\end{subequations}

\noindent According to Corollary \ref{curve3}, the functions $G_{(p)}^Y(z)$ then satisfy

\begin{equation}
\begin{aligned} 
&F_{(1)}\left(z,G_{(1)}^Y(z)\right)&=0\ ,\quad F_{(1)}\left(G_{(1)}^Y(z)-z,G_{(1)}^Y(z)\right)=0\ ,\\
&F_{(2)}\left(z,G_{(2)}^Y(z)\right)&=0\ ,\quad F_{(2)}\left(G_{(0)}^Y(z)-z,G_{(0)}^Y(z)\right)=0\ .
\end{aligned}
\end{equation} 

\noindent Again we may use the above to compute the analytic structure and asymptotic behaviour of $G_{(p)}^Y(z)$ on all sheets -- see Appendix \ref{app:asympt}. Comparing to \eqref{case2c} and \eqref{case2d}, we conclude that 

\begin{subequations} 
\begin{align}
\label{G-2a}
&G_{(1)}^Y(z)_+=z+G_{X_1}^{X_2}(z)\ ,\quad G_{(1)}^Y(z)_-=t_3z^2+t_2z-W_{(1)}(z)\ ,\\
\label{G-2b}
&G_{(2)}^Y(z)_+=z+W_{(2)}(z)\ ,\quad G_{(2)}^Y(z)_-=t_3z^2/4+t_2z/2-W_{(2)}(z)-W_{(2)}(-z)\ .
\end{align}
\end{subequations}

\noindent Our results reproduce the analytic structure found in \cite{Carroll96,Carroll97,Atkin11} as well as the relation between the $p=1$ and $p=2$ boundary conditions reported in \cite{Atkin12}: at the level of the polynomial equation, the correspondence with the quantities defined therein is

\begin{equation}
W_Y(z) \leftrightarrow W_A(a)\ ,\quad G_Y^{(1)}(z) \leftrightarrow x(a)\ ,\quad G_Y^{(2)}(z) \leftrightarrow m(a)\ .
\end{equation} 

\noindent The polynomial $E(x,y)=-t_3F_{(1)}(x,x+y)$ is of order 3 in both $x$ and $y$, 

\begin{equation}
\begin{aligned}
E(x,y)&=x^3+y^3-t_3x^2y^2-\frac{1-t_2}{t_3}(x^2y+y^2x)-\frac{1-t_2}{t_3}(x^2+y^2)\\
&\quad-\frac{2-2t_2+t_2^2-t_3^2}{t_3}xy-t_3c^{(1)}_{1,1}(x+y)-t_3c^{(1)}_{1,0}\ ,
\end{aligned}
\end{equation}

\noindent and satisfies $E(x,y)=E(y,x)$ and $E(z,G_{X_1}^{X_2}(z))=0$, as follows from comparison of \eqref{case2a} and \eqref{case2b}. This is the usual spectral curve of the two-matrix model introduced by Eynard \cite{Eynard02}.

\subsection{\texorpdfstring{$(q,k)=(3,1)$}{(q,k)=(3,1)} -- 3-states Potts model}
\label{subsec:q=3}

\begin{figure}[h]
\centering
	\begin{pspicture}(10,4)
\psset{unit=.8cm}

\rput(2.8,5){$D_4$}
\psdot[dotsize=.15](2,4)
\psdot[dotsize=.15](3,4)
\psdot[dotsize=.15](4,4.5)
\psdot[dotsize=.15](4,3.5)
\psline(2,4)(3,4)
\psline(3,4)(4,4.5)
\psline(3,4)(4,3.5)

\rput(0,1.5){$A_4$}
\psdot[dotsize=.15](1,3)
\psdot[dotsize=.15](1,2)
\psdot[dotsize=.15](1,1)
\psdot[dotsize=.15](1,0)
\psline(1,3)(1,2)
\psline(1,2)(1,1)
\psline(1,1)(1,0)

\rput(2,3){$\mathbb{I}$}
\rput(3,3){$F$}
\rput(4,3){$\psi,\psi^{\dagger}$}
\rput(2,2){$\varepsilon$}
\rput(3,2){$N$}
\rput(4,2){$\sigma,\sigma^{\dagger}$}
\rput(2,1){$\varepsilon$}
\rput(3,1){$N$}
\rput(4,1){$\sigma,\sigma^{\dagger}$}
\rput(2,0){$\mathbb{I}$}
\rput(3,0){$F$}
\rput(4,0){$\psi,\psi^{\dagger}$}


\psline[linestyle=dashed](0.5,1.5)(5,1.5)

\rput(10,2.5){$
	\begin{tabular}[b]{lll}
		&$A_4\times D_4$&$X_{(p|\sigma)}$\\
		\hline
		$\mathbb{I}$&$(1,1)$&$X_1$\\
		$\psi$&$(1,3)$&$X_2$\\
		$\psi^{\dagger}$&$(1,4)$&$X_3$\\
		$F$&$(1,2)$&$X_1+X_2+X_3$\\
		$\varepsilon$&$(2,1)$&$X_2+X_3$\\
		$\sigma$&$(2,3)$&$X_1+X_3$\\
		$\sigma^{\dagger}$&$(2,4)$&$X_1+X_2$\\
		$N$&$(2,2)$&--\\
		\hline
	\end{tabular}$}
\end{pspicture}
	\vspace{.5cm}
	\caption{Integrable boundary conditions for the 3-states-Potts model ($q=3$) on a fixed lattice are labelled by the nodes of the graph $A_4\times D_4$; the dashed line separates two equivalent choices of a fundamental domain.}
	\label{fig:statespotts}
\end{figure}

\noindent This model is equivalent to the $D_4$ lattice model on planar triangulations, for which $W_{(1)}(z)$ was first calculated by Daul in \cite{Daul94}. The full list of boundary conditions of the $D_4$ lattice model is given in Figure \ref{fig:statespotts} \cite[p.60]{Behrend00}: $W_{(1)}(z)$ captures the $S_3$-triplet $\{\mathbb{I},\psi,\psi^{\dagger}\}$, $W_{(2)}(z)$ the $S_3$-triplet $\{\varepsilon,\sigma,\sigma^{\dagger}\}$, and $W_{(3)}(z)$ the singlet $\{F\}$; thanks to Corollary \ref{curve3}, the spectral curve for the latter also defines another singlet $W_{(0)}(z)$, which may be conjectured to describe the one remaing independent boundary condition $\{N\}$, though herein we will not attempt to prove its equvialence to the microscopic definition given in \cite{Behrend00}. From \eqref{Z1}, we see that the partition function can be written as

\begin{subequations}
\begin{align}
\label{Z-potts-3}
Z_{N,3}&=\int\mathrm{d}X_1\mathrm{d}X_2\mathrm{d}X_3\ \E^{-N\mathrm{tr}[U(X_1)+U(X_2)+U(X_3)]}\E^{N\mathrm{tr}(X_1+X_2+X_2)^2/2}\\
	&=\int\mathrm{d}Y\ \E^{-N\mathrm{tr}Y^2/2}\left(\int\mathrm{d}X\ \E^{-N\mathrm{tr}[U(X)-XY]}\right)^3\ .
\end{align}
\end{subequations}

\noindent Again we may set $X_\pm=X_1\pm X_2+t_2/t_3$ and integrate out $X_-$, which gives

\begin{equation}
\begin{aligned}
\label{lightcone}
Z_{N,3}&=\mathrm{const.}\times\int\frac{\mathrm{d}X_+\mathrm{d}X_3\ \E^{-N\mathrm{tr}[U_+(X_+)+U(X_3)-X_+X_3]}}{\sqrt{\mathrm{Det}(X_+\otimes\mathbb{I}+\mathbb{I}\otimes X_+)}}\ ,
\end{aligned}
\end{equation}

\noindent where $U_+(z)$ is as in the previous section. Using the defintion \eqref{def-g} in the planar limit, the above expressions respectively imply the following set of equations:

\begin{subequations}
\begin{align}
\label{case3a}
z&=W_Y(z)_--2W_Y(z)_++3G_Y^X(z)_+\ ,\\
\label{case3b}
U'(z)&=W_{(1)}(z)_-+G_X^Y(z)_+\ ,\\
\label{case3c}
U'(z)&=W_{(1)}(z)_-+G_{X_3}^{X_1+X_2}(z)_++z\ ,\\
\label{case3d}
U'_+\left(z+t_2/t_3\right)&=W_{(2)}(z)_-+G_{X_1+X_2}^{X_3}(z)_++W_{(2)}(-z)\ .
\end{align}
\end{subequations}

\noindent Once again, the first line is consistent with \eqref{curve2} in Proposition \ref{prop-curve} and the relations \eqref{identificationG} in Subsection \ref{subsec:voiculescu}. The analytic structure and asymptotic behaviour of all relevant functions can be determined as before - cf. Appendix \ref{app:asympt}, Example \ref{ex-potts3}. The resulting spectral curves are

\begin{equation}
\label{polynomial2}
\begin{aligned}
F_{(1)}(x,y)&=x^6+x^5 \left(-\frac{6 t_2}{t_3}-6 y\right)-\frac{4 y^5}{t_3}+x^4 \left(13y^2+\frac{24 t_2 -6 }{t_3}y+\frac{9 t_2^2+2 t_3^2}{t_3^2}\right)\\&\quad+\frac{17-18 t_2}{t_3^2}y^4+x^3 \left(\frac{24-28 t_2}{t_3}y^2+\frac{-12 t_2^2+24 t_2-8 t_3^2}{t_3^2}y-12 y^3-c^{(1)}_{0,0}\right)\\&\quad+x^2 \left(y c^{(1)}_{1,1}+c^{(1)}_{1,0}+\frac{\left(6 t_2-30 \right) y^3}{t_3}+\frac{\left(-15 t_2^2-54 t_2+10 t_3^2+9\right) y^2}{t_3^2}+4 y^4\right)\\&\quad+x \left(-y^2 c^{(1)}_{2,2}-y c^{(1)}_{2,1}-c^{(1)}_{2,0}+\frac{\left(4 t_2 +12 \right) y^4}{t_3}+\frac{\left(18 t_2^2+24 t_2-4 t_3^2-18\right) y^3}{t_3^2}\right)\\&\quad+y^3 c^{(1)}_{3,3}+y^2 c^{(1)}_{3,2}+y c^{(1)}_{3,1}+c^{(1)}_{3,0}\ ,
\end{aligned}
\end{equation}

\begin{equation}
\label{polynomial3}
\begin{aligned}
F_{(3)}(x,y)&=x^6+x^5\left(\frac{18t_2}{t_3}+6y\right)+\frac{108y^5}{t_3}+x^4\left(9y^2+\frac{72t_2-18}{t_3}y+\frac{117t_2^2+6t_3^2}{t_3^2}\right)\\&\quad+\frac{702t_2-243}{t_3^2}y^4+x^3\left(-4y^3+\frac{36t_2-72}{t_3}y^2+\frac{24t_3^2+252t_2^2-216t_2}{t_3^2}y-c^{(3)}_{0,0}\right)\\
&\quad +x^2\left(-12y^4-\frac{90t_2+54}{t_3}y^3+\frac{81-486t_2-135t_2^2+18t_3^2}{t_3^2}y^2+c^{(3)}_{1,1}y+c^{(3)}_{1,0}\right)\\
&\quad +x\left(\frac{36(1-t_2)}{t_3}y^4+\frac{162-234t_2^2-12t_3^2}{t_3^2}y^3-c^{(3)}_{2,2}y^2-c^{(3)}_{2,1}y-c^{(3)}_{2,0}\right)\\
&\quad +c^{(3)}_{3,3}y^3+c^{(3)}_{3,2}y^2+c^{(3)}_{3,1}y+c^{(3)}_{3,0}\ .
\end{aligned}
\end{equation}

\noindent According to Corollary \ref{curve3}, the functions $G_{(p)}^Y(z)$ then satisfy

\begin{equation}
\begin{aligned} 
&F_{(1)}\left(z,G_{(1)}^Y(z)\right)&=0\ , \quad F_{(1)}\left(G_{(2)}^Y(z)-z,G_{(2)}^Y(z)\right)=0\ ,\\
&F_{(3)}\left(z,G_{(3)}^Y(z)\right)&=0\ ,\quad F_{(3)}\left(G_{(0)}^Y(z)-z,G_{(0)}^Y(z)\right)=0\ .
\end{aligned}
\end{equation} 

\noindent As before, the above fixes the analytic structure and asymptotic behaviour of $G_{(p)}^Y(z)$ on all sheets -- see Appendix \ref{app:asympt}. Comparing to \eqref{case3c} and \eqref{case3d}, we conclude that 

\begin{subequations} 
\begin{align}
\label{G-3a}
&G_{(1)}^Y(z)_+=z+G^{X_1+X_2}_{X_3}(z)\ ,\quad G_{(1)}^Y(z)_-=t_3z^2+t_2z-W_{(1)}(z)\ ,\\
\label{G-3b}
&G_{(2)}^Y(z)_+=z+G^{X_3}_{X_1+X_2}(z)\ ,\quad G_{(2)}^Y(z)_-=t_3z^2/4+t_2z/2-W_{(2)}(z)-W_{(2)}(-z)\ .
\end{align}
\end{subequations}

\noindent Similarly, one can show 

\begin{equation}
\label{G-3c}
G_{(3)}^Y(z)_+=z+W_{(3)}(z)\ ,\quad G_{(3)}^Y(z)_-=t_3z^2/9+t_2z/3+\mathcal{O}(z^{-1})\ .
\end{equation}

\noindent $F_{(1)}(x,y)$ corresponds to the spectral curve first described in \cite{Daul94,ZJ99}; the remaining expressions are new results. The polynomials $-t_3F_{(p)}(x,x+y)=4E_{(p)}(x,y)$ are of degree one less in $x$. For example,

\begin{equation}
\begin{aligned}
\label{e1}
E_{(1)}(x,y)&=x^5+y^5+x^4 \left(-\frac{t_3}{4} y^2-\frac{2 t_2 -20}{4 }y-\frac{8-24 t_2}{4 t_3}\right)-\frac{17-18 t_2}{4 t_3}y^4\\&\quad+x^3 \left(\frac{t_3\widetilde{c}_1}{4}-t_3 y^3-\frac{14 t_2 -34}{4 t_3}y^2-\frac{12 t_2^2-84 t_2+32}{4 t_3}y\right)\\&\quad+x^2 \left(\frac{t_3\widetilde{c}_2}{4 }y-\frac{\widetilde{c}_3}{4}-t_3 y^4-\frac{11\left(t_2 -1\right)}{2}y^3-\frac{39 t_2^2-90 t_2-2 t_3^2+57}{4 t_3}y^2\right)\\&\quad+x \left(\frac{ t_3\left(3  c^{(1)}_{2,2}- c^{(1)}_{3,3}\right)}{4}y^2-\frac{t_3 \left(2 c^{(1)}_{3,2}- c^{(1)}_{2,1}\right)}{4}y-\frac{4 t_2 t_3-8}{4}y^4-\frac{9t_2^2-24 t_2-2 t_3^2+25}{2 t_3}y^3\right)\\&\quad-\frac{1}{4} t_3 y^3 c^{(1)}_{3,3}-\frac{1}{4} t_3 y^2 c^{(1)}_{3,2}-\frac{t_3(c^{(1)}_{3,1}-c^{(1)}_{2,0})}{4}x-\frac{1}{4} t_3 y c^{(1)}_{3,1}-\frac{1}{4} t_3 c^{(1)}_{3,0}\ ,
\end{aligned}
\end{equation}

\noindent where 

\begin{subequations}
\begin{align}
\widetilde{c}_1&=c^{(1)}_{2,2}-c^{(1)}_{1,1}-c^{(1)}_{3,3}+c^{(1)}_{0,0}\ ,\\
\widetilde{c}_2&=c^{(1)}_{2,2}-2c^{(1)}_{1,1}-3c^{(1)}_{3,3}\ ,\\
\widetilde{c}_3&=c^{(1)}_{1,0}-c^{(1)}_{2,1}+c^{(1)}_{3,2}\ .
\end{align}
\end{subequations}

\noindent This expression satisfies $E_{(2)}(x,y)=E_{(1)}(y,x)$ and is equivalent to the polynomial $Q(x_3,x_+)$ reported previously by the author in \cite{ksm13}. Upon inspection of \eqref{case3b}, \eqref{case3c} and \eqref{case3d} we conclude that 

\begin{equation}
0=E_{(2)}\left(z,G_{X_1+X_2}^{X_3}(z)\right)=E_{(2)}\left(G_{X_3}^{X_1+X_2}(z),z\right)\ .
\end{equation}

\section{Critical behaviour}
\label{sec:crit}

\noindent This section discusses the critical behaviour of $W_{(p)}(z)$ for $0< q< 4$. The existence of a second-order phase transition for the Potts model in this regime has been demonstrated on a fixed lattice by Baxter \cite{Baxter71,Baxter73}; here we describe their random-lattice counterparts\footnote{When $q>4$, these critical points do not exist, though presumably another critical point emerges as for the $O(n)$ model on planar triangulations, for which $\gamma_s=1/2$ when $n>2$ \cite{Eynard95,Durhuus96}.}. According to Proposition \ref{prop-curve}, it suffices to determine the critical behaviour of $W_Y(z)$ for $1>\nu>0$. Its possible critical exponents are determined by the multiplicity of the singularity at the left edge $z_-$ of the spectral density $\rho_Y(z)$, which controls the large-order behaviour of the generating function $W_Y(z)$. 

\begin{figure}
\label{fig:phasediagram}
\centering
\includegraphics[width=.5\textwidth]{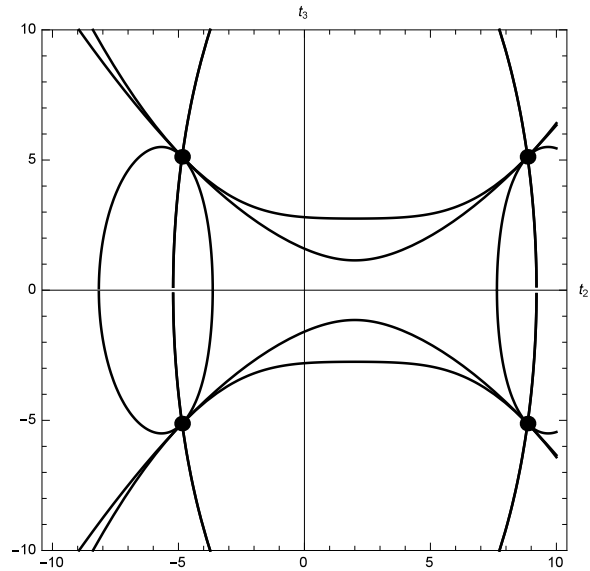}
\caption{A portion of the phase diagram of the 3-states Potts model on planar triangulations. Along the critical lines, $\partial_y^4 E_{(1)}$, $\partial_x\partial_y^3 E_{(1)}$ and $\partial_x^3\partial_y E_{(1)}$ vanish, with the polynomial $E_{(1)}$ as in \eqref{e1}; at the critical points, $\partial_x^4 E_{(1)}$ vanishes in addition.}
\end{figure}

Let us begin with the case of triangulations covered in Proposition \ref{prop-ycurve}. Then $z_-=\delta_U$ when both $t_2$ and $t_3$ are at their critical values $t_{2,c}$, $t_{3,c}$, with $t_{m>3}=0$. When $\nu$ is rational, exact expressions for the critical lines and points can be obtained easily by requiring sufficiently many derivatives of the polynomial \eqref{pol} to vanish; the result is depicted for the example $(q,k)=(3,1)$ in Figure \ref{fig:phasediagram}. For example, from expressions \eqref{polynomial0}, \eqref{polynomial1} and \eqref{polynomial2}, we find

\begin{equation}
(t_{2,c},t_{3,c})=
\begin{cases}
\left(1\pm2\sqrt{3},\pm\sqrt{2}\right)\ ,\quad& q=1\ ,\\
\left(2\pm2\sqrt{7},\pm\sqrt{10}\right)\ ,\quad& q=2\ ,\\
\left(3\pm\sqrt{47},\pm\sqrt{105}/2\right)\ ,\quad& q=3\ .
\end{cases}
\end{equation}

 \noindent Let us parametrise the vicinity of this point by eliminating $\delta_U$ in favour of the scaling parameter $\varepsilon=z_--\delta_U$ and investigate the limit $\varepsilon\to 0$. We would like to expand $W_Y(z)$ in powers of $\varepsilon$, keeping $(z-z_-)/\varepsilon$ finite. Setting again $w(z)=\sqrt{z-\delta_U}$, this requires the expansion of $f(w)$ in \eqref{lambdares} in powers of $\sqrt{\varepsilon}\equiv w_-$, keeping $w/\sqrt{\varepsilon}$ finite. As we show in Appendix \ref{app:auxprob}, 
equation \eqref{f-scaling}, the terms of $\mathcal{O}(\varepsilon^{n\pm\nu/2})$ in the expansion of $f_{\mathrm{sing.}}(w)$ can be written as

\begin{equation}
	\label{Gscaling}
	\varepsilon^{n\pm\nu/2}\left
(t_n^{(\pm)}T_{2n\pm\nu}(-w/\sqrt{\varepsilon})+u_n^{(\pm)}U_{2n\pm\nu}(-w/\sqrt{\varepsilon})\right)\ ,
\end{equation}

\noindent where $T_\nu(w)$ (resp. $U_\nu(w)$) is the Chebyshev function of the first (resp. second) kind as defined in equation \eqref{chebyshevfct}. Using \eqref{Chebyshevdiscont}, the term of same order in the expansion of the discontinuity $f(w)_+-f(w)_-$ across $w/\sqrt{\varepsilon}\in[1,\infty)$ becomes

\begin{equation}
	\label{discont}
	\begin{aligned}		-2\I\sin(\pi\nu)\varepsilon^{n\pm\nu/2}\large\left(u_n^{(\pm)}\frac{T_{2n+1\pm\nu}(w/\sqrt{\varepsilon})}{\sqrt{1-w^2/\varepsilon}}+t_n^{(\pm)}\sqrt{1-w^2/\varepsilon}U_{2n-1\pm\nu}(w/\sqrt{\varepsilon})\large\right)\ .\\
	\end{aligned}
\end{equation}

\noindent Comparing to \eqref{lambdadens} and requiring that $\rho_Y(z)\to 0$ as $z\to z_-$ reveals that this expression must vanish as $w\to\pm\sqrt{\varepsilon}$, which implies that $u_n^{(\pm)}=0$ for all $n$. Using $(z-z_-)/\varepsilon\equiv w^2/\varepsilon-1$ and the relation \eqref{lambdares} together with

\begin{equation}
T_{2-\nu}\left(\sqrt{1-\eta}\right)+T_{2-\nu}\left(-\sqrt{1-\eta}\right)=2\cos\left(\frac{\pi\nu}{2}\right)T_{2-\nu}\left(\sqrt{\eta}\right)\ ,
\end{equation}

\noindent gives the following expansion of $W_Y(z)$:

\begin{equation}
\begin{aligned}
W_Y(z_--\varepsilon \eta)&=W_Y(z_-)+C\varepsilon^{1-\nu/2}T_{2-\nu}\left(\sqrt{\eta}\right)-\varepsilon\frac{2\eta}{4-q}+\mathcal{O}(\varepsilon^{1+\nu/2})\ ,
\end{aligned}
\end{equation}

\noindent where $W_{Y}(z_-)=(2z_-+qt_{2,c}/t_{3,c})/(4-q)$ and $C$ is a normalisation constant. The expansion of $G_Y^{(p)}(z)$ now follows immediately from Proposition \ref{prop-curve}; the leading non-analytic term reads

 \begin{equation}
 C\varepsilon^{1-\nu/2}\left[T_{2-\nu}\left(\sqrt{\eta}\right)-\frac{2p}{q}\cos\left(\frac{\pi\nu}{2}\right)T_{2-\nu}\left(\sqrt{1-\eta}\right)\right]\ .
 \end{equation}
 
\noindent The string exponent $\gamma_s$ predicted by $W_{(p)}(z-z_c)\sim (z-z_c)^{1-\gamma_s}$ is in agreement with previous findings \cite{Daul94,ZJ99,Eynard99}, namely

\begin{equation}
	\gamma_s=\frac{\nu}{\nu-2}\ .
\end{equation}
 
\noindent In particular, $\gamma_s=-1/2$, $-1/3$, $-1/5$ and $0$ for $q=1$, $2$, $3$ and $4$ respectively, which is consistent with Liouville theory interacting with conformal matter of central charges $c_M=0$, $1/2$, $4/5$ and $1$ according to \eqref{KPZ}. Whilst in the first two cases the conformal field theory is unique, there exist two distinct modular invariants at $c_M=4/5$, corresponding to the $(A_4,A_5)$ Virasoro minimal model and the $(A_4,D_4)$ minimal model, which admits a conserved spin-3 current is diagonal under the extended $\mathcal{W}_3$-algebra \cite{Cappelli86,Caldeira03}. In light of the $S_3$-symmetry of the partition function \eqref{Z-potts-3} and the resulting spectrum of boundary conditions \cite{Affleck98} -- cf. Figure \ref{fig:statespotts} -- we expect our equations to describe the latter coupled to Liouville theory, not the former. 

\section{Discussion}
\label{sec:pottsdiscussion}

\noindent Let us summarise our results. Starting from the matrix integral representation of the Potts model on a random lattice in Lemma \ref{lem-Z}, we employed the saddle point apprixomation to express $W_{(p)}(z)$ via the $p$-independent average $W_Y(z)$ in Proposition \ref{prop-curve}. For the case of planar triangulations, Proposition \ref{prop-ycurve} provides an explicit elliptic parametrisation of the latter for arbitrary $q\neq 4$. Just as equation \eqref{Z1} defines an analytic continuation of the partition function to the complex $q$-plane, equation \eqref{curve2} may thus be used to define the analytic continuation of $W_{(p)}(z)$ in the complex $p$- and $q$-plane. What is more, Corollary \ref{curve3} showed that $W_{(p)}(z)$ and $W_{(q-p)}(z)$ can be related algebraically -- in the case studies in Section \ref{sec:cases}, this resulted in the $p$- and $(q-p)$-boundary conditions being described by a single spectral curve defined by the zero locus of \eqref{pol}. Remarkably, equations \eqref{G-2a}, \eqref{G-3a} and \eqref{G-3b} indicate that $G_{(p)}^Y(z)-z$ and $G_{(q-p)}^Y(z)-z$ are functional inverses, generalising the well-known duality\footnote{Note that this involution is in general distinct from the Kramers-Wannier duality \cite{Kramers41} on the dynamical lattice: e.g. for $q=3$, the latter interchanges $p=1$ with $p=3$, and $p=2$ with $p=0$, mixing singlets and triplets \cite{Affleck98}.} interchanging the two matrices of the symmetric Hermitian two-matrix model \cite{BEH01a,BEH01b}. 

Our results naturally pave the way for a number of further developments: Firstly, going beyond the planar limit, as was done in \cite{eynard11b} for the $O(n)$ model on random lattices, it would be interesting to explore if and when the curves defined by \eqref{pol} can be used as a valid part of the initial data of the topological recursion algorithm \cite{Eynard07}, which allows to compute averages to all orders in $1/N$. Secondly, for general values of $h$ in Lemma \ref{lem-Z}, the remarkably simple result in Lemma \ref{lem-loop} should enable us to investigate the boundary renormalisation group flow relating boundary conditions with different $p$. This flow is expected to induce a partial order on the spectrum of boundary states in accordance with the boundary analogue of the $c$-theorem \cite{Zamolodchikov86}, as conjectured in \cite{Affleck91} and finally proven by Friedan and Konechny \cite{Friedan03}; it would be interesting to derive this fact directly from the matrix model, thus extending the work of \cite{Carroll96,Carroll97,Atkin12}.

Finally, it would be instructive to check if the universal results of Section \ref{sec:crit} can be reproduced by other means, e.g. by explicitly constructing the corresponding conformal field theory. Remarkably, as exemplified by the case of the $D_4$ model, this appears to require a non-diagonal partition function in the Liouville sector in general. From this perspective, various other corners of the $(q,k)$-parameter space also warrant more detailed investigations. Of particular interest would be the computation of the scaling behaviour for strongly coupled models with $q>4$: one might wonder if there exist analogues of the critical points of the $O(n)$ model on a random lattice with $n>2$ reported in \cite{Eynard95,Durhuus96}. For the models with $q=5^2-6n$ for $n\in\{1,2,3\}$ in the infinite $k$ limit, $c_M=18$, $12$ and $6$, respectively, and Liouville theory allows a truncation to a tachyon-free spectrum \cite{Gervais84,Das89}; the matrix model might help in reconciling the conflicting CFT predictions \cite{Bilal92} and \cite{Gervais94}. Finally, it might be of interest to enquire about the existence of the `t Hooft limit $q\to\infty$, $k\to\infty$, $q/(q+k)$ fixed, which has been studied for the pure $\mathcal{W}_q$ minimal models in \cite{Gaberdiel13}.

\chapter{The Critical Potts Model Coupled to Liouville Theory}
\label{chap:cft}
\thispagestyle{title}
\section{Overview}

\noindent In this chapter, we investigate the spectrum of the $\mathcal{W}_q$ conformal minimal models coupled to gravity in two dimensions. We shall consider the theory on the sphere and the disk. Requiring the overall conformal anomaly to vanish allows for the interpretation of these theories as a family of bosonic string backgrounds. A new feature with respect to the so-called minimal string (i.e., a Virasoro minimal model coupled to Liouville theory) is the presence of conversed currents with integer spins up to $q$ on the worldsheet which -- with the exception of the stress-energy tensor -- remain \em ungauged\em. The matter sector consequently enjoys an extended non-linear symmetry admitting additional globally conserved charges given by the generators of the so-called $\mathcal{W}_q$ algebra. A major motivation for invoking the latter is that it can be understood as a continuous extension of a discrete symmetry arising from the continuum limit of a critical statistical model on a random lattice. In particular, when $q\leq 4$, such theories are expected to provide a description of the scaling limit of the Potts model on a random planar lattice with discrete symmetry group $S_q$. Unlike two-dimensional $\mathcal{W}$-gravity, which is obtained from these systems by coupling the remaining currents to higher-spin gauge fields, and for which the critical target space dimension increases with $q$ \cite{Bouwknegt92b}, the former enter a strong-coupling regime with tachyonic instabilities when the central charge of the minimal model exceeds one.

The unitary $\mathcal{W}_q$ CFTs also provide a dual description of three-dimensional spin-$q$ gravity coupled to scalar matter, with Newton coupling $G_N=\frac{3}{2}\ell c_M$, where $c_M$ denotes the central charge \eqref{cmatter} of the minimal model and $-1/\ell^2$ is the cosmological constant: for example, the torus partition function of the CFT equals the semiclassical partition function of the gravity theory in thermal Anti-de Sitter space with radius $\ell$ \cite{Gaberdiel13}. The perturbative excitations of the latter are described by Chern-Simons theory with an $\mathfrak{sl}(q,\mathbb{R})\oplus\mathfrak{sl}(q,\mathbb{R})$-valued connection; coupling the boundary CFT to Liouville theory such that the overall conformal anomaly vanishes corresponds to switching from Dirichlet to Neumann boundary conditions for the metric on the boundary of AdS \cite{Compere08}. In this way, our computations also solve a problem in three-dimensional spin-$q$ gravity with negative cosmological constant, with the boundary metric allowed to fluctuate but the asymptotic behaviour of the higher spin gauge fields held fixed. The presence of tachyonic excitations for $c_M\geq 1$ indicates the perturbative instability of the boundary condition for large enough $G_N/\ell$. 

 This chapter is organised at follows: After determining the spectrum of theory on the sphere in Section \ref{sec:bulkspectrum} from the cohomology of the BRST operator associated with the diffeomorphism symmetry, we proceed to introduce a family of conformally invariant boundary conditions from the tensor product of the Liouville and minimal model boundary states Section \ref{sec:boundarystates}. We present evidence that upon analytic continuation of the boundary cosmological constant, this construction in fact overcounts the number of distinct boundary conditions, as has previously been observed for the minimal string in \cite{Seiberg03}. Lastly, a summary and discussion of our results is provided in Section \ref{sec:discussioncft}.

\section{Bulk states}
\label{sec:bulkspectrum}

\noindent Here we discuss the spectrum of physical states on $\mathbb{CP}^1$. For $q=2$, our results reduce to those for the Virasoro minimal models coupled to gravity, for which the spectrum was first determined in \cite{LZ91,LZ92}. These results were rederived in \cite{Bouwknegt92} in a more elementary free field formalism and our approach will be close in spirit. 

The reparametrisation invariance requires the sum $c_M+c_L+c_{gh}$ to vanish, which according to \eqref{cmatter} and \eqref{cLiouville} gives a condition on the background charges, fixing the Liouville coupling $b$ for given $q$, $p$ and $p'$:

\begin{equation}
\label{c=0}
Q_L^2-Q_M^2=\frac{26-q}{12}\;,
\end{equation}

\noindent where here and in what follows we abbreviated $Q_M^2=Q_0^2\rho\cdot\rho$. When $c_M\leq 1$, the susceptibility exponent $\gamma_s=1-b^{\pm 2}$ then follows from \eqref{KPZ}. For later comparison with the matrix model with cubic potential, we print this relationship for the simplest model with $k=1$ in \eqref{affinelevel}, i.e. $p'-p=p-q=1$:

\begin{equation}
\label{stringexpcft}
\gamma_s=\frac{1}{12}\left(1-\frac{6}{(q+2)}\pm\frac{\sqrt{(4-q)(52+23q)}}{(q+2)}\right)
\end{equation}

\noindent The above formula produces the values $\gamma_s=-1/2$, $-1/3$,$-1/5$ and $0$ for $q=1$, $2$, $3$ and $4$, respectively if we pick the negative branch of the square root such that the weak coupling regime $2/3\leq b^2 \leq 1$ corresponds to $1\leq q\leq 4$\footnote{This is also the only choice for which the cosmological constant operator obeys the Seiberg bound.}; the strong-weak duality $b \to1/b$ permutes these two branches. When \eqref{c=0} holds, the BRST charge $\D$ defined in \eqref{brst} becomes nilpotent and denoting the irreducible $\mathcal{W}_q$-module with conformal dimension \eqref{matterweight} by $\mathcal{M}(\lambda)$, we define the holomorphic part of the physical Hilbert space of the $\mathcal{W}_q$ minimal model coupled to gravity as

 \begin{equation}
 \bigoplus_{\lambda\in\mathcal{B}^{(q)}_{p,p'}}\bigoplus_{n\in\mathbb{Z}}H^n(\mathcal{M}(\lambda)\otimes\mathcal{F}_L(P)\otimes\mathcal{F}_{gh},\D)\ ,
 \end{equation}
 
\noindent where $\mathcal{F}_L(P)$ and $\mathcal{F}_{gh}$ are defined in \eqref{FockLiouville} and \eqref{Fockghost}, respectively and $H^n$ denotes the subspace of $\mathrm{Ker}\ \D/\mathrm{Im}\ \D$ with ghost number $n$. In what follows, we shall argue that for a suitable choice of the fundamental domain $\mathcal{B}^{(q)}_{p,p'}$, the following result holds:

\begin{prop}
\label{prop:cohomology} Let the complex $\mathcal{C}^\perp(\lambda)$ be defined as

\begin{equation}
\begin{aligned}
\label{Cperp}
\mathcal{C}^\perp(\lambda)&=\bigoplus_{N^i\in\mathbb{Z}}\bigoplus_{w\in S_q}\mathcal{F}^{\perp}\left(P_{\lambda^w-pp'N^ie_i},\lambda^w-pp'N^ie_i\right)\ ,\\
P_{\lambda^w-pp'N^ie_i}^2&=\frac{q-2}{12}-\frac{1}{pp'}\left(\lambda^w-pp'N^ie_i\right)^2\ .
\end{aligned}
\end{equation}

\noindent Then for any $\lambda\in\mathcal{B}^{(q)}_{p,p'}$,
\begin{equation}
\label{cohomologyresult}
H^n\left(\mathcal{M}(\lambda)\otimes\mathcal{F}_L(P)\otimes\mathcal{F}_{gh},\D\right)\simeq H^n\left(\mathcal{C}^{\perp}(\lambda),\D'\right)\ .
\end{equation}
\end{prop}

\noindent The Fock space of transverse oscillations $\mathcal{F}^\perp(P,\lambda)$ in the above is defined in \eqref{Fperp} and $\lambda^w$ is as in \eqref{felder}. In particular, the highest-weight states created by the `tachyon' operators

\begin{equation}
\label{tachyons}
\mathcal{T}_{\lambda}(z)=c(z)V_{Q_L+\I P_{\lambda}}^L(z)V_{Q_0\rho-\frac{1}{\sqrt{pp'}}\lambda}^M(z)\ .
\end{equation}

\noindent are always contained in the cohomology of $\D'$. When $P_{\lambda^w-pp'N^ie_i}^2<0$, these states are non-normalisable and the sign of the square root can be fixed by the Seiberg bound $\I P>0$. To see explicitly when this prescription breaks down, note $P_{\lambda^w-pp'N^ie_i}^2\leq P_{\rho}^2=(c_M-1)/12$, where $\lambda=\rho$ labels the dressed identity field, for which

\begin{equation}
\frac{1}{12}\frac{q-4}{q+2}\leq P_{\rho}^2\leq\frac{q-2}{12}\ .
\end{equation}

\noindent Hence, ``macroscopic'' states with finite real $P$ are absent from the spectrum only for $c_M\leq 1$, signalling the well-known Kosterlitz-Thouless transition of Liouville theory at $c_L=25$. When $c_M>1$, we expect the complex values of $\gamma_s$ yielded by \eqref{stringexpcft} -- and, more generally, the KPZ relations -- to be unreliable. 

The result \eqref{cohomologyresult} follows from an application of the following useful result on doubly graded complexes:

\begin{lem}[\cite{Bershadsky89,Bouwknegt90}]\label{lem:doublecomplex} Let $\D$ and $\D'$ be commuting, nilpotent differentials on a complex $\mathcal{C}$ and suppose $H^n(\mathcal{C},\D)=0$ for $n\neq 0$ and $H^n(\mathcal{C},\D')=0$ for $n\neq 0$. Then

\begin{equation}
H^n\left(H^0(\mathcal{C},\D'),\D\right)\simeq H^n\left(H^0(\mathcal{C},\D),\D'\right)\ .
\end{equation}
\end{lem}

\noindent The remainder of this section is devoted to demonstrating that in the case at hand, the conditions in the above lemma are indeed satisfied. To this end, we introduce the complex 

\begin{equation}
\mathcal{C}(P,\lambda)=\bigoplus_{N^i\in\mathbb{Z}}\bigoplus_{w\in S_q}\mathcal{F}(P,\lambda^w-pp'N^ie_i)\ ,\quad \lambda\in\mathcal{B}^{(q)}_{p,p'}\ ,
\end{equation}

\noindent where we defined an extended Fock space from the tensor product of \eqref{Fockmatter}, \eqref{FockLiouville} and \eqref{Fockghost},

\begin{equation}
\label{fock-ext}
\mathcal{F}(P,\lambda)=\mathcal{F}_M(\lambda)\otimes\mathcal{F}_L(P)\otimes\mathcal{F}_{gh}\;.
\end{equation} 

\noindent Firstly, on this complex, we must have $\D^2=(\D')^2=0$ and also $[\D,\D']=0$, since $\D'$ acts nontrivially only on $\mathcal{F}_M(\lambda)$ and $[\D',L_0^M]=0$ by construction. Secondly, recall from the introduction in Subsection \ref{subsec:cft} that $H^n(\mathcal{C}(P,\lambda),\D')=0$ for $n\neq 0$ is already implied in the free-field resolution of $\mathcal{M}(\lambda)$ along the lines of \cite{Felder88,Bouwknegt90,Bouwknegt91}. To show applicability of Lemma \ref{lem:doublecomplex}, it thus remains to characterise $H^{n}(\mathcal{F}(P,\lambda),\D)$. The result is 

\begin{lem} Pick a basis in root space such that  $\rho=(|\rho|,0,\dots,0)$, and let
\begin{equation}
\label{Fperp}
\mathcal{F}^{\perp}(P,\lambda)=\mathrm{span}\left\{|P\rangle_L\otimes |0\rangle_{gh}\otimes \prod_{i=2}^{q-1}\prod_{n_j^{(i)}=1}^{k_i}a^i_{-n_j^{(i)}}|\lambda\rangle_M\large\left.\large\right|k_i\geq0,\ 0<n_1^{(i)}\leq\dots\leq n_{k_i}^{(i)}\right\}\;.
\end{equation}

\noindent Then $H^{n}\left(\mathcal{F}(P,\lambda),\D\right)=\delta_{n,0}\mathcal{F}^{\perp}(P_\lambda,\lambda)$, where $P_\lambda$ is the solution to

 \begin{equation}
\label{physweights}
P^2_{\lambda}=\frac{q-2}{12}-\frac{\lambda^2}{pp'} \;.
 \end{equation}
\end{lem}

\begin{proof} Let us begin by considering those states that satisfy $b_0|\psi\rangle=0$. On this subspace, the total energy $L_0=L_0^M+L_0^L+L_0^{gh}=\{b_0,\mathrm{d}\}$ also annihilates physical states\footnote{This is sometimes called the Hamiltonian constraint in quantum gravity.} and moreover commutes with $\D$ so that we may restrict our attention to the subspace 

\begin{equation}
\mathcal{F}_{\text{rel}}(P,\lambda)=\mathcal{F}(P,\lambda)\cap\mathrm{Ker}\ L_0\cap\mathrm{Ker}\ b_0
\end{equation}

\noindent and determine the so-called relative cohomology of the restriction $\mathrm{d}_{\text{rel}}$ of $\mathrm{d}$ to this subspace, which is \cite{Bouwknegt92}

\begin{equation}
\label{restricted}
\mathrm{d}_{\text{rel}}=\mathrm{d}+b_0\sum_{n\neq0}n\ :c_n c_{-n}:-c_0L_0\ .
\end{equation}

\noindent Since $b_0$ (anti-)commutes with all modes besides $c_0$, the full -- or ``absolute'' -- cohomology is given by

\begin{equation}
\label{abs-rel}
H^{n}\left(\mathcal{F}(P,\lambda\right),\mathrm{d})=H^{n}\left(\mathcal{F}_{\text{rel}}(P,\lambda),\mathrm{d}_{\text{rel}}\right)\oplus c_0H^{n-1}\left(\mathcal{F}_{\text{rel}}(P,\lambda),\mathrm{d}_{\text{rel}}\right)\;.
\end{equation}

\noindent To expose the physical modes, we pick a basis in root space such that $\rho=(\sqrt{\rho\cdot\rho},0,\dots,0)$ and transform the fields into `lightcone' variables,

\begin{equation}
\begin{aligned}
q^{\pm}&=\frac{1}{\sqrt{2}}(\phi_0^1\pm \I \varphi_0)\;,\\
p^{\pm}_n&=\frac{1}{\sqrt{2}}\left(a_0^1\pm\I \alpha_0-(n+1)(Q_M\mp Q_L)\right)\;,\\
\alpha_n^{\pm}&=\frac{1}{\sqrt{2}}(a^1_n\pm \I \alpha_n)\;,\quad n\neq 0\;.
\end{aligned}
\end{equation}

\noindent We note the resulting commutation relations

\begin{equation}
[q^{\pm},p^{\mp}_0]=\I\;,\qquad[\alpha^{\pm}_n,\alpha^{\mp}_m]=n\delta_{m+n,0}\;,
\end{equation}

\noindent and refer to the remaining $a_n^i$, $i=2\dots q-1$ as the transverse modes. In these variables, the restriction to the kernel of $L_0$ reads

\begin{equation}
\label{mass-shell}
L_0|\psi\rangle=(L_0^{\perp}+L_0^{\parallel})|\psi\rangle=0\;,
\end{equation}

\noindent where, using the expressions \eqref{Lnmatter}, \eqref{LnLiouville} and \eqref{Lnghost},

\begin{equation}
\begin{aligned}
L_0^{\parallel}&=\ p_0^+p_0^- +\sum_{n\neq0}\ (\ :\alpha_n^+\alpha_{-n}^-:+\ n:c_{-n}b_n:)+\frac{2-q}{24}\;,\\
L_0^{\perp}&=\frac{1}{2}\sum_{n\in\mathbb{Z}}\sum_{i=2}^{q-1}\ :a_n^ia_{-n}^i:\;,
\end{aligned}
\end{equation}

\noindent and \eqref{restricted} can be decomposed as $\mathrm{d}_{\text{rel}}=\mathrm{d}^{\parallel}_{+}+\mathrm{d}^{\parallel}_{-}+\mathrm{d}^{\parallel}+\mathrm{d}^{\perp}$, where

\begin{subequations}
\begin{align}
\mathrm{d}^{\parallel}_{\pm}&=\sum_{n\neq0}p_n^{\pm}\ :c_{-n}\alpha_n^{\mp}:\;,\\
\mathrm{d}^{\parallel}&=\sum_{n,m\neq0\atop m+n\neq 0}:c_{-n}\left(\alpha^+_{-m}\alpha^-_{m+n}+\frac{1}{2}(m-n)c_{-m}b_{m+n}\right):\;,\\
\mathrm{d}^{\perp}&=\frac{1}{2}\sum_{i=2}^{q-1}\sum_{n\neq0}\ :c_{-n} a_m^ia_{n-m}^i:\;.
\end{align}
\end{subequations}

\noindent We observe that $\mathrm{d}^{\parallel}_{+}$ is in fact nilpotent on the extended Fock space and (anti-)commutes with both $b_0$ and the total energy. We can therefore determine the cohomology $\mathrm{d}^{\parallel}_{+}$ on $\mathcal{F}$ first and thereafter restrict to $\mathcal{F}_{\text{rel}}$. The procedure to determine the cohomology of $\mathrm{d}^{\parallel}_{+}$ on $\mathcal{F}$ is analogous to that for the critical bosonic string: We distinguish the cases

\begin{enumerate}
\item Either $p_n^+|\lambda\rangle_M\otimes|P\rangle_L\neq 0$ or $p_n^-|\lambda\rangle_M\otimes|P\rangle_L\neq 0$ $\forall\ n\in\mathbb{Z}\setminus \{0\}$,

\item Otherwise, i.e. $p_{n_\pm}^\pm|\lambda\rangle_M\otimes|P\rangle_L=0$ for a pair of non-zero integers $(n_+,n_-)$. 
\end{enumerate}

\noindent In the first case, we may assume w.l.o.g. that the operator

\begin{equation}
\label{op}
\mathcal{O}=\sum_{n\neq0}(p_n^+)^{-1}:\alpha_{-n}^+b_n:\;
\end{equation}

\noindent exists. We observe that only states annihilated by

\begin{equation}
\{\mathcal{O},\mathrm{d}_+^{\parallel}\}=\sum_{n\neq 0}(\ :\alpha_n^+\alpha_{-n}^-:\ +n\ :c_{-n}b_n:\ )
\end{equation}

\noindent may be physical since any other state closed under $\mathrm{d}^{\parallel}_{+}$ is also $\mathrm{d}^{\parallel}_{+}$-exact. But the above expression is the level operator for the modes $\alpha_n^{\pm}$, $b_n$ and $c_n$. This implies

\begin{equation}
\label{cohomologyresult2}
H^n\left(\mathcal{F}(P,\lambda),\mathrm{d}_+^\parallel\right)=\delta_{n,0}\mathcal{F}^{\perp}(P,\lambda)\ ,
\end{equation}

\noindent where $\mathcal{F}^{\perp}(P,\lambda)$ is given by \eqref{Fperp}. 
\noindent Finally, we need to restrict to the subspace $\mathcal{F}_{\text{rel}}$ by imposing \eqref{mass-shell}, i.e.

\begin{equation}
\left(p_0^+p_0^-+L_0^{\perp}+\frac{2-q}{24}\right)|\lambda\rangle_M\otimes|P\rangle_L\otimes|0\rangle_{gh}=0\;.
\end{equation}

\noindent From the expressions of the Liouville and matter conformal weights and using \eqref{c=0}, we see that the above equation holds iff $P=\pm P_{\lambda}$, where $P_\lambda$ solves the equation \eqref{physweights}.

The second case occurs iff the following equations hold simultaneously: 

\begin{subequations}
\label{zeroespm}
\begin{align}
\frac{n_+-n_-}{2}Q_M-\frac{n_++n_-}{2}Q_L&=\I P\ ,\\
\frac{n_++n_-}{2}Q_M-\frac{n_+-n_-}{2}Q_L&=-\frac{1}{\sqrt{pp'}}\frac{\lambda\cdot\rho}{\sqrt{\rho\cdot\rho}}\ .
\end{align}
\end{subequations}

\noindent Such a state would survive the projection onto $\mathrm{Ker}\ L_0$ when $\frac{1}{24}(n_+n_-(26-q)+q-2)$ is a positive integer. However, for each such $\lambda\in\mathcal{B}^{(q)}_{p,p'}$ there exists another choice of fundamental domain $\widetilde{\mathcal{B}}^{(q)}_{p,p'}$ and  $\widetilde{\lambda}\in\widetilde{\mathcal{B}}^{(q)}_{p,p'}$ such that 
\begin{subequations}
\begin{align}
&L_0^M|\lambda\rangle_M=L_0^M|\widetilde{\lambda}\rangle_M\ ,\\
&\lambda^w-pp'N^ie_i\neq\widetilde{\lambda}^w-pp'N^ie_i\quad\forall w\in S_q\ ,\quad N^i\in\mathbb{Z}\ .
\end{align}
\end{subequations}
\noindent For example, $\mathcal{B}^{(q)}_{p,p'}$ may be chosen such that $\lambda\cdot e_i\equiv p'r^i-ps^i\geq 1$ \cite{Caldeira04}; another choice can be obtained from it by imposing $\widetilde{\lambda}\cdot e_i\leq-1$ for some $i$ instead\footnote{For $q=2$, this simply corresponds to the reflection symmetry of the Kac table.}. We may therefore always choose a resolution of $\mathcal{M}(\lambda)=H^0(\mathcal{C}(\widetilde{\lambda}),d')$ for which \eqref{zeroespm} is never satisfied. On the latter, the operator \eqref{op} is well defined, thus yielding the same result as in the previous case.
\end{proof} 

\begin{rem} It is instructive to see how the known results for the Virasoro minimal model coupled to gravity are recovered from the above results when $q=2$: then the Liouville coupling satisfies $b^2=p/p'$ and we obtain the particularly simple relationship $\gamma_s=1-p'/p$ since $c_M\leq 1$. The Fock module of transverse oscillations $\mathcal{F}^{\perp}(\lambda)$ reduces to $\mathbb{C}$, with $|\lambda\rangle_M\otimes|P_{\lambda}\rangle\otimes|0\rangle_{gh}$ the only state. Introducing the Kac labels $1\leq r< p$, $1\leq s<p'$ by setting $ e_1\cdot\lambda^w=p'wr-ps$, we find that the transverse complex \eqref{Cperp} becomes

\begin{equation}
\begin{aligned}
\mathcal{C}^{\perp}(r,s)&=\bigoplus_{w\in\{-1,1\}}\bigoplus_{N\in\mathbb{Z}}\mathrm{span}\left\{\left|\lambda^w-Npp' e_1;P_{\lambda^w-Npp' e_1}\right\rangle\right\}\;,\\
P_{\lambda^w-Npp' e_1}&=\pm\frac{\I}{\sqrt{2pp'}}\left(wp'r-p s+2pp'N\right)\;.
\end{aligned}
\end{equation}

\noindent The physical states are then given by the cohomology of $\mathrm{d}^{\prime}$ on $\mathcal{C}^{\perp}(r,s)$, which is exactly the procedure first used in \cite{Bouwknegt92} to determine the spectrum; hence we recover the familiar result of Lian and Zuckerman \cite{LZ91} for the physical states of the Virasoro minimal model coupled to gravity. 
\end{rem}

\section{Boundary states}
\label{sec:boundarystates}

\noindent  Using the operator-state correspondence and modular invariance, we associate to each conformal boundary condition a ``boundary state'' in the  of  $L_n-\bar{L}_{-n}$ of the physical Hilbert space; identities in this section are therefore implied to hold modulo BRST exact terms. Because the Virasoro algebra acts diagonally on the tensor factors in \eqref{fock-ext}, conformal invariance has to be preserved in each sector independently. Our ansatz for a conformally invariant boundary state, given $\lambda\in\mathcal{B}^{(q)}_{p,p'}$, is thus the tensor product of \eqref{mattercardy}, \eqref{BLiouville} and \eqref{Bghost},

\begin{equation}
\label{bdryansatz}
|\sigma\rangle_{\lambda}=|\lambda\rangle_C\otimes|\sigma\rangle_{\text{FZZT}}\otimes|B\rangle_{gh}\;.
\end{equation}

\noindent We now present some evidence that the above definition actually overcounts the number of independent boundary conditions if we analytically continue the boundary cosmological constant \eqref{bdrycc}, as was observed in the case $q=2$ in \cite{Seiberg03}. Consider the one-point function of a tachyon operator \eqref{tachyons} on the upper half plane with matter boundary condition $|\sigma\rangle_{\lambda'}$ on the real line,

\begin{equation}
\langle\mathcal{T}_{\lambda}\rangle_{\lambda'}(\sigma)=\lim_{z,\bar{z}\to \infty}\ \langle 0|\mathcal{T}_{\lambda}(z)\bar{\mathcal{T}}_{\bar{\lambda}}(\bar{z})|\sigma\rangle_{\lambda'}\;.
\end{equation}

\noindent In particular, the one-point function of the dressed identity computes the first derivative of the partition function on the disk with boundary condition $\lambda$,

\begin{equation}
\langle\mathcal{T}_{\rho}\rangle_{\lambda}(\sigma)=\left.\frac{\partial D(\mu,\mu_B;\lambda)}{\partial\mu}\right|_{\mu_B}\ .
\end{equation}

\noindent The factorisation of the one-point function into a product of matter and Liouville contributions implies 

\begin{equation}
\label{one-pt-ratio}
\langle\mathcal{T}_{\lambda}\rangle_{\lambda^{\prime}}(\sigma)=\frac{S_{\lambda\lambda^{\prime}}}{S_{\lambda\rho}}\langle\mathcal{T}_{\lambda}\rangle_{\rho}(\sigma)\ ,
\end{equation}

\noindent where the modular $S$-matrix is as in \eqref{modsmatrix}. The ratio of $S$-matrix elements can be written in terms of $SU(q)$ characters,

\begin{equation}
\frac{S_{\lambda\lambda^{\prime}}}{S_{\rho\lambda^{\prime}}}=\chi_{r^i\omega_i-\rho}\left(\frac{2\pi\lambda^{\prime}}{p}\right)\chi_{s^i\omega_i-\rho}\left(-\frac{2\pi\lambda^{\prime}}{p'}\right)\ .
\end{equation}

\noindent Applying the Weyl character formula to the above results in

\begin{equation}
\begin{aligned}
\frac{S_{\lambda\lambda^{\prime}}}{S_{\rho\lambda^{\prime}}}&=\sum_{\mu\in\Omega_{r^i\omega_i-\rho}}\text{mult}_{r^i\omega_i-\rho}(\mu)\E^{2\pi \I\lambda^{\prime}\cdot\mu/p}\sum_{\nu\in\Omega_{s^i\omega_i-\rho}}\text{mult}_{s^i\omega_i-\rho}(\nu)\E^{-2\pi \I\lambda^{\prime}\cdot\nu/p'}\ ,
\end{aligned}
\end{equation}

\noindent where $\Omega_{r^i\omega_i-\rho}$ (resp. $\Omega_{s^i\omega_i-\rho}$) denotes the set of weights of the $\widehat{\mathfrak{su}}(q)_{k}$ (resp. $\widehat{\mathfrak{su}}(q)_{k+1}$) representation of highest weight $r^i\omega_i-\rho$ (resp. $s^i\omega_i-\rho$) and $\text{mult}_{r^i\omega_i-\rho}(\mu)$ denotes the $S_q$-invariant multiplicity of the corresponding state. We summarise the above with the abbreviated notation

\begin{equation}
\label{smatrixcharacter}
\frac{S_{\lambda\lambda^{\prime}}}{S_{\rho\lambda^{\prime}}}=\sum_{\mu,\nu}\text{mult}_{\lambda-(p'-p)\rho}(\mu,\nu) \E^{2\pi \I \lambda^{\prime}\cdot(\mu/p-\nu/p')}\ .
\end{equation}

\noindent To obtain relations between different boundary states, we introduce operators that change the boundary conditions $\lambda$ and $\sigma$ when acting on $|\sigma\rangle_{\lambda}$: 

\begin{equation}
D_L(\sigma^{\prime})|\sigma\rangle_{\lambda}=|\sigma^{\prime}+\sigma\rangle_{\lambda}\ ,\qquad D_M(\lambda)|\sigma\rangle_{\rho}=|\sigma\rangle_{\lambda}\ .
\end{equation}

\noindent The $\sigma$-translation operator can be represented explicitly as

\begin{equation}
D_L(\sigma)=\E^{\pi\sigma(\bar{\alpha}_0-\alpha_0)}\ .
\end{equation} 

\noindent We claim the operator changing the matter boundary condition can be written as

\begin{equation}
D_M(\lambda)=\sum_{\mu,\nu}\text{mult}_{\lambda-(p'-p)\rho}(\mu,\nu)\E^{\frac{2\pi \I}{\sqrt{pp'}} (Q_0\rho- a_0)\cdot(p'\mu-p\nu)}\ .
\end{equation}

\noindent To see this, note first that the coherent states are eigenstates of $D_M( \lambda)$,

\begin{equation}
D_M(\lambda^{\prime})|B(\lambda^w)\rangle_\Lambda=\left|B(\lambda^w)\right\rangle_\Lambda\sum_{\mu,\nu}\text{mult}_{\lambda^{\prime}-(p'-p)\rho}(\mu,\nu) \E^{2\pi \I (\lambda+(1-w)r^i\omega_i)\cdot(\mu/p-\nu/p')}\ .
\end{equation}

\noindent To determine the action of $D_M(\lambda)$ on the Ishibashi states, we need to sum this expression over the Felder complex according to \eqref{matterishibashi}. To this end, we write $w$-dependent phase contribution as 

\begin{equation}
r^i(w\omega_i)\cdot\left(p'\mu-\nu\right)=r^i\left(p'\mu^j-p\nu^j\right)\omega_i\cdot (w^{-1}\omega_j)\ . 
\end{equation}

\noindent Noting that the infinite sum in \eqref{matterishibashi} over the $N^i$ produces the irrelevant phase $2\pi \I\sum_iN^i(p'\nu^i-p\mu^i)$, we thus find

\begin{equation}
\begin{aligned}
D_M(\lambda^{\prime})|\lambda;\Lambda\rishi_M
	&=\sum_{w\in S_q}\sum_{N^j\in\mathbb{Z}}\kappa^w_N D_M(\lambda^{\prime})\left|B(\lambda^{w}-pp'N^j e_j)\right\rangle_\Lambda\\
	&=\frac{S_{\lambda^{\prime}\lambda}}{S_{\rho\lambda}}\sum_{w\in S_q}\sum_{N^j\in\mathbb{Z}}\kappa^w_N\left|B(\lambda^{w}-pp'N^j e_j)\right\rangle_\Lambda\\
	&=\frac{S_{\lambda^{\prime}\lambda}}{S_{\rho\lambda}}|\lambda;\Lambda\rishi_M\ ,
\end{aligned}
\end{equation}

\noindent where we used the identity \eqref{smatrixcharacter}. In conjunction with \eqref{mattercardy} it follows that $D_M(\lambda)$ takes the identity Cardy state to the state $|\lambda\rangle_C$ as advertised. Now, by construction, the bulk tachyons \eqref{tachyons} create eigenstates $\langle\mathcal{T}_{\lambda}|=\lim_{z,\bar{z}\to\infty}\langle 0|\mathcal{T}_{\lambda}(z)\bar{\mathcal{T}}_{\bar{\lambda}}(\bar{z})$ of both $D_L(\sigma)$ and $D_M(\lambda)$,

\begin{subequations}
\begin{align}
\label{cond1}
&\langle\mathcal{T}_{\lambda}|D_L(\sigma')=\E^{\I\pi\sigma'(P_{\lambda}-P_{\bar{\lambda}})}\langle\mathcal{T}_{\lambda}|\ ,\\
\label{cond2}
&\langle\mathcal{T}_{\lambda}|D_M(\lambda)=\sum_{\mu,\nu}\text{mult}_{\lambda^{\prime}-(p'-p)\rho}(\mu,\nu)\E^{2\pi \I \lambda \cdot(\mu/p-\nu/p')} \langle\mathcal{T}_{\lambda}|\ .
\end{align}
\end{subequations}

\noindent where $P_{\lambda}$ is given by the on-shell Liouville weight \eqref{physweights}. Using the above observations, we may express a given tachyon one-point function as a sum over insertions of $D_L(\sigma)$,

\begin{equation}
\begin{aligned}
\langle\mathcal{T}_{\lambda}\rangle_{\lambda^{\prime}}(\sigma)&=\langle\mathcal{T}_{\lambda}|D_M(\lambda^{\prime})|\sigma\rangle_{\rho}\\
&=\sum_{\mu,\nu}\text{mult}_{\lambda^{\prime}-(p'-p)\rho}(\mu,\nu)\E^{2\pi \I \lambda \cdot(\mu/p-\nu/p')}\langle\mathcal{T}_{\lambda}|\sigma\rangle_\rho\\
&=\sum_{\mu,\nu}\text{mult}_{\lambda^{\prime}-(p'-p)\rho}(\mu,\nu)\langle\mathcal{T}_{\lambda}|D_L\left(\Delta_{\mu,\nu}^{(\lambda)}\right)|\sigma\rangle_{\rho}\ ,
\end{aligned}
\end{equation}

\noindent where we abbreviated 

\begin{equation}
\label{bcctranslations}
\Delta_{\mu,\nu}^{(\lambda)}=\pm\frac{2}{(P_{\lambda}-P_{\bar{\lambda}})}\lambda\cdot(\mu/p-\nu/p')\ .
\end{equation}

\noindent Indeed, for any such solution we can replace the corresponding boundary state as a sum over states with trivial matter configurations when inserting an arbitrary tachyon in the bulk:

\begin{equation}
\label{seibergshih-general}
\langle\mathcal{T}_{\lambda}\rangle_{\lambda^{\prime}}(\sigma)=\sum_{\mu,\nu}\text{mult}_{\lambda^{\prime}-(p'-p)\rho}(\mu,\nu)\langle\mathcal{T}_{\lambda}\rangle_{\rho}(\sigma+\Delta_{\mu,\nu}^{(\lambda)})\ ,\quad \lambda,\ \lambda^{\prime}\in\mathcal{B}^{(q)}_{p,p'}\ .
\end{equation}

\noindent Note that at this stage, the sum rule for this decomposition can depend on the bulk insertion. Together with the factorisation property \eqref{one-pt-ratio}, the above relations place constraints on the $\sigma$-dependence of disc one-point functions: Setting $\lambda=\rho$ in the sum rule above and comparing with \eqref{one-pt-ratio}, we find a set of functional equations labelled by $\lambda$ for the one-point function of the area operator $\mathcal{T}_{\rho}$:

\begin{equation}
\begin{aligned}
\frac{S_{\rho\lambda}}{S_{\rho\rho}}\langle\mathcal{T}_{\rho}\rangle_{\rho}(\sigma)&=\sum_{\mu\in\Omega_{r^i\omega_i-\rho}}\text{mult}_{r^i\omega_i-\rho}(\mu)\exp\left(-\frac{\rho\cdot\mu}{pP_{\rho}}\frac{\partial}{\partial\sigma}\right)\\
&\quad\times\sum_{\nu\in\Omega_{s^i\omega_i-\rho}}\text{mult}_{s^i\omega_i-\rho}(\nu)\exp\left(\frac{\rho\cdot\nu}{p'P_{\rho}}\frac{\partial}{\partial\sigma}\right)\langle\mathcal{T}_{\rho}\rangle_{\rho}(\sigma)\ .\\ 
\end{aligned}
\end{equation}

\noindent where $P_{\rho}^2=(c_M-1)/12$ and $\lambda=(p'r^i-ps^i)\omega_i\in\mathcal{B}^{(q)}_{p,p'}$ as before. More compactly, in terms of $SU(q)$ characters,

\begin{equation}
\label{one-pt-eqn}
0=\left[\chi_{r^i\omega_i-\rho}\left(-\frac{2\pi\rho}{pP_\rho}\frac{\partial}{\partial\sigma}\right)\chi_{s^i\omega_i-\rho}\left(\frac{2\pi\rho}{p'P_\rho}\frac{\partial}{\partial\sigma}\right)-d_{\lambda}\right]\langle\mathcal{T}_{\rho}\rangle_{\rho}(\sigma)\ ,\quad \lambda\in\mathcal{B}^{(q)}_{p,p'}\ ,
\end{equation}

\noindent where $d_{\lambda}=S_{\rho\lambda}/S_{\rho\rho}$ is sometimes called the ground state degeneracy or quantum dimension of the state $\lambda$. We can then use \eqref{one-pt-ratio} to obtain the disc one-point of $\mathcal{T}_{\rho}$ for all other matter configurations on the boundary. Note that in general, not all the above equations are independent.

\begin{rem} Once again, it is instructive to consider how our results reduce to those for the Virasoro minimal model coupled to gravity upon setting $q=2$. The root space of $SU(2)$ is one-dimensional, and the Weyl character formula yields a sum over $\widehat{\mathfrak{su}}(2)_{k}\times \widehat{\mathfrak{su}}(2)_{k+1}$ representation weights. Explicitly, let the Kac indices $(r,s)$ and $(k,l)$ be defined by $ e_1\cdot \lambda =p'r-ps$ and $ e_1\cdot\lambda^{\prime}=p'k-pl$. The ratio of S-matrix elements \eqref{smatrixcharacter} can be written in terms of $SU(2)$ characters $\chi_j(\theta)=\mathrm{tr}_j\exp(2\I \theta J_3)$,

\begin{equation}
\begin{aligned}
\frac{S_{(r,s),(k,l)}}{S_{(1,1),(k,l)}}
&=\chi_{\frac{r-1}{2}}\left(\frac{\pi\lambda^{\prime}}{p}\right)\chi_{\frac{s-1}{2}}\left(\frac{-\pi\lambda^{\prime}}{p'}\right)\\
&=(-1)^{k(s-1)+r(l-1)}\chi_{\frac{r-1}{2}}\left(\frac{\pi kp'}{p}\right)\chi_{\frac{s-1}{2}}\left(-\frac{\pi lp}{p'}\right)\ .\\
\end{aligned}
\end{equation}

\noindent The dependence of the $\sigma$-translations \eqref{bcctranslations} on the bulk insertion cancels and we find from \eqref{seibergshih-general}

\begin{equation}
\begin{aligned}
\label{ss0}
\langle\mathcal{T}_{r,s}\rangle_{k,l}(\sigma)&=\sum_{m=1-k, 2}^{k-1}\sum_{n=1-l, 2}^{l-1}\langle\mathcal{T}_{r,s}\rangle_{1,1}\left(\sigma+\I m/b+\I n b\right)\ ,
\end{aligned}
\end{equation}

\noindent where we increment $m$ and $n$ in steps of $2$. Hence we conclude that all boundary states can be replaced with superpositions of the identity Cardy state, in agreement with \cite{Seiberg03,Martinec05}. 

Consider \eqref{one-pt-eqn} for $(r,s)=(1,2)$ and $(2,1)$, respectively: these imply that as a function of the variables $\zeta$ and $\eta$ defined by the relations \eqref{bdrycc2} and \eqref{bdrycc}, $\langle\mathcal{T}_{1,1}\rangle$ satisfies

\begin{equation}
\begin{aligned}
\mathrm{Re}\ \langle\mathcal{T}_{1,1}\rangle(\zeta)&=\cos(\pi p'/p)\langle\mathcal{T}_{1,1}\rangle_{1,1}(-\zeta)\ ,\quad &\zeta&\in [1,\infty)\ ,\\
\mathrm{Re}\ \langle\mathcal{T}_{1,1}\rangle(\eta)&=\cos(\pi p/p')\langle\mathcal{T}_{1,1}\rangle_{1,1}(-\eta)\ ,\quad &\eta&\in[1,\infty)\ ,
\end{aligned}
\end{equation} 

\noindent which is the same equation as \eqref{spectral-conf-backgr} arising at the critical point of the Hermitian two-matrix matrix model, the solutions to which are studied in the second part of Appendix \ref{app:auxprob}.
\end{rem}

\section{Discussion}
\label{sec:discussioncft}

\noindent Let us summarise the results of this chapter: in Section \ref{sec:bulkspectrum}, we determined the the spectrum of physical states on the sphere using a generalisation of the free-field formalism used in \cite{Bouwknegt91,Bilal92}. Our main result as summarised in Proposition \ref{prop:cohomology} demonstrated the absence of states that would arise in the cohomology of the usual bosonic string -- i.e. free bosons coupled to gravity -- thanks to the symmetries of the model. Nevertheless, the cohomology includes operators that create boundaries in the worldsheet when the central charge of the minimal model exceeds one, signalling the expected Kosterlitz-Thouless transition of Liouville theory. Another notable feature that distinguishes the spectrum of the model from that of the minimal string is that though no state can carry overall spin, when $q>2$, the full Hilbert space does include states with non-zero spins in both the matter and the Liouville sector, a phenomenon which resembles the ``deconfinement of chirality'' described in \cite{Gervais84,Gervais94}. 

We then introduced a family of conformally invariant boundary conditions \eqref{bdryansatz} in Section \ref{sec:boundarystates}, parametrised by the primary fields $\lambda\in\mathcal{B}^{(q)}_{p,p'}$ of the $\mathcal{W}_q$ minimal model and the cosmological constant on the boundary. The relation \eqref{seibergshih-general} following from the subsequent analysis of the one-point function of tachyon operators \eqref{tachyons} revealed that on the disk, we can replace any boundary state with a sum over boundary states with $\lambda=\rho$ and complex values of the boundary cosmological constant, lending evidence to the fact that the tensor product \eqref{bdryansatz} overcounts physically distinct boundary conditions. This generalises the observation made in \cite{Seiberg03} for the minimal string, to which our results reduce for $q=2$. Moreover, in conjunction with the factorisation of matter and Liouville contributions, this provided an immediate derivation of a series of functional difference equations \eqref{one-pt-eqn} obeyed by the tachyon one-point functions.  

Our results are also of relevance to the holographic description of higher-spin gravity with negative cosmological constant alluded to in the introduction. In particular, the CFT on the disk $D$ defines the holographic dual of a 3-manifold $\mathcal{M}_3$ with $\partial\mathcal{M}_3=D\cup \mathcal{M}_2$, where $\partial D=\partial \mathcal{M}_2$ \cite{Takayanagi11}. Notably, this construction has been invoked in \cite{Verlinde15} for a proposal of local observables on $\mathcal{M}_3$. Unlike \cite{Takayanagi11}, where the usual Dirichlet boundary condition is imposed on the metric on $D$, here we impose Neumann boundary conditions on both $D$ and $\mathcal{M}_2$. An important consequence of this modification is the emergence of the relation \eqref{seibergshih-general}, rendering boundary conditions corresponding to excited matter states semiclassically indistinguishable from a quantum superposition of boundary conditions corresponding to the matter ground state.

A central question raised by this analysis is whether the degeneracy implied by \eqref{seibergshih-general} persists in more complicated amplitudes and thus holds on the entire physical Hilbert space, as conjectured for $q=2$ in \cite{Seiberg03} and subsequently challenged in \cite{Atkin11,Atkin12}: Here we have only considered one-point functions of tachyons on the disk; to see if the identification of boundary states holds generally on the Hilbert space requires more work. This leads us to the investigation in the following chapter: there we will consider the case $q=2$ and show how this degeneracy is lifted upon inclusion of `infinite-genus' worldsheets, or more precisely, effects contributing non-perturbatively in the string coupling constant.

\chapter{Wronskians, Duality and Cardy Branes}
\label{chap:wronskians}
\thispagestyle{title}
\section{Overview}
\label{sec:overview}

\noindent Here we consider the double scaling limit of the ensemble \eqref{pottsmeasure} with $q=2$, where $V_1$ and $V_2$ are polynomials of degree $p$ and $p'$ respectively. As discussed in the introduction, Section \ref{int:scaling}, the universality classes of the critical points in the phase diagram spanned by the coefficients of $V_1$ and $V_2$ are labelled by pairs $(p,p')$ of coprime integers and are described by Liouville theory coupled to a Virasoro minimal model \cite{Douglas89,kazakov89,Daul93}. In Subsection \ref{subsec:cft}, we saw that the conformally invariant boundary states of Liouville theory fall into two classes: the discrete set of Zamolodchikov-Zamolodchikov (ZZ) \cite{ZZ01} branes, and the Fateev-Zamolodchikov-Zamolodchikov-Teschner \cite{FZZ00,ZZ01,Teschner02} (FZZT) brane $|\sigma\rangle_{\mathrm{FZZT}}$ defined in \eqref{BLiouville}. Their tensor product with the Cardy boundary states $|r,s\rangle_\mathrm{C}$ of the minimal model yields the complete brane spectrum of the theory. As Seiberg and Shih pointed out \cite{Seiberg03}, the resulting set of $(p-1)(p'-1)/2$ distinct FZZT branes -- one per primary field of the minimal model -- appears to be at odds with the merely two obvious boundary conditions that can be imposed in the matrix model description, corresponding to the resolvents of the matrices $X_1$ and $X_2$, which compute the partition function of a worldsheet with a single connected boundary. The solution to this paradox put forward in \cite{Seiberg03} is based on the conjecture that all boundary states can be written as superpositions of a single boundary state with analytically continued values of the boundary cosmological constant $\eqref{bdrycc}$:

\begin{equation}
\label{SS-rel}
|\sigma\rangle_{\mathrm{FZZT}}\otimes|r,s\rangle_\mathrm{C}=\sum_{m=-(r-1)}^{r-1}\sum_{n=-(s-1)}^{s-1}|\sigma+\I m/b+\I n b\rangle_{\mathrm{FZZT}}\otimes|1,1\rangle_\mathrm{C}\ .
\end{equation} 

\noindent In the above, $b^2=p/p'$ and we increment $m$ and $n$ in steps of $2$. Indeed, the relation \eqref{ss0} for the one-point functions on the disk derived in the previous chapter is consistent with this proposal; see also \cite{Martinec05,Gesser10,IR10,BIR10,BIR12} for an extensive amount of evidence. It was hence concluded that the resolvents of the matrices $X_1$ and $X_2$ suffice to capture all boundary conditions and there is no contradiction. Later, the $X_1+X_2$-resolvent was computed directly from the matrix model in \cite{Atkin11,Atkin12} for the unitary $(p,p+1)$ series of critical points and found to describe the $(r,s)=(1,p-1)$ boundary condition, where the validity of \eqref{SS-rel} was challenged for worldsheets of non-planar topology. However, the lack of an independent construction of the complete brane spectrum in the matrix model has until now obstructed attempts at a satisfactory solution of these debates. 

 Here we point out that generally, the analytic continuation of an asymptotic expansion need not coincide with the asymptotic expansion of the analytically continued function, which is the well-known Stokes' phenomenon \cite{Stokes47,Stokes58}, and we are led to wonder about the fate of this observation beyond perturbation theory in the string coupling $g_s$. Indeed, in the operator formalism, the non-perturbative differential equations \eqref{psi-system} and \eqref{chi-system} allow complete sets of $p$ and $p'$ \em independent \em solutions for the Baker-Akhiezer functions, respectively, only \em one \em of which describes the double-scaling limit of the expectation value of the resolvent of $X_1$ resp. $X_2$. It was later discovered that the remaining independent solutions in fact provide a consistent set of boundary conditions for normal matrices with eigenvalues supported on appropriate arcs away from the real axis \cite{Chan10,Chan12a}, suggesting their relevance for Stokes' phenomenon displayed by the resolvent operator. This motivates our study of the Wronskian for the non-perturbative linear differential equations \eqref{psi-system} and \eqref{chi-system}. The main purpose of this investigation is to answer the questions

\begin{enumerate}
\item What differential equations does the Wronskian satisfy? 
\item Which new observables are captured by the Wronskian? 
\end{enumerate}

\noindent The remainder of this chapter is organised as follows: In Section \ref{sec:W}, we will detail the reasoning for considering the Wronskian associated with the system of differential equations \eqref{psi-system}. Employing a mild generalisation thereof\footnote{see also \cite{Towse00,Anderson05,GattoScherbak12a,GattoScherbak12b} for recent work.} dating back to Schmidt \cite{Schmidt39}, we subsequently answer the first question. We then use the results to turn to the second question, providing evidence for the conjecture that the set of independent Wronskians is organised in a Kac table whose entries are in one-to-one correspondence with the primary fields of the minimal model. In Section \ref{sec:semiclassical}, we then show how this table reproduces the relation \eqref{SS-rel} in the semiclassical limit. Together, these observations strongly suggest that the Wronskian provides a non-perturbative description of the general FZZT brane with $(r,s)\neq (1,1)$, and that the degeneracy \eqref{SS-rel} is resolved by additional degrees of freedom whose independence is invisible in perturbation theory in $g_s$. We close with a discussion of results and possible further developments in Section \ref{sec:discussionwronsk}.

\section{Generalised Wronskian}
\label{sec:W}

\noindent This section proceeds as follows: In Subsection \ref{subsec:wronsk-cardy}, we explain how the relation \eqref{SS-rel} hints at the Wronskian associated with the linear differential equation for the Baker-Akhiezer function. In Subsection \ref{subsec:diff}, we derive the analogues of \eqref{psi-system} and \eqref{chi-system}, allowing us to introduce an isomonodromy system akin to \eqref{isomonodromysystem}, each of which defines a spectral curve. In Subsection \ref{subsec:duality}, we use the properties of the duality transformation $(p,p')\to(p',p)$ to determine the complete set of observables defined by the Wronskian; the fact that a Kac table for the latter emerges directly from the matrix model without reference to the worldsheet conformal field theory provides the first piece of evidence that our construction provides a non-perturbative description of the FZZT branes with general Cardy labels $(r,s)\neq (1,1)$, which we will refer to as \em Cardy branes \em for short.

\subsection{Wronskians and Cardy branes}
\label{subsec:wronsk-cardy}

\noindent Recall from the introduction that the single-trace operator $\mathrm{tr}(x-X)$ creates a connected boundary in the worldsheet corresponding to the state $|\sigma\rangle_{\mathrm{FZZT}}\otimes|1,1\rangle_{\mathrm{C}}$, and the determinant operator $\det(x-X)$ creates the associated brane at target space position $x$. From the expansion \eqref{determinantgraphical}, it can be seen that a linear combination of such boundary states as in \eqref{SS-rel} indicates that the matrix model operator corresponding to the general \em Cardy brane \em $|\sigma\rangle_{\mathrm{FZZT}}\otimes|r,s\rangle_{\mathrm{C}}$ factorises into a product of more elementary operators. Indeed, such a relation is expected from Polchinski's general combinatorial picture applied to the present context \cite{Polchinski94,Fukuma99}. It was proven by Morozov in \cite{Morozov94} that the average of a general product of characteristic polynomials

\begin{equation}
\begin{aligned}
\alpha_n^{(M)}(x_1,x_2,\dots x_M)&=\left\langle\prod_{k=1}^M\det(x_k-X)\right\rangle_{n\times n}\ \\ \beta_n^{(M)}(y_1,y_2,\dots y_M)&=\left\langle\prod_{k=1}^M\det(y_k-Y)\right\rangle_{n\times n}\ 
\end{aligned}
\end{equation}

\noindent can be written in terms of the orthogonal polynomials $\{\alpha_n\}_{n=1}^N$ defined in \eqref{charpol} as

\begin{equation}
\label{morozov}
\alpha_n^{(M)}(x_1,x_2,\dots x_M)=\frac{\det_{1\leq k, l\leq M}\alpha_{n+1-k}(x_l)}{\det_{1\leq k,l \leq M}x_k^{l-1}}\ ,
\end{equation}

\noindent and similarly for $\beta_n^{(M)}$. As shown in \cite{Maldacena04}, in the double-scaling limit, equation \eqref{morozov} can be written in terms of the alternating polynomial $a_\lambda(z)=\det_{1\leq a, b\leq n}z_b^{a-1+\lambda_a}$,

\begin{equation}
\label{proddsl}
\left\langle\prod_{k=1}^M\det(x_k-X)\right\rangle_{n\times n}=\frac{a_\varnothing(\partial)}{a_\varnothing(\zeta)}\prod_{k=1}^M\psi^{(1)}(t;\zeta_k)\quad \mathrm{as}\ N\to\infty\ ,\ \varepsilon\to 0\ ,
\end{equation}

\noindent with $\zeta_k=(x_k-x_c)/\varepsilon$ finite. In the above, $\partial_{i}\psi^{(j)}=\partial_t\psi^{(j)}\delta_{ij}$.

To proceed, we make the further observation that the translations $\sigma\to\sigma+\I m/b+\I n b$ in \eqref{SS-rel} move between sheets of the spectral curve defined by the zero locus of the semiclassical limit of the polynomials \eqref{spectralcurvedsl} -- a fact which has been widely discussed, including \cite{Martinec05,Gesser10,IR10,BIR10,BIR12}. This semiclassical spectral curve provides the initial data for the topological recursion algorithm described in \cite{Eynard07}, which computes the asymptotic expansion of arbitrary correlation function to any finite order in $g_s$ -- it thus appears that to all orders in the perturbative expansion, all $p$ branches $\{\psi^{(j)}(t;\zeta)\}_{j=1}^p$ of the solution can indeed be obtained from a single principal branch $\psi^{(1)}(t;\zeta)$ by mere analytic continuation $\zeta\to\E^{2\pi\I}\zeta$. This indicates that by performing the asymptotic expansion, we lose the information required to distinguish one solution from the other. We therefore generalise the expression \eqref{proddsl} to account for the complete set of independent solutions $\{\psi^{(j)}(t;\zeta)\}_{j=1}^p$ resp. $\{\chi^{(j)}(t;\zeta)\}_{j=1}^{p'}$ to \eqref{psi-system} resp. \eqref{chi-system},  

\begin{equation}
\label{branestack}
\begin{aligned}W^{(M)}_\varnothing[\psi](t;\zeta)&=a_\varnothing(\partial)\left.\prod_{k=1}^M\psi^{(j_k)}(t;\zeta_k)\right|_{\zeta_1=\zeta_2=\dots\zeta_M=\zeta}\ ,\\
W^{(M)}_\varnothing[\chi](t;\eta)&= a_\varnothing(\partial)\left.\prod_{k=1}^M\chi^{(j_k)}(t;\eta_k)\right|_{\eta_1=\eta_2=\dots\eta_M=\eta}\ .
\end{aligned}
\end{equation}

\noindent We may regard the above as the antisymmetrised ground state wave function of $M$ coincident branes and their duals. Notably, due to the fermionic statistics, the existence of such ``brane stacks'' requires the presence of an additional quantum number -- the label $j$ distinguishing the independent solutions. 

\subsection{Differential equations and spectral curve}
\label{subsec:diff}

\noindent Here we derive the differential equations satisfied by the observables \eqref{branestack}. We also introduce the corresponding spectral curves and define an extension of the charge conjugation, which will turn out useful when we consider the duality transformation \eqref{duality} in the next subsection. To this end, we need to keep track of derivatives of $W_\varnothing^{(n)}(t;\zeta)$ with respect to the spectral parameters $\zeta_k$ in \eqref{branestack}. This is conveniently achieved by the following 

\begin{defn}
Denote the set of Young diagrams with $n$ rows by $\Lambda_n$ and the subset of $\binom{p}{n}$ diagrams with at most $p-n$ boxes in each row by $\Lambda_{p,n}$; denote the number of boxes in the $a^{\mathrm{th}}$ row of $\lambda$ by $\lambda_a$, and call $|\lambda|=\sum_{a=1}^n\lambda_a$ the \em size \em of the diagram $\lambda$.
\end{defn}

\begin{defn} Let $\{\psi^{(j)}\}_{j=1}^p$ denote the $p$ solutions to \eqref{psi-system}. Given $\lambda\in\Lambda_n$, we define the \em generalised Wronskian\em 

\begin{equation}
\label{defn-wronskian}
W^{(n)}_{\lambda}(t;\zeta)=\det_{1\leq a, b \leq n}\left.\left(\partial_t^{n-a+\lambda_{n-a+1}}\psi^{(j_b)}(t;\zeta_b)\right)\right|_{\zeta_1=\zeta_2=\dots\zeta_n=\zeta}\ .
\end{equation}
\end{defn}

\noindent For notational simplicity we keep the dependence on $\{j_b\}_{b=1}^n$ implicit. Note that from the properties of the determinant it follows immediately that $\partial_tW^{(n)}_\lambda(t;\zeta)= 0$ for $n\geq p$. We note another useful representation of $W_\lambda^{(n)}$ also reported in \cite{GattoScherbak12a,GattoScherbak12b}:

\begin{lem}
\label{lem1}
Let $S_{[\lambda_1,\lambda_2,\dots\lambda_n]}(z)\equiv S_\lambda(z)$ denote the Schur polynomial in $n$ variables $\{z_k\}_{k=1}^n$. Then the generalised Wronskian can be expressed as

\begin{equation}
W^{(n)}_{\lambda}(t;\zeta)=S_\lambda(\partial)
W^{(n)}_{\varnothing}(t;\zeta)\ ,\quad \partial_k=\frac{1}{k}\sum_{i=1}^n\partial_{(i)}^k\ ,
\end{equation}

\noindent where $\partial_{(i)}\psi^{(j)}=\partial_t\psi^{(j)}\delta_{ij}$ and $\varnothing$ denotes the diagram with $\lambda_a=0\ \forall a$.
\end{lem}

\begin{proof} It suffices to note that $W_\lambda^{(n)}$ can be expressed in terms of the alternating polynomial $a_\lambda(\partial)=\det_{1\leq a,b\leq n}\partial_{(b)}^{a-1+\lambda_a}$ in the derivatives $\partial_{(i)}\psi^{(j)}=\delta_{ij}\partial_t\psi^{(j)}$:

\begin{equation}
\begin{aligned}
W^{(n)}_{\lambda}(t;\zeta)&= (-1)^{n(n-1)/2}\det_{1\leq a, b\leq n}\left(\partial_{(j_b)}^{\lambda_a+a-1}\right)\prod_{k=1}^n\psi^{(j_k)}(t;\zeta)\\
&=a_\lambda(\partial)\prod_{k=1}^n\psi^{(j_k)}(t;\zeta)\ .
\end{aligned}
\end{equation}

\noindent Using the defining relation $S_\lambda(z)=a_\lambda(z)/a_\varnothing(z)$ then proves the statement.
\end{proof}

\noindent In light of the discussion in the preceeding subsection, we may think of $W^{(n)}_{\lambda\neq\varnothing}$ as the excited state created by the operator $S_\lambda(\partial)$ acting on the ground state \eqref{branestack}. We can now state the differential equations satisfied by $W_\lambda^{(n)}$ by analogy with \eqref{psi-system}:

\begin{prop}
\label{Prop-I}
The functions $W^{(n)}_\lambda(t;\zeta)$ satisfy

\begin{subequations}
\begin{align}
\label{W-P}
\zeta W^{(n)}_\lambda(t;\zeta)&=\mathbb{P}^{(n)}_{\lambda,j}(t;\partial)W^{(n)}_\varnothing(t;\zeta)\ ,\quad j=1,2,\dots n\ ,\\
\label{W-Q}
\partial_\zeta W^{(n)}_\lambda(t;\zeta)&=\beta_{p,p'}\sum_{j=1}^n\mathbb{Q}^{(n)}_{\lambda,j}(t;\partial)W^{(n)}_\varnothing(t;\zeta)\ ,
\end{align}
\end{subequations}

\noindent with the $p^{\mathrm{th}}$ and $(p')^{\mathrm{th}}$ order differential operators

\begin{equation}
\begin{aligned}
\label{PQn-diff-eqn}
\mathbb{P}^{(n)}_{\lambda,j}(t;\partial)=&\ 2^{p-1}S_{[\lambda_1,\dots,\lambda_j+p,\dots,\lambda_n]}(\partial)+\sum_{m=2}^{p+\ell_j}U_m^{(\ell_j)}(t)S_{[\lambda_1,\dots,\lambda_j+p-m,\dots,\lambda_n]}(\partial)\ ,\\
\mathbb{Q}^{(n)}_{\lambda,j}(t;\partial)=&\ 2^{p'-1}S_{[\lambda_1,\dots,\lambda_j+p',\dots,\lambda_n]}(\partial)+\sum_{m=2}^{p'+\ell_j}V_m^{(\ell_j)}(t)S_{[\lambda_1,\dots,\lambda_j+p'-m,\dots,\lambda_n]}(\partial)\ ,
\end{aligned}
\end{equation}
\noindent where $\ell_j=\lambda_j+j-1$ is the hook length of the first box in each row and
\begin{equation}
U_m^{(\ell)}(t)=\sum_{k=0}^{\min[m-2,\ell]} \binom{\ell}{k} \left(\partial_t^k u_{m-k}(t)\right)\ ,\quad 
V_m^{(\ell)}(t)=\sum_{k=0}^{\min[m-2,\ell]} \binom{\ell}{k} \left(\partial_t^k v_{m-k}(t)\right)\ . 
\end{equation}

\end{prop}

\begin{proof} We first demonstrate \eqref{W-P}. Using \eqref{Ppsi} and expanding $\partial_t^nf(t)=\sum_{k=0}^n\binom{n}{k}(\partial_t^{k}f(t))\partial^{n-k}$, we can express

\begin{equation}
\begin{aligned}
2^{p-1}S_{[\lambda_1,\dots,\lambda_n+p]}
&=-\sum_{m=2}^{p}\sum_{k=0}^{\ell_n}\binom{\ell_n}{k}\large\left[\left(\partial_t^k u_m(t)\right)-\delta_{mp}\delta_{k0}\zeta\large\right]S_{[\lambda_1,\dots,\lambda_n+p-m-k]}(\partial)\ \\&=-\sum_{m=2}^{p+\ell_n}\sum_{k=0}^{\min[m-2,\ell]}\binom{\ell_n}{k}\large\left[\left(\partial_t^k u_{m-k}(t)\right)-\delta_{mp}\delta_{k0}\zeta\large\right]S_{[\lambda_1,\dots,\lambda_n+p-m]}(\partial)\ ,
\end{aligned}
\end{equation}

\noindent with $u_{n<0}(t)=v_{n<0}(t)=0$, which implies \eqref{W-P} for $j=n$. For $j\neq n$, we first use 

\begin{equation}
\begin{aligned}
S_{[\lambda_1,\dots\lambda_j,\lambda_{j+1},\dots \lambda_n]}(\partial)&=-S_{[\lambda_1,\dots\lambda_{j+1}+1,\lambda_j-1,\dots \lambda_n]}(\partial)\\&=(-1)^{n-j}S_{[\lambda_1,\dots,\lambda_{j+1}+1,\lambda_{j+2}+1\dots\lambda_n,\lambda_j-n+j]}(\partial)
\end{aligned}
\end{equation}

\noindent and then apply the previous result to $S_{[\lambda_1,\dots\lambda_j+p\dots \lambda_n]}(\partial)$. Equation \eqref{W-Q} can be obtained by explicit evaluation of the derivative of $W^{(n)}_\lambda$ w.r.t. $\zeta$, which gives a sum of $n$ terms, one for the action of $\partial_\zeta$ on each row of the matrix $\partial_t^{a-1+\lambda_a}\psi^{(j_b)}$. In each term, we may use \eqref{Qpsi} and subsequently commute the derivatives to the right of $v_n(t)$ using the same procedure as for \eqref{W-P}, which immediately yields \eqref{W-Q}.
\end{proof}

\begin{cor}
\label{cor1}
$\mathcal{M}_{p,n}=(\mathrm{span}\{W^{(n)}_\lambda\}_{\lambda\in\Lambda_{p,n}},+)$ is a module of the ring $\mathcal{R}_n$ of symmetric polynomials in $n$ variables over $\mathbb{R}$.
\end{cor}

\begin{proof} Note first that every $r\in\mathcal{R}_n$ can be expanded in Schur polynomials. Hence Lemma \ref{lem1} provides a map $\mathcal{R}_n\times\mathcal{M}_{p,n}\longrightarrow\mathcal{M}_{p,n}$,
\begin{equation}
\label{ring}
S_{\lambda}(\partial)W^{(n)}_\mu(t;\zeta)=\sum_{\nu\in\Lambda_{p,n}}f^{(n)\nu}_{\lambda\mu}(t;\zeta)W^{(n)}_\nu(t;\zeta)\ ,\quad\lambda,\mu\in\Lambda_{p,n}\ ,
\end{equation}
\noindent where the $f^{(n)\nu}_{\lambda\mu}(t;\zeta)$ are determined by the Littlewood-Richardson rule 
\begin{equation}
S_\lambda S_\mu=\sum_{|\nu|=|\lambda|+|\mu|\atop \nu\in \Lambda_n} c^\nu_{\lambda\mu}S_\nu\ .
\end{equation}
\noindent Whenever $\nu_a>p-n$ for some $a$ on the right-hand side, we apply Proposition \ref{Prop-I} repeatedly to obtain a linear combination of $S_\lambda$ with $\lambda\in\Lambda_{p,n}$. Since $\mathcal{M}_{p,n}$ is an abelian group under addition, it is an $\mathcal{R}_n$-module.
\end{proof}

\noindent We now look for a suitable generalisation of the isomonodromy description \eqref{isomonodromysystem} and the spectral curve \eqref{spectralcurvedsl}. To this end, one first chooses an ordering on $\Lambda_{p,n}$, for example

\begin{equation}
\label{youngordering}
\lambda < \lambda'\quad\Leftrightarrow\quad|\lambda|<|\lambda'|\quad\mathrm{or}\quad\sum_{a=0}^{n-1}\lambda_{n-a}(p-n)^a<\sum_{a=0}^{n-1}\lambda'_{n-a}(p-n)^a\ .
\end{equation}

\noindent We then have the following result:

\begin{prop}
\label{Prop-II}
Let $\vec W^{(n)}(t;\zeta)=(W^{(n)}_\varnothing,W^{(n)}_{\tiny \yng(1)},\dots)^T$ be the $\binom{p}{n}$-vector with entries ordered according to \eqref{youngordering}. Then there exist $\binom{p}{n}\times\binom{p}{n}$ matrices $\mathcal{B}^{(n)}(t;\zeta)$ and $\mathcal{Q}^{(n)}(t;\zeta)$ such that 

\begin{subeqnarray}
\label{Lax1}
\partial_t\vec W^{(n)}(t;\zeta)&=\mathcal{B}^{(n)}(t;\zeta)\vec W^{(n)}(t;\zeta)\ ,\\
\label{Lax2}
\partial_\zeta\vec W^{(n)}(t;\zeta)&=\mathcal{Q}^{(n)}(t;\zeta)\vec W^{(n)}(t;\zeta)\ .
\end{subeqnarray}
\end{prop}

\begin{proof} Since $\partial_t=S_{\tiny \yng(1)}(\partial)$, we can use Corollary \ref{cor1} to find $\mathcal{B}^{(n)}$: 

$$\partial_tW^{(n)}_\lambda=\sum_{\mu\in\Lambda_{p,n}}f^{(n)\mu}_{{\tiny \yng(1)},\lambda}W^{(n)}_\mu\ ,\quad\lambda,\mu\in\Lambda_{p,n}\quad \Rightarrow\quad(\mathcal{B}^{(n)})^\mu_\lambda=f^{(n)\mu}_{{\tiny \yng(1)},\lambda}\ .$$

\noindent To show the existence of $\mathcal{Q}^{(n)}$, we use first use \eqref{W-Q} to expand the right hand side in Schur polynomials and thereafter apply Corollary \ref{cor1}.
\end{proof}

\begin{defn}
\label{def:spectralcurve}
\noindent We introduce the characteristic polynomials
\begin{equation}
\label{def:spectralpol}
\begin{aligned}
F^{(n)}(t;\zeta,z)=\det\left(z\mathbb{I}_{\binom{p}{n}\times \binom{p}{n}}-\mathcal{B}^{(n)}(t;\zeta)\right)\ ,\\
G^{(n)}(t;\zeta,Q)=\det\left(Q\mathbb{I}_{\binom{p}{n} \times \binom{p}{n}}-\mathcal{Q}^{(n)}(t;\zeta)\right)\ ,\\
\end{aligned}
\end{equation}
\noindent and define the \em spectral curve \em of the system \eqref{psi-system},
\begin{equation}
\mathcal{C}_{p,p'}^{(n)}(t)=\{(P,Q)\in\mathbb{C}^2|G^{(n)}(t;P,Q)=0\}\ .
\end{equation}
\end{defn}

\noindent Observe that for given $n$, the spectral curves for $\mathcal{C}^{(n)}_{p,p'}(t)$ and $\mathcal{C}^{(p-n)}_{p,p'}(t)$ are of the same degree. To pave the way for a definition of the duality transformation, it is useful to relate these two systems by extending the definition of the charge conjugation \eqref{chargeconj} as follows: For given $\lambda\in\Lambda_{p,n}$, we define complement $\lambda^\perp$ and conjugate $\lambda^\vee$ via

\begin{subequations}
\begin{align}
\lambda_a^\perp&=(p-n)-\lambda_{n-a+1}\ ,\quad &a=1,\dots n\\ \lambda_a^\vee&=\mathrm{max}_{1\leq b\leq r}\left\{b|\lambda_{r-b+1}\geq a\right\}\ ,\quad &a=1,\dots p-n
\end{align}
\end{subequations}

\noindent and make the following

\noindent \begin{defn} We define the charge conjugation
\begin{equation}
\label{conj-wronsk}
\begin{aligned}
\mathcal{C}:\mathcal{M}_{p,n}&\longrightarrow\mathcal{M}_{p,p-n}\ ,\\
W^{(n)}_\lambda&\longmapsto \mathcal{C}[W^{(n)}_\lambda]=W^{(p-n)}_{\mathcal{C}(\lambda)}\ ,
\end{aligned}
\end{equation} 
where $\mathcal{C}(\lambda)=(-1)^{|\lambda|}(\lambda^{\perp})^\vee$.
\end{defn}

 \noindent We close this subsection with a few examples illustrating the above construction; to facilitate the presentation, we relegate the explicit equations for the Lax operators to Appendix \ref{app:Q}. These immediately determine the full non-perturbative spectral curve via Definition \ref{def:spectralcurve}. Below, we print the corresponding polynomials \eqref{def:spectralpol} in the semiclassical limit $g_s\to 0$. For later comparison with the conformal field theory prediction \eqref{SS-rel}, we evaluate this limit in the conformal background, in which \eqref{spectral-conf-backgr} holds.

\begin{ex}
\label{ex-puregrav}
$(p,p')=(3,2)$. The only allowed cases $n=1,2$ are equivalent to the $3\times 3$ Lax systems discussed in \cite{Bergere13} and references therein. In the bases

\begin{equation} 
\begin{aligned}
\vec W^{(1)}(t;\zeta)=\large\left(W^{(1)}_\varnothing,W^{(1)}_{\tiny \yng(1)},W^{(1)}_{\tiny \yng(2)}\large\right)^T\ ,\quad
\vec W^{(2)}(t;\zeta)=\large\left(W^{(2)}_\varnothing,W^{(2)}_{\tiny \yng(1)},W^{(2)}_{\tiny \yng(1,1)}\large\right)^T\ ,
\end{aligned}
\end{equation}

\noindent the Lax operators satisfy $\mathcal{B}^{(2)}=-\mathcal{C}^{-1}\left(\mathcal{B}^{(1)}\right)^T\mathcal{C}$ and $\mathcal{Q}^{(2)}=-\mathcal{C}^{-1}\left(\mathcal{Q}^{(1)}\right)^T\mathcal{C}$, where the charge conjugation matrix $\mathcal{C}: \mathcal{M}_{3,1}\longrightarrow\mathcal{M}_{3,2}$ has components $\mathcal{C}_{ab}=(-1)^{a+1}\delta_{4-a,b}$ and  $\mathcal{B}^{(n)}$ and $\mathcal{Q}^{(n)}$ are given in Appendix \ref{app:Q}, Example \ref{app:wronsk1}. The eigenvalues of $\mathcal{B}^{(n)}$, $\mathcal{Q}^{(n)}$ are given by the zeroes of

\begin{equation}
\begin{aligned}
\label{puregrav-curve}
\pm F^{(n)}(t;\zeta,z)&=\pm z^3-\frac{\zeta}{4}\pm\frac{3v_2}{4}z+\frac{3\dot{v}_2}{8}\ ,\\
\pm G^{(n)}(t;\zeta,Q)&=\pm Q^3-\frac{\zeta^2}{2}\mp Q\left(\frac{3v_2^2}{4}+\frac{\ddot{v}_2}{2}\right)-\frac{v_2^3}{4}-\frac{v_2\ddot{v}_2}{2}+\frac{\dot{v}_2^2}{8}\ ,
\end{aligned}
\end{equation}

\noindent where the upper (resp. lower) sign holds for $n=1$ (resp. $n=2$), in agreement with \cite{Bergere13}. 
\end{ex}

\begin{ex}
\label{ex:ising}
$(p,p')=(4,3)$. For $n=1$ and $n=3$, we again recover the familiar $4\times 4$ Lax systems discussed in \cite{Bergere13}. On the other hand, in the nontrivial case $n=2$ the system \eqref{Lax1} is 6-dimensional. In the basis
\begin{equation} 
\vec W^{(2)}(t;\zeta)=\left(W^{(2)}_\varnothing,W^{(2)}_{\tiny \yng(1)},W^{(2)}_{\tiny \yng(2)},W^{(2)}_{\tiny \yng(1,1)},W^{(2)}_{\tiny \yng(2,1)},W^{(2)}_{\tiny \yng(2,2)}\right)^T\ ,
\end{equation} 

\noindent the charge conjugation matrix $\mathcal{C}: \mathcal{M}_{4,2}\longrightarrow\mathcal{M}_{4,2}$ is given by

\begin{equation}
\begin{pmatrix}
0 & 0 & 1 \\
0 &-1 & 0 \\
1 & 0 & 0
\end{pmatrix}
\otimes
\begin{pmatrix}
0 & 1 \\
1 & 0 
\end{pmatrix}\ .
\end{equation}

\noindent We find $\mathcal{B}^{(2)}=-\mathcal{C}^{-1}\left(\mathcal{B}^{(2)}\right)^T\mathcal{C}+\mathcal{O}(g_s)$, $\mathcal{Q}^{(2)}=-\mathcal{C}^{-1}\left(\mathcal{Q}^{(3)}\right)^T\mathcal{C}$, with $\mathcal{B}^{(2)}$ and $\mathcal{Q}^{(2)}$ given in Appendix \ref{app:Q}, Example \ref{app:wronsk2}. Now consider the limit $g_s\to 0$. In the conformal background, $v_2(t)\to-1$ and $u_4(t)\to1$. The eigenvalues of $\mathcal{B}^{(n)}$, $\mathcal{Q}^{(n)}$ are then given by the zeroes of
\begin{equation}
\begin{aligned}
\label{ex:isingcurve}
 F^{(2)}(t;\zeta,z)&=z^6-\frac{2}{3}z^4+\frac{1}{2}z^2\zeta-\frac{7}{18}z^2\ ,\\
 G^{(2)}(t;\zeta,Q)&=Q^6-\frac{2}{27}Q^4+2Q^2\zeta^3-\frac{16}{3}Q^2\zeta^2+\frac{85}{18}Q^2\zeta-\frac{2023}{1458}Q^2 .
\end{aligned}
\end{equation}
\end{ex}

\begin{ex}
\label{ex:yanglee}
$(p,p')=(5,2)$. For the nontrivial cases $n=2,3$, we pick a basis
\begin{equation} 
\begin{aligned}
\vec W^{(2)}(t;\zeta)&=\large\left(W^{(2)}_\varnothing,W^{(2)}_{\tiny \yng(1)},W^{(2)}_{\tiny \yng(2)},W^{(2)}_{\tiny \yng(1,1)},W^{(2)}_{\tiny \yng(3)},W^{(2)}_{\tiny \yng(2,1)},W^{(2)}_{\tiny \yng(3,1)},W^{(2)}_{\tiny \yng(2,2)},W^{(2)}_{\tiny \yng(3,2)},W^{(2)}_{\tiny \yng(3,3)}\large\right)^T\ ,\\
\vec W^{(3)}(t;\zeta)&=\large\left(W^{(3)}_\varnothing,W^{(3)}_{\tiny \yng(1)},W^{(3)}_{\tiny \yng(1,1)},W^{(3)}_{\tiny \yng(2)},W^{(3)}_{\tiny \yng(1,1,1)},W^{(3)}_{\tiny \yng(2,1)},W^{(3)}_{\tiny \yng(2,1,1)},W^{(3)}_{\tiny \yng(2,2)},W^{(3)}_{\tiny \yng(2,2,1)},W^{(3)}_{\tiny \yng(2,2,2)}\large\right)^T\ ,
\end{aligned}
\end{equation}

\noindent in which the charge conjugation matrix $\mathcal{C}: \mathcal{M}_{5,3}\longrightarrow\mathcal{M}_{5,2}$ is given by

\begin{equation}
\mathcal{C}=\begin{pmatrix}
0 & \dots & & & \I\sigma_2 \\
\vdots & & & \mathbb{I}_{2\times 2} & \\
& & -\mathbb{I}_{2\times 2} & \\
& \mathbb{I}_{2\times 2} & & & \vdots \\
-\I\sigma_2 & & & \dots & 0
\end{pmatrix}\ ,
\end{equation}

\noindent where $\sigma_2$ denotes the $2^{\mathrm{nd}}$ Pauli matrix. We have $\mathcal{B}^{(2)}=-\mathcal{C}^{-1}\left(\mathcal{B}^{(3)}\right)^T\mathcal{C}+\mathcal{O}(g_s)$, $\mathcal{Q}^{(2)}=-\mathcal{C}^{-1}\left(\mathcal{Q}^{(3)}\right)^T\mathcal{C}$, with $\mathcal{B}^{(2)}$ and $\mathcal{Q}^{(2)}$ given in Appendix \ref{app:Q}, Example \ref{app:wronsk3}. Now consider the limit $g_s\to 0$. In the conformal background, and $v_2(t)\to1$, $u_4(t)\to -5/2$ and $u_3(t)$ and $u_5(t)$ both vanish. The eigenvalues of $\mathcal{B}^{(n)}$, $\mathcal{Q}^{(n)}$ are given by the zeroes of
\begin{equation}
\begin{aligned}
 F^{(n)}(t;\zeta,z)&=\frac{1}{256}(256z^{10}-960z^8+960z^6\pm 176z^5\zeta-300z^4+25z^2\pm5z\zeta-\zeta^2),\\
 G^{(n)}(t;\zeta,Q)&=Q^{10}-\frac{15}{4}Q^8+\frac{15}{4}Q^6\pm Q^5\left(\frac{11}{8}\zeta^2-\frac{11}{16}\right)-\frac{75}{64}Q^4\\&\quad\pm Q^3\left(\frac{5}{16}-\frac{5}{8}\zeta^2\right)+\frac{25}{256}Q^2\pm Q\left(\frac{5}{128}-\frac{5}{256}\right)\ ,
\end{aligned}
\end{equation}

\noindent where the upper (resp. lower) sign holds for $n=2$ (resp. $n=3$).
\end{ex} 

\subsection{Kac table and duality}
\label{subsec:duality}

\noindent In this section, we provide evidence that the set of differential equations that characterise the generalised Wronskians for given $(p,p')$ fall into a Kac table whose entries are in one-to-one correspondence with the $(p-1)(p'-1)/2$ Cardy states of the  $(A_{p-1},A_{p'-1})$ minimal model. To keep track of which Baker-Akhiezer function we take the Wronskian of, we refine our notation as follows:

\begin{equation}
\label{defn-wronskian-2}
W^{(n)}_\lambda[f](\cdot)=\det_{1\leq a, b \leq n}\left(\partial_t^{n-a+\lambda_{n+a-1}}f^{(j_b)}(t;\cdot)\right)\ .
\end{equation}

\begin{figure}
\vspace{-2.5cm}
\centering
\begin{pspicture}(10,5)
\psset{unit=.7cm}

\psframe(0,0)(10,5)

\psline(0,1)(10,1)
\psline(2,0)(2,5)
\psline(0,4)(10,4)
\psline(8,0)(8,5)

\psline[linestyle=dotted](1,-.5)(9,5.5)

\psline{->}(-.5,.5)(-.5,4.5)
\rput[b]{90}(-1,2.5){\tiny{$1\leq s\leq p'-1$}} 
\psline{->}(3,-.5)(7,-.5)
\rput(5,-1){\tiny{$1\leq r\leq p-1$}}

\psframe[linestyle=dashed,linecolor=blue](.1,.1)(9.9,.9)
\rput(1,0.5){\tiny{$\psi$}}
\rput(5,0.5){\tiny{$W^{(r)}[\psi]$}}
\rput(9,0.5){\tiny{$W^{(p-1)}[\psi]$}}

\psframe[linestyle=dashed,linecolor=red](-.1,-.1)(2.1,5.1)
\rput(1,4.5){\tiny{$\mathcal{L}\chi$}}
\rput[b]{90}(1,2.5){\tiny{$\mathcal{L}W^{(p'-s)}[\chi]$}}

\psframe[linestyle=dashed,linecolor=green](.1,4.9)(9.9,4.1)
\rput(5,4.5){\tiny{$W^{(r)}[\mathcal{L}\chi]$}}
\rput(9,4.5){\tiny{$\mathcal{L}^2\psi$}}

\psframe[linestyle=dashed,linecolor=gray](7.9,5.1)(10.1,-.1)
\rput[b]{90}(9,2.5){\tiny{$\mathcal{L}W^{(p'-s)}[\mathcal{L}\psi]$}}

\end{pspicture}\nolinebreak\begin{pspicture}(6,8)
\psset{unit=.7cm}

\psframe(0,0)(7,8)

\psline(2,0)(2,8)
\psline(0,1)(7,1)
\psline(5,0)(5,8)
\psline(0,7)(7,7)

\psline[linestyle=dotted](-.5,0)(7.5,8)

\psline{->}(-.5,2)(-.5,6)
\rput[b]{90}(-1,4){\tiny{$1\leq r\leq p-1$}}
\psline{->}(1.5,-.5)(5.5,-.5)
\rput(3.5,-1){\tiny{$1\leq s\leq p'-1$}}

\psframe[linestyle=dashed,linecolor=red](.1,.1)(6.9,0.9)
\rput(1,0.5){\tiny{$\chi$}}
\rput(3.5,0.5){\tiny{$W^{(s)}[\chi]$}}
\rput(6,0.5){\tiny{$W^{(p'-1)}[\chi]$}}

\psframe[linestyle=dashed,linecolor=blue](-.1,-.1)(2.1,8.1)
\rput(1,7.5){\tiny{$\mathcal{L}\psi$}}
\rput[b]{90}(1,4){\tiny{$\mathcal{L}W^{(p-r)}[\psi]$}}

\psframe[linestyle=dashed,linecolor=gray](.1,7.1)(6.9,7.9)
\rput(3.5,7.5){\tiny{$W^{(s)}[\mathcal{L}\psi]$}}
\rput(6,7.5){\tiny{$\mathcal{L}^2\chi$}}

\psframe[linestyle=dashed,linecolor=green](4.9,-.1)(7.1,8.1)
\rput[b]{90}(6,4){\tiny{$\mathcal{L}W^{(p-r)}[\mathcal{L}\chi]$}}

\end{pspicture}
\caption{Definition of the boundary of the Kac table (left) and its image under the duality transformation (right). Regions of the same color are related by a Laplace transform.}
\label{fig:kactable}
\end{figure}
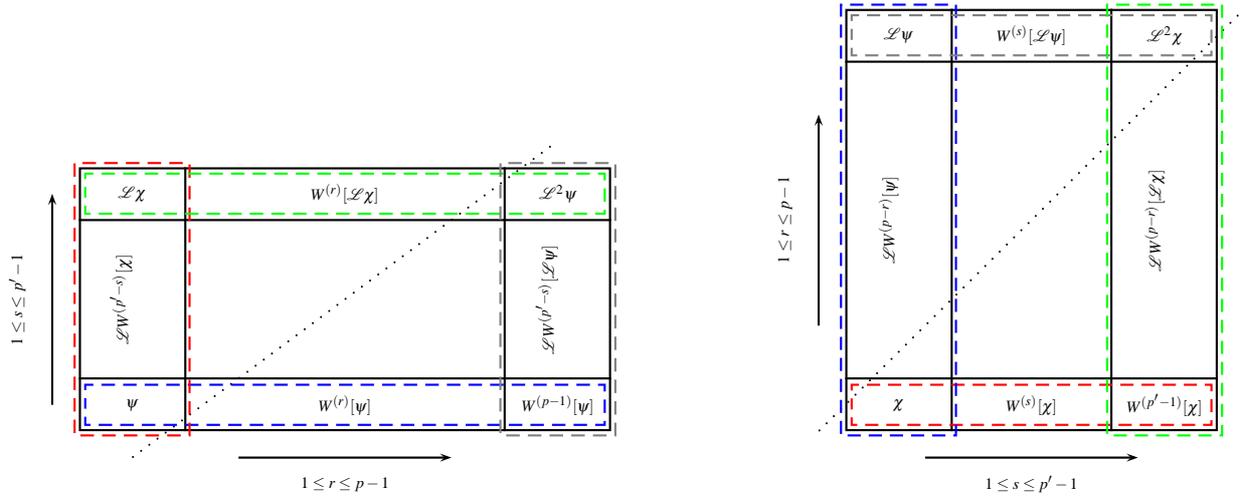

\noindent Moreover, because we are only interested in relations between entire modules $\{\mathcal{M}_{p,n}\}_{n=1}^p$ whose elements satisfy the same differential equations, we shall at times omit the subscript $\lambda$ in $W^{(n)}_\lambda$. Our derivation is based on the property of the Virasoro minimal model that the duality transformation \eqref{duality} takes the boundary state $|r,s\rangle_\mathrm{C}$ of the $(p,p')$ model to the equivalent state $|s,r\rangle_\mathrm{C}$ of the $(p',p)$ model: Under the assumption that the composition of the Laplace transform \eqref{laplace} with the charge conjugation \eqref{conj-wronsk} extends this duality transformation non-perturbatively, we can fill the entries on the boundary of the Kac table shown to the left of Figure \ref{fig:kactable} via

\begin{equation}
\begin{aligned}
\label{cardybranes}
\Psi^{(r,1)}(\zeta)&=W^{(r)}[\psi](\zeta)\ ,\qquad &\Psi^{(r,p'-1)}(\zeta)&=W^{(r)}\left[\mathcal{L}\chi\right](\zeta)\ ,\\
\Psi^{(1,s)}(\zeta)&=\mathcal{L}W^{(p'-s)}[\chi](\zeta)\ ,\qquad &\Psi^{(p-1,s)}(\zeta)&=\mathcal{L} W^{(p'-s)}[\mathcal{L}\psi](\zeta)\ .
\end{aligned}
\end{equation}

\noindent Introducing the dual wave functions $\widetilde{\Psi}^{(s,r)}(\eta)=\mathcal{L}\mathcal{C}[\Psi^{(r,s)}](\eta)$, we can complete the entries on the boundary of the image of the Kac table under the duality transformation shown to the right of Figure \ref{fig:kactable} in the analogous manner:

\begin{equation}
\begin{aligned}
\widetilde{\Psi}^{(1,r)}(\eta)&=\mathcal{L} W^{(p-r)}[\psi](\eta)\ ,\qquad &\widetilde{\Psi}^{(p'-1,r)}(\eta)&=\mathcal{L} W^{(p-r)}[\mathcal{L}\chi](\eta)\ ,\\
\widetilde{\Psi}^{(s,1)}(\eta)&=W^{(s)}[\chi](\eta)\ ,\qquad &\widetilde{\Psi}^{(s,p-1)}(\eta)&=W^{(s)}\left[\mathcal{L}\psi\right](\eta)\ .
\end{aligned}
\end{equation} 
 
\noindent Let us discuss some evidence in favour of this proposal. Firstly, note that our extension of the duality transformation $\mathcal{C}\circ\mathcal{L}$ evidently preserves the string equation \eqref{stringeqn} and hence the bulk physics. Secondly, since $W^{(r)}[\psi]$ and $W^{(s)}[\chi]$ are constant for $r\geq p$ resp. $s\geq p'$, the corresponding differential equations are trivial and the table in Figure \ref{fig:kactable} is bounded in the appropriate way. Thirdly, since the $(p-1)^{\mathrm{th}}$-degree Wronskian for a linear differential equation of order $p$ satisfies the transpose of the originial differential equation, we deduce from the definition of the charge conjugation \eqref{conj-wronsk} that 

\begin{equation}
\begin{aligned}
\mathcal{L}\mathcal{C} [\chi](\zeta)&=\mathcal{L}W^{(p'-1)}[\chi](\eta)=\psi(\zeta)\ ,\\
\mathcal{L}\mathcal{C} [\psi](\eta)&=\mathcal{L}W^{(p-1)}[\psi](\eta)=\chi(\eta)\ , 
\end{aligned}
\end{equation}

\noindent where equality means that the corresponding modules are characterised by the same differential equations; such a relation has also been pointed in \cite{Bergere13}. In the same way, it follows that

\begin{equation}
\begin{aligned}
\mathcal{C}[\mathcal{L}\chi](\zeta)&=W^{(p-1)}[\mathcal{L}\chi](\zeta)=\psi(-\zeta)\ , \\ \mathcal{C}[\mathcal{L}\psi](\eta)&=W^{(p'-1)}[\mathcal{L}\psi](\eta)=\chi(-\eta)\ .
\end{aligned}
\end{equation}

\noindent These relations prove the consistency of our proposal at the corner entries of the table. Lastly, a consistency check for other entries on the boundary of the table is provided by a comparison of the spectral curves: after performing a Laplace transform \eqref{laplace} and charge conjugation \eqref{conj-wronsk} on \eqref{Lax1}, we obtain a new isomondromy system of \em different \em size, with operators $\widetilde{\mathcal{B}}^{(n)}(t;\eta)$ and $\widetilde{\mathcal{Q}}^{(n)}(t;\eta)$, where now $0<n<p'$. If this system is to provide a dual description of the same brane, we expect the zero locus of their characteristic polynomial to define the same spectral curve, i.e.

\begin{equation}
\label{spectraldual}
\widetilde{G}^{(n)}(t;P,Q)\propto G^{(p-n)}(t;Q,P)\ .
\end{equation}

\noindent Below we provide a check of the above relation for a few simple examples.

\begin{ex}
\noindent $(p,p')=(3,2)$. After a Laplace transform, a complete basis is given by $\vec{\widetilde{W}}^{(n)}(t;\eta)=\left(\mathcal{L}[W^{(n)}_\varnothing],\partial_t\mathcal{L}[W^{(n)}_\varnothing]\right)$, with associated $2\times 2$ Lax pair

\begin{equation}
\begin{aligned}
\widetilde{\mathcal{B}}^{(2)}&=-\mathcal{C}(\widetilde{\mathcal{B}}^{(1)})^T\mathcal{C}^{-1}&=&\frac{1}{2}
\begin{pmatrix}
0 & 2   \\
-v_2-\eta & 0   \\
\end{pmatrix}\ ,\\
\widetilde{\mathcal{Q}}^{(2)}&=-\mathcal{C}(\widetilde{\mathcal{Q}}^{(1)})^T\mathcal{C}^{-1}&=&\frac{1}{2}
\begin{pmatrix}
-\dot{v}_2 & 2v_2+4\eta  \\
-(v_2+\eta)(v_2-2\eta)+\ddot{v}_2 & \dot{v}_2  \\
\end{pmatrix}\ .
\end{aligned}
\end{equation}

\noindent The spectral curve reads

\begin{equation}
\begin{aligned}
\widetilde{G}^{(n)}(t;\eta,P)&=\pm 2\eta^3+P^2\pm \eta\left(\frac{3v_2^2}{2}+\ddot{v}\right)+\frac{v_2^3}{2}+\ddot{v}_2v_2-\frac{\dot{v}_2^2}{4}\ ,
\end{aligned}
\end{equation}

\noindent where the upper (resp. lower) sign holds for $n=1$ (resp. $n=2$). Comparison with \eqref{puregrav-curve} shows that \eqref{spectraldual} is indeed satisfied.
\end{ex}

\begin{ex}
\noindent $(p,p')=(4,3)$. After a Laplace transform, a complete basis is given by $\vec{\widetilde{W}}^{(2)}(t;\eta)=\left(\mathcal{L}[W^{(2)}_\varnothing],\mathcal{L}[W^{(2)}_{\tiny \yng(1)}],\mathcal{L}[W^{(2)}_{\tiny \yng(1)^2}]\right)$ as a consequence of \eqref{stringeqn}. To leading order in $g_s$, the associated $3\times 3$ Lax pair is given by

\begin{equation}
\begin{aligned}
\widetilde{\mathcal{B}}^{(2)}&=-\mathcal{C}(\widetilde{\mathcal{B}}^{(1)})^T\mathcal{C}^{-1}&=&\frac{1}{2}
\begin{pmatrix}
0 & 2 & 0 \\
0 & 0 & 2 \\
-\eta & -v_2& \\
\end{pmatrix}\ ,\\
\widetilde{\mathcal{Q}}^{(2)}&=-\mathcal{C}(\widetilde{\mathcal{Q}}^{(1)})^T\mathcal{C}^{-1}&=&\frac{1}{2}
\begin{pmatrix}
\frac{v_2^2}{9}-\frac{1}{2\eta} & -\eta-\frac{1}{9\eta}v_3^2 & \frac{1}{3}v_2-\frac{v_2^2}{9\eta} \\
-\frac{1}{6}v_2\eta+\frac{v_2^2}{18} & -\frac{v_2^2}{18} & \frac{2v_3}{9\eta}-\eta \\
\frac{1}{2}\eta^2+\frac{1}{18}v_3^2 & \frac{v_2}{3}\eta & \frac{v_2^2}{18} \\
\end{pmatrix}\ .
\end{aligned}
\end{equation}

\noindent Taking the semiclassical limit and evaluating in the conformal background, the spectral curve simplifies to
\begin{equation}
\begin{aligned}
\lim_{g_s\to 0}\widetilde{G}^{(2)}(t;\eta,P)=\frac{1}{2}\left[T_3\left(-P\right)-T_4(\frac{\eta}{\sqrt{2}})\right]\ .\\
\end{aligned}
\end{equation}

\noindent Comparing the latter to \eqref{ex:isingcurve}, it follows that $P^2\times\widetilde{G}^{(2)}(t;Q,P)\propto G^{(2)}(t;P,Q)$ and \eqref{spectraldual} is indeed satisfied.
\end{ex}

\section{Semiclassical limit}
\label{sec:semiclassical}

\noindent In the previous section, we saw that the generalised Wronskians allow us to define a set of averages involving independent degrees of freedom that are in one-to-one correspondence with the entries on the boundary of the Kac table. Here we study how this table reproduces the relation \eqref{SS-rel} in the semiclassical limit $g_s\to 0$, providing another piece of evidence in favour of our definition in Section \ref{subsec:duality}. Inspection of \eqref{spectral-conf-backgr} reveals that in the conformal background, the $p$ solutions to $\lim_{g_s\to 0}G^{(1)}(t;\zeta,Q)=0$ can be parametrised as $\zeta(\tau)=\cosh(p\tau)$ and $Q=Q^{(j)}(\tau)$, where 

\begin{equation}
 Q^{(j)}(\tau)=\cosh[p'(\tau-2\pi\I(j-1)/p)]\ ,\quad \ 1\leq j \leq p\ .
\end{equation}

\noindent In general, the zeroes of $\lim_{g_s\to 0}G^{(n)}(t;\zeta,Q)=G^{(n)}_{\mathrm{cl.}}(\zeta,Q)$ are then obtained from the set of linear combinations

\begin{equation}
Q^{(j_1,j_2,\dots j_n)}(\tau)=\sum_{k=1}^nQ^{(j_k)}(\tau)
\end{equation}

\noindent on the fundamental domain $\left\{(j_1,j_2,\dots j_n)\ |\ 1\leq j_1<j_2< \dots < j_n\leq p\right\}$. This is to be compared with the relation \eqref{SS-rel}, which involves the following analytic continuations of $\zeta$ and $\eta$:

\begin{subequations}
\begin{align}
\zeta_{l,k}&=\cosh\left[p(\tau+\I\pi l/p'+\I\pi k/p)\right]\ ,\\
\eta_{k,l}&=\cosh\left[p'(\tau+\I\pi l/p'+\I\pi k/p)\right]\ .
\end{align}
\end{subequations}

\noindent Comparing the explicit form of the modular $S$-matrix of the $(p,p')$ minimal model

\begin{equation}
S_{(r,s)(m,n)}=2\sqrt{\frac{2}{pp'}}(-1)^{sm+rn+1}\sin(\pi r mp'/p)\sin(\pi s np/p')\ 
\end{equation}

\noindent with the relations

\begin{equation}
\label{bdryentropy}
\sum_{m=-(s-1)}^{s-1}\frac{\zeta_{0,m}}{\zeta}=\frac{\sin(\pi s p/p')}{\sin(\pi p/p')}\ , \quad \sum_{n=-(r-1)}^{r-1}\frac{\eta_{0,n}}{\eta}=\frac{\sin(\pi rp'/p)}{\sin(\pi p'/p)}\ ,
\end{equation}

\noindent where $m$ and $n$ are incremented in steps of 2, we conclude that the product of the above two quantities can be written as

\begin{equation}
d_{r,s}=\frac{S_{(r,s)(1,1)}}{S_{(1,1)(1,1)}}\ .
\end{equation}

\noindent The numbers $d_{r,s}$ are sometimes called the quantum dimension of the state $(r,s)$; The definition of Cardy states \eqref{mattercardy} gave rise to the interpretation of these numbers as ground state degeneracies, and their logarithm as ``boundary entropies'' -- see also the end of Section \ref{sec:boundarystates} of the previous chapter. From the latter, it is evident that the branch $j_k=k-(n+1)/2$ mod $p$ gives a description of the $(r,s)=(1,n)$ Cardy brane consistent with the conformal field theory prediction \eqref{SS-rel} when evaluated in the conformal background\footnote{This result is also consistent with the relation \eqref{ss0} derived in the previous chapter.}, since

\begin{equation}
Q^{(j_1,j_2,\dots j_n)}(\tau)= d_{1,n} Q^{(0)}(\tau)\ . 
\end{equation}

\noindent However, because the isomonodromy system \eqref{Lax1} and \eqref{Lax2} has size $\binom{p}{n}$, the spectral curve will in general contain other factors besides the above branch. For $(r,s)=(1,2)$, this can be seen in the Examples \ref{ex:ising} and \ref{ex:yanglee}, where the spectral curve factorises as

\begin{subequations}
\begin{align}
(p,p')&=(4,3)\ :\quad G^{(2)}_{\mathrm{cl.}}(\zeta,Q)=\mathrm{const.}\times Q^2\left(T_4\left(\frac{Q}{\sqrt{2}}\right)-T_3(-\zeta)\right)\ , \\
(p,p')&=(5,2)\ :\quad G^{(2)}_{\mathrm{cl.}}(\zeta,Q)=\mathrm{const.}\times \prod_{\pm}\left(T_5\left(\frac{2}{1\pm \sqrt{5}}Q\right)-T_2(\mp\zeta)\right)\ .
\end{align}
\end{subequations}

 \noindent We close this section by demonstrating that this observation generalises to arbitrary $(p,p')$ according to the following

\begin{prop}
\label{prop:semiclassicaln=2}
Let $T_p(\cosh\tau)=\cosh(p\tau)$ denote the $p^{\mathrm{th}}$ Chebyshev polynomial of the first kind. Then up to normalisation, the semiclassical curve for $n=2$ can be written as
\begin{equation}
\label{2-pt-curve}
G^{(2)}_{\mathrm{cl.}}(\zeta,Q)=
\begin{cases} \prod_{a=1}^{(p-1)/2}\left(T_p\left(\frac{Q}{2\cos(\pi p' a/p)}\right)-T_{p'}((-1)^a\zeta)\right)\ ,\quad & p\ \mathrm{odd}\ ,\\
 Q^{p/2}\prod_{a=1}^{(p-2)/2}\left(T_p\left(\frac{Q}{2\cos(\pi p' a/p)}\right)-T_{p'}((-1)^a\zeta)\right)\ ,\quad & p\ \mathrm{even}\ .
\end{cases}
\end{equation}
\end{prop}

\begin{proof} The zeroes of $G^{(2)}_{\mathrm{cl.}}(\zeta,Q)$ are parametrised by $p'(p'-1)/2$ functions $Q^{(j_1,j_2)}(\tau)$ on the fundamental domain $\left\{(j_1,j_2)\ |\ 1\leq j_1<j_2\leq p\right\}$. Eliminating $j_2$ in favour of $a=j_2-j_1$, we find

\begin{equation}
Q^{(j_1,j_2)}(\tau)=2\cos(2\pi p' a/p)Q^{(j_1+a)}(\tau)\ ,
\end{equation}

\noindent which is the solution to

\begin{equation}
\label{n=2-factor}
T_p\left(\frac{Q}{2\cos(\pi p' a/p)}\right)-T_{p'}(\zeta)=0\ .
\end{equation} 

\noindent We can now distinguish the following two cases:

\begin{enumerate}

\item When $p$ is odd, we may choose $1\leq j_1\leq p$, $1\leq a\leq (p-1)/2$ as a fundamental domain, giving the first line in \eqref{2-pt-curve}.

\item When $p$ is even, we may choose $1\leq j_1\leq p$, $1\leq a\leq (p-2)/2$, together with $1\leq j_1\leq p/2$ for $a=p/2$ as a fundamental domain. Since $p'$ is odd, $Q^{(j_1,j_2)}(\tau)=0$ for $a=p/2$, giving the factor $Q^{p/2}$ in the second line of \eqref{2-pt-curve}. For $1\leq a\leq (p-2)/2$, $Q^{(j_1,j_2)}(\tau)$ again solves \eqref{n=2-factor}, giving the remainder of the second line in \eqref{2-pt-curve}. 

\end{enumerate}
\end{proof}

\section{Discussion}
\label{sec:discussionwronsk}

\noindent Let us summarise our results. In Section \ref{sec:W}, we first motivated our definition of the Wronskian \eqref{defn-wronskian} to describe the independent degrees of freedom arising in the non-perturbative description outlined in the introduction, Subsection \ref{int:doublescaling}. We then derived the non-perturbative differential equations satisfied by the latter in Proposition \ref{Prop-I}. The construction of the isomonodromy system using Proposition \ref{Prop-II} consequently allowed for the construction of the spectral curves according to Definition \ref{def:spectralcurve}. A Kac table of independent branes with entries in one-to-one correspondence with the primary fields of the minimal model then emerged naturally from the properties of the Wronskian in conjunction with the duality transformation \eqref{duality}. Finally, in Section \ref{sec:semiclassical}, we then showed how the semiclassical limit $g_s\to 0$ of the Wronskian includes a branch consistent with the degeneracy \eqref{SS-rel} predicted by conformal field theory. Altogether, these results provide ample evidence that due to Stokes' phenomenon, the non-perturbative the general Cardy brane with $(r,s)\neq(1,1)$ can \em not \em be described by analytic continuation of a single principal solution $(r,s)=(1,1)$, but is instead a bound state of \em independent \em degrees of freedom, whose wave function is given by the Wronskian functions \eqref{defn-wronskian}, and more generally \eqref{cardybranes}.

There are many possible extensions of this work that we have not touched upon. Various computations and consistency checks have only been performed for particular examples; a more general proof of these statements would surely provide deeper insight. We have also omitted the entries in the bulk of the Kac table. To define the wave functions for the corresponding branes, additional successive Wronskian operations must be performed on products of the Wronskians considered herein; we leave an investigation of this more complicated case for future work \cite{Wronskians}. It would also be interesting to extend our results to the non-diagonal theories with $q>2$ studied in Chapters \ref{chap:potts} and \ref{chap:cft}, which would allow for a check of the results reported in \cite{Wilshin08}.

Finally, a potential application of our results pertains to the analogy of the Baker-Akhiezer function $\psi(t;\zeta)$ with the correlator of a gauge theory dual to a spacetime with a horizon \cite{Fidkowski03}. Our results suggest that whilst perturbatively, we can obtain a description of the physics behind the horizon by analytic continuation through the branch cut in the complex $\zeta$-plane, the non-perturbative correlator exhibits Stokes' phenomenon signalling the presence of independent degrees of freedom, as also alluded to in \cite{Seiberg03}. It would be interesting to explore the implications of our results for this topic, which may pave the way for an extension of these considerations to more complicated backgrounds such as Witten's black hole captured by the $SL(2,\mathbb{R})/U(1)$ coset model \cite{Witten91}, which also has a matrix description \cite{Kazakov00}.

\chapter{Summary}
\label{chap:sum}
\thispagestyle{empty}

\noindent In this thesis, we have developed novel descriptions of boundary conditions for statistical models on random surfaces employing the measure \eqref{pottsmeasure} in the planar, scaling, and double scaling limit. We began by introducing each of these limits in Chapter \ref{chap:intro}, detailing their connection to statistical physics on planar lattices, conformal field theory and finally non-perturbative string theory in a low target space dimension. In each of these cases, we paid particular attention to the description of boundaries in the emsembles of random surfaces that arise in these limits. Following this compressed review came the three chapters containing the bulk of the author's original work,   the main results of which we henceforth summarise:
 
 In Chapter \ref{chap:potts}, we derived the large-$N$ spectral density of sums of random matrices of the form $X_1+X_2+\dots X_p$, $1\leq p\leq q$, distributed according to \eqref{pottsmeasure} by generalising Voiculescu's formula \eqref{Voicu} to a situation beyond free probability (Proposition \ref{prop-curve}) and explained the interpretation of these quantities as disk partition functions of the $q$-states Potts model with $p$ allowed, equally weighted colors on a connected boundary. Besides finding a remarkable algebraic relation between the boundary conditions with $p$ and $q-p$ colors (Corollary \ref{curve3}) and providing an elliptic parametrisation of the general solution for arbitrary $q\neq 4$ (Proposition \ref{prop-ycurve}), we derived the explicit polynomial equations satisfied by the latter for specific examples with $q=1,2,3$. The scaling relations obtained for these cases were found to be consistent with a description in terms of Liouville theory coupled to a minimal model with central charge $c_M=0,1/2$ and $4/5$, setting the stage for Chapter \ref{chap:cft}.
 
  Therein, we considered the non-diagonal $\mathcal{W}_q$ minimal model with conserved higher-spin currents coupled to Liouville theory as a description for the universality classes corresponding to the critical points in the phase diagram of  the model \eqref{pottsmeasure}. Using the free-field resolution of the $\mathcal{W}_q$-modules, we considered the cohomology of the nilpotent BRST operator associated with worldsheet diffeomorphisms (Proposition \ref{prop:cohomology}) and deduced the presence of the tachyon operators \eqref{tachyons} in the spectrum of observables. We proceeded to consider the one-point function of the latter on the disk and showed that the na\"{i}ve tensor product of the Liouville FZZT and matter Cardy states appears to overcount the number of physically distinct boundary conditions if one allows for complex values of the boundary cosmological constant. This degeneracy provided a simple way to derive a functional difference equation for the tachyon one-point functions that turned out to agree with the equations obtained from the scaling limit of the matrix model. 
  
 Finally, in Chapter \ref{chap:wronskians}, we investigated the double scaling limit of \eqref{pottsmeasure} for $q=2$ to understand the fate of this degeneracy beyond perturbation theory in the string coupling. We argued that the resolvent operator may exhibit Stokes' phenomenon and proposed a generalised Wronskian as an observable that can resolve the degeneracy non-perturbatively. Without reference to conformal field theory, we determined the differential equations that govern the Wronskians (Propositions \ref{Prop-I} and \ref{Prop-II}) and found a maximum of $(p-1)(p'-1)/2$ independent Wronskians, one per Cardy state of the minimal model. Moreover, we could explicitly show for various examples that each entry of the resulting Kac table consistently reproduces the relation \eqref{SS-rel} predicted by conformal field theory (Proposition \ref{prop:semiclassicaln=2}). We argued that this is strong evidence that the above-mentioned degeneracy is an artefact of the asymptotic expansion and the usual determinant operator is insufficient to capture all information about the theory. Instead, the degeneracy is resolved non-perturbatively by the independent degrees of freedom comprising the Wronskian.
 
Altogether, these developments have led to the description of a multitude of nontrivial, well-defined boundary conditions whose properties, to the knowledge of the author, have not been previously described in the otherwise vast existing literature on the subject. This suggests that despite the excellent understanding we have of these models -- largely thanks to their intimate connection to the theory of integrable systems -- many of their properties remain to be worked out. Indeed, as seen from the discussions in Section \ref{sec:pottsdiscussion}, \ref{sec:discussioncft} and \ref{sec:discussionwronsk}, this work has also prompted numerous follow-up questions that warrant further investigation, including the scaling behaviour of strongly coupled models with $c_M>1$, the inclusion of magnetic fields on the boundary and the extension of the Wronskian to the interior of the Kac table. It has also hinted at diverse connections to other fields, such as free probability theory, higher-spin gravity in three dimensions and physics behind black hole horizons. Extending the insights of this thesis more comprehensively to some of the more complicated models may be more challenging; not all of them may share the simplicity of the Hermitian matrix model. Nevertheless, it is reasonable to expect that some features discovered herein may persist in more generality and it is the hope of the author that this thesis enticed the reader about the potential of these avenues to further our general understanding of the mathematical description of random geometry, boundaries and branes.

\appendix
\chapter{Auxiliary Saddle Point Problem}
\thispagestyle{title}
\label{app:auxprob}

\noindent Given $|\nu|\leq 1$, consider a function $f(w)$ holomorphic on $\mathbb{C}\setminus[\alpha,\beta]$ for some connected $[\alpha,\beta]\subset\mathbb{R}$, satisfying

\begin{equation}
	\label{linear}
	\mathrm{Re}\ f(w)=\cos (\pi \nu)f(-w)\ ,\quad w\in[\alpha,\beta]\ .
\end{equation} 

\noindent The general solution to this equation was first described in \cite{Eynard95} and we will derive it below; thereafter we investigate the limit $\alpha/\beta\to 0$.
\\

\noindent \em General solution. \em We begin by showing that any function satisfying \eqref{linear} is uniquely specified by the behaviour at its singularites. To this end, it is useful to introduce a new coordinate $\sigma$ by

\begin{equation}
	\label{defnphi}
	\begin{aligned}
		\mathbb{C}\setminus[\pm\alpha,\pm\beta]&\longrightarrow(0,1)\times[0,\tau)\subset\mathbb{C}\;,\\
		w&\longmapsto\sigma(w)=\frac{1}{2K}\int_{1}^{w/\alpha}\mathrm{d}t\left((1-t^2)(1-(\alpha t/\beta)^2)\right)^{-1/2}\ .
	\end{aligned}
\end{equation}

\noindent By definition of the Jacobi elliptic function\footnote{Our conventions for elliptic functions are those of Gradshtein and Ryzhik \cite{Gradshteyn07}.} $\mathrm{sn}(u|k)$ of elliptic modulus $k$, the inverse map is

\begin{equation}
	\label{ellipticparam}
	\begin{aligned}
		(0,1)\times[0,\tau)&\longrightarrow\mathbb{C}\setminus[\pm\alpha,\pm\beta]\;,&\\
		\sigma&\longmapsto w(\sigma)
		=\alpha\ \mathrm{sn}\left(2K\sigma+K\right|\alpha/\beta)\ .
	\end{aligned}
\end{equation}

\noindent Here, $K$ and $K'$ are given by the complete elliptic integrals of the first and second kind, respectively:

\begin{equation}
\label{ellipticK}
	\begin{aligned} 
		K&=\int_0^1\mathrm{d}t\left((1-t^2)(1-(\alpha t/\beta)^2)\right)^{-1/2}\;,\\
		K'&=\int_0^{\infty}\mathrm{d}t\left((1+t^2)(1+(\alpha t/\beta)^2)\right)^{-1/2}\ .
	\end{aligned}
\end{equation}

\noindent This change of variables correpsonds to parametrising the two-cut complex $w$-plane on the torus $\mathbb{C}/(\mathbb{Z}+\tau\mathbb{Z})$ with modular parameter 

\begin{equation}
	\label{tau}
	\tau=\I \frac{K'}{K}\;.
\end{equation}

\noindent The coordinate $w$ is invariant under $\sigma\to-\sigma$ and (anti-) periodic along the respective cycles of the torus:

\begin{equation}
	w(\sigma+m+n\tau)=(-1)^m w(\sigma)\;,\qquad(m,n)\in\mathbb{Z}^2\;.
\end{equation}

\noindent We also require the Jacobi theta functions

\begin{equation}
	\begin{aligned}
		\vartheta_1(u|\tau)&=\frac{1}{\I}\sum_{n\in\mathbb{Z}}(-1)^n \E^{\I\pi\tau(n+1/2)^2}\E^{\I u(2n+1)}\ ,\\
		\vartheta_2(u|\tau)&=\sum_{n\in\mathbb{Z}} \E^{\I\pi\tau(n+1/2)^2}\E^{\I u(2n+1)}\ ,\\
		\vartheta_3(u|\tau)&=\sum_{n\in\mathbb{Z}}\E^{\I\pi\tau n^2}\E^{2\I u n}\ .
	\end{aligned}
\end{equation}

\noindent In particular, $\vartheta_1$ is an entire function with a unique simple zero at $u=0\ \mathrm{mod}\ \mathbb{Z}\oplus\pi\tau\mathbb{Z}$, satisfying

\begin{equation}
\label{theta1}
	\begin{aligned}
		\vartheta_1\left(u+\pi(m+n\tau)|\tau\right)&=(-1)^{mn} \E^{-\I n(\pi\tau+2u)}\vartheta_1(u|\tau)\;,\qquad(m,n)\in\mathbb{Z}^2\;,\\
		\vartheta_1(-u|\tau)&=-\vartheta_1(u|\tau)\ ,\\
\vartheta_1(u/\tau|-1/\tau)&=-\sqrt{\I\tau}\E^{\I\pi u^2/\tau}\vartheta_1(u|\tau)\ .
	\end{aligned}
\end{equation}

\noindent We also note the equivalent representation of $w(\sigma)$ in terms of $\vartheta_i$,

\begin{equation}
\label{theta23}
	w(\sigma)
	=\sqrt{\alpha\beta}\frac{\vartheta_2(\pi\sigma|\tau)}{\vartheta_3(\pi\sigma|\tau)}\ .
\end{equation}

\noindent Analytic continuation of $f(w(\sigma))$ requires boundary conditions on the rectangle $(0,1)\times[0,\tau)$:

\begin{enumerate}
	\item Analyticity across $[0,\alpha]\cup[\beta,\infty]$ allows us to continue $f(w)$ to the infinite strip $(0,1)\times \I\mathbb{R}$ by
	\begin{equation}
		\label{period}
		f(w(\tau+\sigma))=f(w(\sigma))\ , \quad \sigma\in(0,1)\times[0,\tau)\ .
	\end{equation}
	\item Analyticity across $[-\beta,-\alpha]$ allows us to extend this definition to $(0,2)\times \I\mathbb{R}$ using
	\begin{equation} 
		\label{refl}
		f(w(1+\sigma))=f(w(1-\sigma))\ , \quad \sigma\in(0,1)\times \I\mathbb{R}\ .		
	\end{equation}
	
	\item Finally, using all the above, the functional equation \eqref{linear} implies 
	\begin{equation}
		\label{spe-ell}
		f(w(1+\sigma))=\frac{f(w(\sigma))+f(w(2+\sigma))}{2\cos(\pi\nu)}\ , \quad\sigma\in(0,2)\times \I\mathbb{R}\ .
	\end{equation}
\end{enumerate}

\begin{figure}[t]
	\centering
	\begin{pspicture}(7,6)
\psset{unit=1.6cm}

\psdot[dotsize=.1](1,1)
\rput[tl](1.1,0.9){$0$}
\rput[tl](1.1,1.3){$[\alpha]$}
\psdot[dotsize=.1](1,2)
\rput[tl](1.1,1.9){$\tau/2$}
\rput[tl](1.1,2.3){$[\beta]$}
\psdot[dotsize=.1](1,3)
\rput[tl](1.1,2.9){$\tau$}
\rput[tl](1.1,3.3){$[\alpha]$}

\psdot[dotsize=.1](2,1)
\rput[tl](2.1,0.9){$1/2$}
\rput[tl](2.1,1.3){$[0]$}
\psdot[dotsize=.1](2,2)
\rput[tl](2.1,2.3){$[\infty]$}
\psdot[dotsize=.1](2,3)
\rput[tl](2.1,3.3){$[0]$}

\psdot[dotsize=.1](3,1)
\rput[tl](3.1,1.3){$[-\alpha]$}
\rput[tl](3.1,0.9){$1$}
\psdot[dotsize=.1](3,2)
\rput[tl](3.1,2.3){$[-\beta]$}
\psdot[dotsize=.1](3,3)
\rput[tl](3.1,3.3){$[-\alpha]$}

\psline{->}(0.75,1)(3.5,1)
\rput(4,1){$\mathrm{Re}\ \sigma$}
\psline{->}(1,0.75)(1,3.25)
\rput(1,3.5){$\mathrm{Im}\ \sigma$}
\end{pspicture}
	\vspace{-1cm}
	\caption{A fundamental domain for $w\in\mathbb{C}\setminus[\pm\alpha,\pm\beta]$ is given by $\sigma\in(0,1)\times[0,\tau)$. The images of special points under the map $z(\sigma)$ are indicated in square brackets.}
	\label{fig:sheets}
\end{figure}
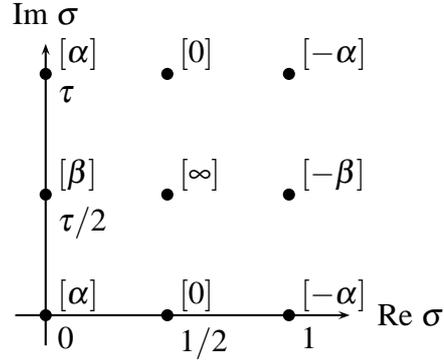

\noindent Solving the latter condition allows us to continue $f(w(\sigma))$ to a meromorphic function on the entire complex $\sigma$-plane, on which $f(w)$ satisfies two (quasi-)periodicity conditions:

\begin{subequations}
\begin{align}
	\label{periodicity1}
	0&=(\E^{-\partial_{\sigma}}-\E^{\I\pi\nu})(\E^{-\partial_{\sigma}}-\E^{-\I\pi\nu})f(w(\sigma))\ ,\\
\label{periodicity2}
	0&=(\E^{-\tau\partial_{\sigma}}-1)f(w(\sigma))\ .
\end{align}
\end{subequations}

\noindent We find it convenient to follow \cite{Borot12} in introducing the unique function in $\mathrm{Ker}(\E^{-\partial_\sigma}-\E^{\I\pi\nu})$ with a simple pole of unit residue at $\sigma=0$ and no other singularities mod $\mathbb{Z}\oplus\tau\mathbb{Z}$ as

\begin{equation}
	\label{g(phi;nu)}
	g(\sigma;\nu)=\frac{\vartheta_1'(0|\tau)}{\vartheta_1( \pi\nu\tau/2|\tau)}\frac{\vartheta_1(\pi\sigma+\pi\nu\tau/2|\tau)}{\vartheta_1(\pi\sigma|\tau)}\E^{\I\pi\nu\sigma}\ ,
\end{equation}

\noindent which has a simple zero at $\sigma=-\nu\tau/2$; any solution to equations \eqref{periodicity1} and \eqref{periodicity2} may be expressed as a linear combination of derivatives $g(\sigma;\pm\nu)$ with shifted argument. The reflection relation \eqref{refl} fixes the relative coefficient, so that the general solution to \eqref{linear} can be expressed as

\begin{equation}
f(w)=\sum_{n\geq 0}\frac{a_n}{n!}\frac{\partial^n}{\partial\sigma_0^n}\left(\E^{-\I \pi\nu/2}g(\sigma(w)-\sigma_0;\nu)-\E^{\I\pi\nu/2}g(\sigma(w)-\sigma_0;-\nu)\right)\ ,
\end{equation}

\noindent where the requirement that $f(w)$ be free of singularities on $\mathbb{C}\setminus[\alpha,\beta]$ demands $\sigma_0=(\tau+1)/2$, and the coefficients $a_n$ are to be determined by boundary conditions supplementing the problem \eqref{linear}, using

 \begin{equation}
 \lim_{z\to\infty}(\sigma-\sigma_0)^{n+1}\frac{1}{n!}\frac{\partial^n}{\partial \sigma_0^n}g(\sigma-\sigma_0;\nu)=1\ .
 \end{equation}
 
  \noindent In particular, if $f(w)$ has a pole of order $m$ at $w=\infty$, then $a_n=0$ for $n>m$.
\\

\noindent \em The limit $\alpha/\beta \searrow 0$. \em In Section \ref{sec:crit} we will be interested in the limit $\alpha/\beta\searrow 0$, in which $\tau\to \I \infty$, and thus

\begin{equation}
	\label{scalinglimit}
	\lim_{\tau\to \I \infty}w(\sigma)/\alpha=\cos(\pi\sigma)\;,\qquad \lim_{\tau\to \I \infty}g(\sigma;\nu)=\frac{\pi\ \E^{\I\pi(\nu-1)\sigma}}{\sin(\pi\nu)}\;.
\end{equation}

\noindent In this limit, $f(w)$ is holomorphic on $w/\alpha\in\mathbb{C}\setminus[1,\infty)$, and the equation \eqref{linear} becomes

 \begin{equation}
 (\E^{-\partial_{\sigma}}-\E^{\I\pi\nu})(\E^{-\partial_{\sigma}}-\E^{-\I\pi\nu})f\left(\alpha\cos(\pi \sigma)\right)=0\ .
 \end{equation}
 
 \noindent A convenient basis for the solution space is given by the Chebyshev functions. These are represented on the unit disk as

\begin{equation}
	\begin{aligned}
		\label{chebyshevfct}
		T_{\nu}(x)&=\cos(\pi\nu\phi)\ ,\\
		U_{\nu}(x)&=\frac{\sin(\pi(\nu+1)\phi)}{\sin(\pi\phi)}\ ,\quad\ x=\cos(\pi\phi)\ .\\
	\end{aligned}
\end{equation}

\noindent From the above definition it is easy to verify that both $T_{\nu}(x)$ and $U_{\nu}(x)$ satisfy equation \eqref{linear}, and $T_{1/\nu}(x)$ is the functional inverse of $T_{\nu}(x)$. For non-integer $\nu$, these functions have a branch cut on $x\in [-1,-\infty)$, with discontinuity

\begin{equation}
	\begin{aligned}
		\label{Chebyshevdiscont}
		T_{\nu}(x)_+-T_{\nu}(x)_-&=-2\I\sin(\pi\nu)\sqrt{1-x^2}\ U_{\nu-1}(-x)\ ,\\
		U_{\nu}(x)_+-U_{\nu}(x)_-&=-2\I\frac{\sin(\pi\nu)}{\sqrt{1-x^2}}T_{\nu+1}(-x)\ .
	\end{aligned}
\end{equation}

\noindent When $\nu\in\mathbb{N}$, the right-hand side vanishes and we recover the definition of the Chebyshev polynomials of the first and second kind. As a result when $\nu=p/q$ is rational, $y=T_{\nu}(x)$ is the solution to the polynomial equation $T_q(y)-T_p(x)=0$. Since \eqref{linear} restricts the scaling exponents $f(w)\sim(-w)^\kappa$ to the form $\kappa=2n\pm\nu$, $n \in\mathbb{Z}$ we can expand $f(w)$ as

\begin{equation}
\label{f-scaling}
	f(w)=\sum_{n\geq0}\sum_{\pm}\alpha^{2n\pm\nu}(t_n^{(\pm)}T_{2n\pm\nu}(-w/\alpha)+u_n^{(\pm)}U_{2n\pm\nu}(-w/\alpha))\ ,\quad |w/\alpha|\leq 1\ ,
\end{equation}

\noindent with constants $t_n^{(\pm)}$, $u_n^{(\pm)}$ to be determined by boundary condtions. 

\chapter{Analytic Structure and Asymptotics}
\label{app:asympt}

\noindent We illustrate the analytic structure of $G_Y^{(p)}(z)$ and $G_{(p)}^Y(z)$ by graphs in which nodes depict sheets and lines between nodes depict branch cuts that connect the sheets.  Of the latter, double lines represent finite cuts and single lines represent cuts that extend to infinity.
\\
\begin{ex}
\label{ex-dimers} $(q,k)=(1,2)$. From equations \eqref{case1a} and \eqref{case1b}, we compute the analytic structure and asymptotic behaviour of $G_{(p)}^Y(z)$,

\begin{pspicture}(-1,0)(10,4)
\psset{unit=.8cm}

\psline(1,1)(1,2)
\psline(1,2)(1,3)
\psline[doubleline=true](1,3)(1,4)

\psdot[dotsize=.15](1,4)
\psdot[dotsize=.15](1,3)
\psdot[dotsize=.15](1,2)
\psdot[dotsize=.15](1,1)

\rput[Br](.5,4){$G_Y^{(1)}(z)_-$}
\rput[Br](.5,3){$G_Y^{(1)}(z)_+$}
\rput[Bl](1.5,4){$z-z^{-1}+\mathcal{O}(z^{-2})$}
\rput[Bl](1.5,3){$\omega^2t_4^{-1/3}z^{1/3}-\frac{t_3}{3t_4}+\omega\frac{t_3^2-3t_2t_4}{9t_4^{5/3}}z^{-1/3}-\omega^2\frac{2t_3^3-9t_2t_3t_4}{81t_4^{7/3}}z^{-2/3}+\frac{1}{3}z^{-1}+\mathcal{O}(z^{-4/3})$}
\rput[Bl](1.5,2){$\omega t_4^{-1/3}z^{1/3}-\frac{t_3}{3t_4}+\omega^2\frac{t_3^2-3t_2t_4}{9t_4^{5/3}}z^{-1/3}-\omega\frac{2t_3^3-9t_2t_3t_4}{81t_4^{7/3}}z^{-2/3}+\frac{1}{3}z^{-1}+\mathcal{O}(z^{-4/3})$}
\rput[Bl](1.5,1){$t_4^{-1/3}z^{1/3}-\frac{t_3}{3t_4}+\frac{t_3^2-3t_2t_4}{9t_4^{5/3}}z^{-1/3}-\frac{2t_3^3-9t_2t_3t_4}{81t_4^{7/3}}z^{-2/3}+\frac{1}{3}z^{-1}+\mathcal{O}(z^{-4/3})$}
\end{pspicture}

\noindent where $\omega=\E^{\I\pi/3}$. From \eqref{polynomial0}, we may compute the asymptotic behaviour of $G_{(p)}^Y(z)$ on all sheets: 

\begin{pspicture}(-1,0)(10,4)
\psset{unit=.8cm}

\rput[Br](0.5,3){$G^Y_{(0)}(z)_-$}
\rput[Br](0.5,4){$G^Y_{(0)}(z)_+$}
\rput[Bl](1.5,1){$z+t_4^{-1/3}z^{1/3}-\frac{t_3}{3t_4}+\frac{t_3^2-3(t_2-1)t_4}{9t_4^{5/3}}z^{-1/3}-\frac{2t_3^3-9(t_2-1)t_3t_4}{81t_4^{7/3}}z^{-2/3}+\frac{1}{3}z^{-1}+\mathcal{O}(z^{-4/3})$}
\rput[Bl](1.5,2){$z+\omega t_4^{-1/3}z^{1/3}-\frac{t_3}{3t_4}+\omega^2\frac{t_3^2-3(t_2-1)t_4}{9t_4^{5/3}}z^{-1/3}-\omega \frac{2t_3^3-9(t_2-1)t_3t_4}{81t_4^{7/3}}z^{-2/3}+\frac{1}{3}z^{-1}+\mathcal{O}(z^{-4/3})$}
\rput[Bl](1.5,3){$z+\omega^2 t_4^{-1/3}z^{1/3}-\frac{t_3}{3t_4}+\omega\frac{t_3^2-3(t_2-1)t_4}{9t_4^{5/3}}z^{-1/3}-\omega^2\frac{2t_3^3-9(t_2-1)t_3t_4}{81t_4^{7/3}}z^{-2/3}+\frac{1}{3}z^{-1}+\mathcal{O}(z^{-4/3})$}
\rput[Bl](1.5,4){$z^{-1}+\mathcal{O}(z^{-2})$}

\psline(1,1)(1,2)
\psline(1,2)(1,3)
\psline[doubleline=true](1,3)(1,4)
\psdot[dotsize=.15](1,1)
\psdot[dotsize=.15](1,2)
\psdot[dotsize=.15](1,3)
\psdot[dotsize=.15](1,4)

\end{pspicture}

\begin{pspicture}(-1,0)(10,2)
\psset{unit=.8cm}

\psline[doubleline=true](1,1)(1,2)

\psdot[dotsize=.15](1,1)
\psdot[dotsize=.15](1,2)

\rput[Br](0.5,2){$G^Y_{(1)}(z)_+$}
\rput[Br](0.5,1){$G^Y_{(1)}(z)_-$}
\rput[Bl](1.5,1){$z+z^{-1}+\mathcal{O}(z^{-2})$}
\rput[Bl](1.5,2){$t_4z^3+t_3z^2+t_2z-z^{-1}+\mathcal{O}(z^{-2})$}

\end{pspicture}

\end{ex}

\begin{ex}
\label{ex-ising} $(q,k)=(2,1)$. From equations \eqref{case2a} and \eqref{case2b}, we compute the analytic structure and asymptotic behaviour of $G_{(p)}^Y(z)$,

\begin{pspicture}(-1,0)(10,4)
\psset{unit=.8cm}

\psline(1,1)(1,2)
\psline[doubleline=true](1,2)(1,3)
\psline(1,3)(1,4)

\psdot[dotsize=.15](1,1)
\psdot[dotsize=.15](1,2)
\psdot[dotsize=.15](1,3)
\psdot[dotsize=.15](1,4)

\rput[Br](0.5,3){$G_Y^{(1)}(z)_-$}
\rput[Br](0.5,2){$G_Y^{(1)}(z)_+$}
\rput[Bl](1.5,4){$z+t_3^{-1/2}z^{1/2}+\frac{t_2}{2t_3}+\frac{t_2^2}{8t_3^{3/2}}z^{-1/2}-\frac{1}{2}z^{-1}+\mathcal{O}(z^{-3/2})$}
\rput[Bl](1.5,3){$z-t_3^{-1/2}z^{1/2}+\frac{t_2}{2t_3}-\frac{t_2^2}{8t_3^{3/2}}z^{-1/2}-\frac{1}{2}z^{-1}+\mathcal{O}(z^{-3/2})$}
\rput[Bl](1.5,2){$t_3^{-1/2}z^{1/2}-\frac{t_2}{2t_3}+\frac{t_2^2}{8t_3^{3/2}}z^{-1/2}+\frac{1}{2}z^{-1}+\mathcal{O}(z^{-3/2})$}
\rput[Bl](1.5,1){$-t_3^{-1/2}z^{1/2}-\frac{t_2}{2t_3}-\frac{t_2^2}{8t_3^{3/2}}z^{-1/2}+\frac{1}{2}z^{-1}+\mathcal{O}(z^{-3/2})$}
\end{pspicture}

\begin{pspicture}(-1,0)(10,4)
\psset{unit=.8cm}

\psline[doubleline=true](1,1)(1,2)
\psline(1,2)(1,3)
\psline[doubleline=true](1,3)(1,4)

\psdot[dotsize=.15](1,1)
\psdot[dotsize=.15](1,2)
\psdot[dotsize=.15](1,3)
\psdot[dotsize=.15](1,4)

\rput[Br](0.5,4){$G_Y^{(2)}(z)_-$}
\rput[Br](0.5,3){$G_Y^{(2)}(z)_+$}
\rput[Bl](1.5,4){$z-z^{-1}+\mathcal{O}(z^{-2})$}
\rput[Bl](1.5,3){$2t_3^{-1/2}z^{1/2}-\frac{t_2}{t_3}+\frac{t_2^2}{4t_3^{3/2}}z^{-1/2}+\mathcal{O}(z^{-3/2})$}
\rput[Bl](1.5,2){$-2t_3^{-1/2}z^{1/2}-\frac{t_2}{t_3}-\frac{t_2^2}{4t_3^{3/2}}z^{-1/2}+\mathcal{O}(z^{-3/2})$}
\rput[Bl](1.5,1){$-z-\frac{2t_2}{t_3}+z^{-1}+\mathcal{O}(z^{-2})$}

\end{pspicture}

\noindent From the resulting polynomials $F_{(p)}(x,y)$, we may also compute the asymptotic behaviour of $G_{(p)}^Y(z)$ on all sheets. For example, from \eqref{polynomial1},

\begin{pspicture}(-1,0)(10,4)
\psset{unit=.8cm}

\psline(1,1)(1,2)
\psline[doubleline=true](1,2)(1,3)
\psdot[dotsize=.15](1,1)
\psdot[dotsize=.15](1,2)
\psdot[dotsize=.15](1,3)

\rput[Br](0.5,3){$G^Y_{(1)}(z)_-$}
\rput[Br](0.5,2){$G^Y_{(1)}(z)_+$}
\rput[Bl](1.5,3){$t_3z^2+t_2z-z^{-1}+\mathcal{O}(z^{-2})$}
\rput[Bl](1.5,2){$z+t_3^{-1/2}z^{1/2}+\frac{t_2-1}{t_3}+\frac{(t_2-1)^2}{8t_3^{3/2}}z^{-1/2}+\frac{1}{2}z^{-1}+\mathcal{O}(z^{-3/2})$}
\rput[Bl](1.5,1){$z-t_3^{-1/2}z^{1/2}+\frac{t_2-1}{t_3}-\frac{(t_2-1)^2}{8t_3^{3/2}}z^{-1/2}+\frac{1}{2}z^{-1}-\mathcal{O}(z^{-3/2})$}

\end{pspicture}

\begin{pspicture}(-1,0)(10,3)
\psset{unit=.8cm}

\psline[doubleline=true](1,1)(1,2)
\psline[doubleline=true](1,2)(1,3)
\psdot[dotsize=.15](1,1)
\psdot[dotsize=.15](1,2)
\psdot[dotsize=.15](1,3)

\rput[Br](0.5,2){$G^Y_{(2)}(z)_-$}
\rput[Br](0.5,3){$G^Y_{(2)}(z)_+$}
\rput[Bl](1.5,1){$-z-\frac{2t_2}{t_3}-z^{-1}+\mathcal{O}(z^{-2})$}
\rput[Bl](1.5,2){$\frac{t_3}{4}z^2+\frac{t_2}{2}z+\mathcal{O}(z^{-2})$}
\rput[Bl](1.5,3){$z+z^{-1}+\mathcal{O}(z^{-2})$}

\end{pspicture}
\end{ex}

\begin{ex}
\label{ex-potts3} $(q,k)=(3,1)$. From equations \eqref{case3a} and \eqref{case3b}, we compute the analytic structure and asymptotic behaviour of $G_{(p)}^Y(z)$,

\begin{pspicture}(-1,0)(10,6)
\psset{unit=.8cm}

\psline(1,1)(1,2)
\psline[doubleline=true](1,2)(1,3)
\psline(1,3)(1,4)
\psline[doubleline=true](1,4)(1,5)
\psline(1,5)(1,6)

\psdot[dotsize=.15](1,1)
\psdot[dotsize=.15](1,2)
\psdot[dotsize=.15](1,3)
\psdot[dotsize=.15](1,4)
\psdot[dotsize=.15](1,5)
\psdot[dotsize=.15](1,6)

\rput[Br](0.5,3){$G_Y^{(1)}(z)_-$}
\rput[Br](0.5,2){$G_Y^{(1)}(z)_+$}
\rput[Bl](1.5,1){$-t_3^{-1/2}z^{1/2}-\frac{t_2}{2t_3}-\frac{t_2^2}{8t_3^{3/2}}z^{-1/2}+\frac{1}{2}z^{-1}-\mathcal{O}(z^{-3/2})$}
\rput[Bl](1.5,2){$t_3^{-1/2}z^{1/2}-\frac{t_2}{2t_3}+\frac{t_2^2}{8t_3^{3/2}}z^{-1/2}+\frac{1}{2}z^{-1}+\mathcal{O}(z^{-3/2})$}
\rput[Bl](1.5,3){$ z-2t_3^{-1/2}z^{1/2}+\frac{t_2}{t_3}-\frac{t_2^2}{4t_3^{3/2}}z^{-1/2}-\mathcal{O}(z^{-3/2})$}
\rput[Bl](1.5,4){$ z+2t_3^{-1/2}z^{1/2}+\frac{t_2}{t_3}+\frac{t_2^2}{4t_3^{3/2}}z^{-1/2}+\mathcal{O}(z^{-3/2})$}
\rput[Bl](1.5,5){$2z-t_3^{-1/2}z^{1/2}+\frac{5t_2}{2t_3}-\frac{t_2^2}{8t_3^{3/2}}z^{-1/2}-\frac{1}{2}z^{-1}-\mathcal{O}(z^{-3/2})$}
\rput[Bl](1.5,6){$2z+t_3^{-1/2}z^{1/2}+\frac{5t_2}{2t_3}+\frac{t_2^2}{8t_3^{3/2}}z^{-1/2}-\frac{1}{2}z^{-1}+\mathcal{O}(z^{-3/2})$}

\end{pspicture}

\begin{pspicture}(-1,0)(10,6)
\psset{unit=.8cm}

\psline[doubleline=true](1,1)(1,2)
\psline(1,2)(1,3)
\psline[doubleline=true](1,3)(1,4)
\psline(1,4)(1,5)
\psline[doubleline=true](1,5)(1,6)

\psdot[dotsize=.15](1,1)
\psdot[dotsize=.15](1,2)
\psdot[dotsize=.15](1,3)
\psdot[dotsize=.15](1,4)
\psdot[dotsize=.15](1,5)
\psdot[dotsize=.15](1,6)

\rput[Br](0.5,6){$G_Y^{(3)}(z)_-$}
\rput[Br](0.5,5){$G_Y^{(3)}(z)_+$}
\rput[Bl](1.5,6){$z-z^{-1}+\mathcal{O}(z^{-2})$}
\rput[Bl](1.5,5){$3t_3^{-1/2}z^{1/2}-\frac{3t_2}{2t_3}+\frac{3t_2^2}{8t_3^{3/2}}z^{-1/2}-\frac{1}{2}z^{-1}+\mathcal{O}(z^{-3/2})$}
\rput[Bl](1.5,4){$-3t_3^{-1/2}z^{1/2}-\frac{3t_2}{2t_3}-\frac{3t_2^2}{8t_3^{3/2}}z^{-1/2}-\frac{1}{2}z^{-1}-\mathcal{O}(z^{-3/2})$}
\rput[Bl](1.5,3){$-2z+3 t_3^{-1/2}z^{1/2}-\frac{9t_2}{2t_3}+\frac{3t_2^2}{8t_3^{3/2}}z^{-1/2}+\frac{1}{2}z^{-1}+\mathcal{O}(z^{-3/2})$}
\rput[Bl](1.5,2){$-2z-3 t_3^{-1/2}z^{1/2}-\frac{9t_2}{2t_3}-\frac{3t_2^2}{8t_3^{3/2}}z^{-1/2}+\frac{1}{2}z^{-1}+\mathcal{O}(z^{-3/2})$}
\rput[Bl](1.5,1){$-3z-6\frac{t_2}{t_3}+z^{-1}+\mathcal{O}(z^{-2})$}

\end{pspicture}

\noindent From the resulting polynomials $F_{(p)}(x,y)$, we may also compute the asymptotic behaviour of $G_{(p)}^Y(z)$ on all sheets. For example, from  \eqref{polynomial2},

\begin{pspicture}(-1,0)(10,5)
\psset{unit=.8cm}

\psline(1,1)(1,2)
\psline[doubleline=true](1,2)(1,3)
\psline[doubleline=true](1,3)(1,4)
\psline(1,4)(1,5)

\psdot[dotsize=.15](1,1)
\psdot[dotsize=.15](1,2)
\psdot[dotsize=.15](1,3)
\psdot[dotsize=.15](1,4)
\psdot[dotsize=.15](1,5)

\rput[Br](0.5,3){$G^Y_{(2)}(z)_-$}
\rput[Br](0.5,4){$G^Y_{(2)}(z)_+$}
\rput[Bl](1.5,5){$z-t_3^{-1/2}z^{1/2}-\frac{t_2-1}{2t_3}-\frac{(t_2-1)^2}{8t_3^{3/2}}z^{1/2}+\frac{1}{2}z^{-1}-\mathcal{O}(z^{-3/2})$}
\rput[Bl](1.5,4){$z+t_3^{-1/2}z^{1/2}-\frac{t_2-1}{2t_3}+\frac{(t_2-1)^2}{8t_3^{3/2}}z^{1/2}+\frac{1}{2}z^{-1}+\mathcal{O}(z^{-3/2})$}
\rput[Bl](1.5,3){$\frac{t_3}{4}z^2+\frac{t_2}{2}z+\mathcal{O}(z^{-2})$}
\rput[Bl](1.5,2){$-z+\I t_3^{-1/2}z^{1/2}-\frac{t_2-1}{2t_3}+\I\frac{(t_2-1)^2}{8t_3^{3/2}}z^{1/2}-\frac{1}{2}z^{-1}+\mathcal{O}(z^{-3/2})$}
\rput[Bl](1.5,1){$-z-\I t_3^{-1/2}z^{1/2}-\frac{t_2-1}{2t_3}-\I\frac{(t_2-1)^2}{8t_3^{3/2}}z^{1/2}-\frac{1}{2}z^{-1}\mathcal{O}(z^{-3/2})$}

\end{pspicture}
\end{ex}

\chapter{Lax Operators}
\thispagestyle{title}
\label{app:Q}

\noindent Here we provide the explicit form of the Lax matrices for $(p,p')=(3,2)$, $(4,3)$ and $(5,2)$, respectively, with the abbrevation $u_{p,\zeta}=u_p-\zeta$.

\begin{ex}
\label{app:wronsk1} $(p,p')=(3,2)$. The compatibility condition \eqref{stringeqn} implies $u_2(t)=3v_2(t)$ and $u_3(t)=3\dot{v}_2(t)/2$, where $v_2(t)$ solves the first Painlev\'e equation: $\ddot{v}_2(t)=6v_2(t)+t$. For $n=1,2$, we find
\begin{equation}
\begin{aligned}
\mathcal{B}^{(2)}&=-\mathcal{C}^{-1}\left(\mathcal{B}^{(1)}\right)^T\mathcal{C}=\frac{1}{4}
\begin{pmatrix}
0 & 4 & 0  \\
-u_2 & 0 & 4  \\
u_{3,\zeta} & 0 & 0  \\
\end{pmatrix}\ ,\\
\mathcal{Q}^{(2)}&=-\mathcal{C}^{-1}\left(\mathcal{Q}^{(1)}\right)^T\mathcal{C}=\frac{1}{4}
\begin{pmatrix}
2v_2 & 0 & -8  \\
-\dot{v}_2+2\zeta & 2v_2 & 0  \\
-\ddot{v}_2 & \dot{v}_2+2\zeta& -4v_2  \\
\end{pmatrix}\ .
\end{aligned}
\end{equation}
\end{ex}

\begin{ex} 
\label{app:wronsk2} $(p,p')=(4,3)$. From \eqref{stringeqn}, $u_2(t)=8v_2(t)$, $u_3(t)=8v_3(t)/3+8\dot{v}_2/6$. We consider the case $n=2$: the Lax operators read

\begin{equation}
\begin{aligned}
\mathcal{B}^{(2)}&=-\mathcal{C}^{-1}\left(\mathcal{B}^{(2)}\right)^T\mathcal{C}=\frac{1}{8}
\begin{pmatrix}
0 & 8 & 0 & 0 & 0 & 0 \\
0 & 0 & 8 & 8 & 0 & 0 \\
-u_3 & -u_2 & 0 & 0 & 8 & 0 \\
0 & 0 & 0 & 0 & 8 & 0 \\
u_{4,\zeta} & 0 & 0 & -u_2 & 0 & 8 \\
0 & u_{4,\zeta} & 0 & u_3 & 0 & 0 \\
\end{pmatrix}\ ,
\end{aligned}
\end{equation}

\noindent and $\mathcal{Q}^{(2)}=\mathcal{Q}^{(2)}_{\mathrm{cl.}}+\mathcal{Q}^{(2)[1]}+\mathcal{Q}^{(2)[2]}$, where 

\begin{equation}
\mathcal{Q}^{(2)}_{\mathrm{cl.}}=\frac{1}{8}
\begin{pmatrix}
-8u_3 & -u_2 & 0 & 0 & -32 & 0 \\
-4u_{4,\zeta} & -8u_3 & -u_2 & 3u_2 & 0 & -32 \\
u_2\left(u_3-\frac{3}{8}u_2\right)& 8u_{4,\zeta}+\frac{5}{8}u_2^2 & -8u_3 & 0 & 3u_2 & 0 \\
0 & 8u_{4,\zeta} & 0 & -8u_3 & -u_2 & 0 \\
u_2u_{4,\zeta} & 0 & 0 & 4u_{4,\zeta}+\frac{u_2^2}{8} & -8u_3 & -u_2 \\
0 & \frac{1}{2}\left(\frac{u_2}{8}u_{4,\zeta}\right) & 0 & -\frac{u_2u_3}{8} & -\frac{u_3}{2}-u_{4,\zeta} & -8u_3 \\
\end{pmatrix}\ ,
\end{equation}

\begin{equation}
\mathcal{Q}^{(2)[1]}=\frac{1}{8}
\begin{pmatrix}
\dot{u}_2 & 0 & \dots & & \dots & 0 \\
-12\dot{u}_3 & 2\dot{u}_2 & \ddots & & & \vdots \\
16\dot{u}_4-2\dot{u}_3 & -18\dot{u}_3 & \dot{u}_2 & & & \\
0 & -2\dot{u}_3& 0 & \dot{u}_2 & \ddots & \vdots \\
0 & 0 & 2-4\dot{u}_4 & 14\dot{u}_3& 8\dot{u}_2 & 0 \\
0 & 0 & -4\dot{u}_4 & 4\dot{u}_4& 15\dot{u}_2-8\dot{u}_3 & -16\dot{u}_2 \\
\end{pmatrix}\ ,
\end{equation}

\begin{equation}
\mathcal{Q}^{(2)[2]}=\frac{1}{8}
\begin{pmatrix}
0&\dots & & &\ \dots\ &\ 0\ \\
3\ddot{u}_2& \ddots &  &  &  & \vdots \\
9\ddot{u}_2+8\ddot{u}_3 & \ddot{u}_2 & & & & \\
2\ddot{u}_3 & 0& & & & \\
-2\ddot{u}_3 & 0 & 2\ddot{u}_3 & 5\ddot{u}_2& \ddots & \vdots \\
0 & 4\ddot{u}_4 & -2\ddot{u}_3 & 10\ddot{u}_3-3\dddot{u}_2 & 0 & 0\\
\end{pmatrix}\ .
\end{equation}
\end{ex}

\begin{ex} 
\label{app:wronsk3} $(p,p')=(5,2)$. In this case \eqref{stringeqn} requires $u_2(t)=20v_2(t)$, $u_3(t)=30\dot{v}_2(t)$ and $u_5(t)=v_2(t)\dot{v}_2(t)/2$. We consider the cases $n=2$ and $3$: the Lax operators read

\begin{equation}
\begin{aligned} 
\mathcal{B}^{(2)}=-\mathcal{C}\left(\mathcal{B}^{(3)}\right)^T\mathcal{C}^{-1}=\frac{1}{16}
\begin{pmatrix}
0 & 16 & 0 & \dots & & & & & \dots & 0 \\
\vdots & \ddots& 16 & 16 & & &  &  &  & \vdots \\
 &  &  &  & 16 & 16 &  &  &  &  \\
 &  &  &  &  & 16 &  &  &  &  \\
-u_4 & -u_3 & -u_2 &  &  &  & 16 &  &  &  \\
 &  &  &  &  &  & 16 & 16 &  &  \\
u_{5,\zeta} &  &  & -u_3 &  & -u_2 &  &  & 16 & \vdots \\
 &  &  &  &  &  &  &  & 16 & 0 \\
\vdots & u_{5,\zeta}  &  & u_4 &  &  &  & -u_2 & \ddots & 16 \\
0 & \dots  & u_{5,\zeta}  &  &  & u_4 &  & u_3 & \dots  & 0 \\
\end{pmatrix}\ 
\end{aligned}
\end{equation}

\noindent and $\mathcal{Q}^{(n)}=\mathcal{Q}^{(n)}_{\mathrm{cl.}}+\mathcal{Q}^{(n)[1]}+\mathcal{Q}^{(n)[2]}$, with

\begin{equation}
\begin{aligned} 
\mathcal{Q}^{(2)}_{\mathrm{cl.}}=\frac{1}{8}
\begin{pmatrix}
16v_2 & 0 & 16 & -16 & 0 & \dots &  &  & \dots & 0 \\
0 & 16v_2 & \ddots &  & 16 &  &  &  &  & \vdots \\
-u_4 & \ddots &-4v_2 & & & & & 16 & & \\
0 &  & & 16v_2 &  &  & 16 & -16 &  &  \\
-u_{5,\zeta}  & -u_4 &  &  & -4v_2 &  &  &  & 16 & \vdots \\
u_{5,\zeta} &  &  &  &  & -4v_2 &  & &  & 0 \\
& & & -u_4 &  &  & -4v_2 &  &  & 16 \\
& u_{5,\zeta}  &  & u_4 &  &  &  & -4v_2 & \ddots & -16 \\
\vdots & & & u_{5,\zeta}  & &	&	& \ddots & -4v_2 & 0 \\
0 & \dots &	 &	 &	 -u_{5,\zeta}  & u_{5,\zeta} & -u_4 & u_4 & 0 & -24v_2\\
\end{pmatrix}\ ,
\end{aligned}
\end{equation}

\noindent which satisfies $\mathcal{Q}^{(2)}_{\mathrm{cl.}}=-\mathcal{C}\left(\mathcal{Q}^{(3)}_{\mathrm{cl.}}\right)^T\mathcal{C}^{-1}$, and the leading quantum corrections

\begin{equation}
\begin{aligned} 
\mathcal{Q}^{(2)[1]}=\frac{1}{4}
\begin{pmatrix}
0 & \dots &  &  & \quad & \quad &\quad & \quad & \dots & 0 \\
 4\dot{v}_2 & \ddots  &   &   &   &   &  &  &  & \vdots \\
 & -3\dot{v}_2 &  &  &  &  &  &  &  &  \\
\vdots & 4\dot{v}_2 &  &  &  &  &  &  &  &  \\
 &  & \dot{v}_2 &  &  &  &  & &  &  \\
 &  & 4\dot{v}_2 & -3\dot{v}_2 &  &  &  &  &  &  \\
 &  &  &  & 4\dot{v}_2  & -11\dot{v}_2 &  &  &  &  \\
 &  &  &  &  & 8\dot{v}_2 & &  &  &  \\
\vdots &  &  &  & &  & 4\dot{v}_2 & \dot{v}_2 & \ddots & \vdots \\
0 & \dots &  &  & &  & \dots & & -3\dot{v}_2 & 0 \\
\end{pmatrix}
\end{aligned}
\end{equation}

\begin{equation}
\begin{aligned} 
\mathcal{Q}^{(3)[1]}=\frac{1}{4}
\begin{pmatrix}
0  & \dots  &   &   &   &   &  &  & \dots & 0 \\
-3\dot{v}_2  & \ddots  &   &   &   &   &  &  &  & \vdots \\
 & 4\dot{v}_2 &   &   &   &  &  &  & & \\
 & \dot{v}_2 &   &   &   &  &  &  & & \\
 &   & 4\dot{v}_2  &  &   &  &  &  & & \\
 &   & \dot{v}_2  & 8\dot{v}_2  &   &  &  &  & & \\
 &   &   &   &  \dot{v}_2 & 4\dot{v}_2 &  &  & & \\
 &   &   &   &   & -3\dot{v}_2 &  & & & \\
\vdots &   &   &   &   &  & -3\dot{v}_2 & 4\dot{v}_2 & \ddots & \vdots \\
0 & \dots  &   &   &   &  &  &  & \dots & 0\\
\end{pmatrix}
\end{aligned}
\end{equation}

\noindent and all higher-order corrections subsumed in

\begin{equation}
\begin{aligned} 
\mathcal{Q}^{(2)[2]}&=
\begin{pmatrix}
0 & \dots &  &  & \quad & \quad &\quad & \quad & \dots & 0 \\
0& \ddots  &  &   &   &  & &  &   &\vdots \\
3\ddot{v}_2 & \ddots &  &  &  &  &  &  &  &  \\
-\ddot{v}_2 &  &  &  &  &  &  &  &  &  \\
4\dddot{v}_2 & 6\ddot{v}_2 &  &  &  &  &  &  & &  \\
-\dddot{v}_2 &  &  &  &  &  &  &  &  &  \\
 & & & 6\ddot{v}_2 & &  &  &    &  &  \\
&  & \ddot{v}_2 & -3\dddot{v}_2 &  &  &  &  &  &  \\
\vdots  & &  & -4\dddot{v}_2& \ddot{v}_2 &  & & \ddots & \ddots & \vdots \\
0 & \dots  &   &  &   & -4\dddot{v}_2 & \ddot{v}_2 & -6\ddot{v}_2 &0 &  0\\
\end{pmatrix}
\end{aligned}
\end{equation}

\begin{equation}
\begin{aligned} 
\mathcal{Q}^{(3)[2]}&=
\begin{pmatrix}
 0 &  \dots &   &   &   &   &  &  & \dots & 0 \\
 0 &  \ddots &   &   &   &   &  &  &  & \vdots \\
 3\ddot{v}_2 &  \ddots &   &   &   &   &  &  &  &  \\
 6\ddot{v}_2 &   &   &   &   &   &  &  &  &  \\
 \dddot{v}_2 & -\ddot{v}_2  &   &   &   &   &  &  &  &  \\
 -4\dddot{v}_2 &   &   &   &   &   &  &  &  &  \\
  &   &   &  -\ddot{v}_2 &   &   &  &  &  &  \\
  & -4\dddot{v}_2 & -6\ddot{v}_2 & 3\ddot{v}_2  &   &   &  &  &  &  \\
\vdots  &   &   &   & -6\ddot{v}_2  &   &  & \ddots & \ddots & \vdots \\
 0 & \dots  &   &   & 4\dddot{v}_2 & -\dddot{v}_2  & -3\ddot{v}_2 &  & 0 & 0 \\
\end{pmatrix}
\end{aligned}
\end{equation}
\end{ex}

\addcontentsline{toc}{chapter}{Bibliography}
\bibliography{ref}

\providecommand{\href}[2]{#2}\begingroup\raggedright\begin{thebibliography}{100}

\bibitem{Wigner55}
E.~P. Wigner, \emph{Characteristic vectors of bordered matrices with infinite
  dimensions}, {\emph{Ann. Math.} {\bf 62} (1955) 548}.

\bibitem{akemann15}
G.~Akemann, J.~Baik and P.~Di~Francesco, \emph{The Oxford Handbook of Random
  Matrix Theory}.
\newblock Oxford Handbooks in Mathematics Series. Oxford University Press,
  2015.

\bibitem{Tutte62}
W.~Tutte, \emph{A census of planar triangulations}, {\emph{Canad. J. Math.}
  {\bf 14} (1962) 21}.

\bibitem{Tutte63}
W.~Tutte, \emph{A census of planar maps}, {\emph{Canad. J. Math.} {\bf 15}
  (1963) 249}.

\bibitem{tHooft74}
G.~'t~Hooft, \emph{{A Planar Diagram Theory for Strong Interactions}},
  {\emph{Nucl. Phys.} {\bf B72} (1974) 461}.

\bibitem{Brezin77}
E.~Brezin, C.~Itzykson, G.~Parisi and J.~B. Zuber, \emph{{Planar Diagrams}},
  {\emph{Commun. Math. Phys.} {\bf 59} (1978) 35}.

\bibitem{Bessis80}
D.~Bessis, C.~Itzykson and J.~B. Zuber, \emph{{Quantum field theory techniques
  in graphical enumeration}}, {\emph{Adv. Appl. Math.} {\bf 1} (1980) 109}.

\bibitem{David85}
F.~David, \emph{Planar diagrams, two-dimensional lattice gravity and surface
  models}, {\emph{Nucl. Phys.} {\bf B257} (1985) 45}.

\bibitem{Ambjorn85a}
J.~Ambjorn, B.~Durhuus and J.~Frohlich, \emph{{Diseases of Triangulated Random
  Surface Models, and Possible Cures}}, {\emph{Nucl. Phys.} {\bf B257} (1985)
  433}.

\bibitem{Ambjorn85b}
J.~Ambjorn, B.~Durhuus, J.~Frohlich and P.~Orland, \emph{{The Appearance of
  Critical Dimensions in Regulated String Theories}}, {\emph{Nucl. Phys.} {\bf
  B270} (1986) 457}.

\bibitem{Kazakov85}
V.~A. Kazakov, A.~A. Migdal and I.~K. Kostov, \emph{{Critical Properties of
  Randomly Triangulated Planar Random Surfaces}}, {\emph{Phys. Lett.} {\bf
  B157} (1985) 295}.

\bibitem{Boulatov86}
D.~Boulatov and V.~Kazakov, \emph{{The Ising Model on Random Planar Lattice:
  The Structure of Phase Transition and the Exact Critical Exponents}},
  {\emph{Phys. Lett.} {\bf B186} (1987) 379}.

\bibitem{Kazakov86}
V.~Kazakov, \emph{{Ising model on a dynamical planar random lattice: Exact
  solution}}, {\emph{Phys. Lett.} {\bf A119} (1986) 140}.

\bibitem{Kazakov88}
V.~Kazakov, \emph{Exactly solvable potts models, bond- and tree-like
  percolation on dynamical (random) planar lattice}, {\emph{Nucl. Phys. Proc.
  Suppl.} {\bf B4} (1988) 93}.

\bibitem{Potts52}
R.~B. Potts, \emph{Some generalized order-disorder transformations},
  {\emph{Math. Proc. Cambridge Philos. Soc.} {\bf 48} (1952) }.

\bibitem{Distler88}
J.~Distler and H.~Kawai, \emph{{Conformal Field Theory and 2D Quantum Gravity
  Or Who's Afraid of Joseph Liouville?}}, {\emph{Nucl. Phys.} {\bf B321} (1989)
  509}.

\bibitem{LZ91}
B.~Lian and G.~Zuckerman, \emph{{New selection rules and physical states in 2-D
  gravity: Conformal gauge}}, {\emph{Phys. Lett.} {\bf B254} (1991) 417}.

\bibitem{Daul93}
J.~Daul, V.~Kazakov and I.~Kostov, \emph{{Rational theories of 2-D gravity from
  the two matrix model}}, {\emph{Nucl. Phys.} {\bf B409} (1993) 311},
  [\href{http://arxiv.org/abs/hep-th/9303093}{{\tt hep-th/9303093}}].

\bibitem{Douglas89}
M.~R. Douglas and S.~H. Shenker, \emph{{Strings in Less Than One-Dimension}},
  {\emph{Nucl. Phys.} {\bf B335} (1990) 635}.

\bibitem{Daul94}
J.-M. Daul, \emph{{Q states Potts model on a random planar lattice}},
  \href{http://arxiv.org/abs/hep-th/9502014}{{\tt hep-th/9502014}}.

\bibitem{Speicher93}
R.~Speicher, \emph{Free convolution and the random sum of matrices},
  {\emph{Publ. RIMS} {\bf 29} (1993) 731}.

\bibitem{Speicher94}
R.~Speicher, \emph{Multiplicative functions on the lattice of non-crossing
  partitions and free convolution}, {\emph{Math. Ann.} {\bf 300} (1994) 97}.

\bibitem{Carroll96}
S.~M. Carroll, M.~E. Ortiz and W.~Taylor, \emph{{The Ising model with a
  boundary magnetic field on a random surface}}, {\emph{Phys. Rev. Lett.} {\bf
  77} (1996) 3947}, [\href{http://arxiv.org/abs/hep-th/9605169}{{\tt
  hep-th/9605169}}].

\bibitem{Carroll97}
S.~M. Carroll, M.~E. Ortiz and W.~Taylor, \emph{{Boundary fields and
  renormalization group flow in the two matrix model}}, {\emph{Phys. Rev.} {\bf
  D58} (1998) 046006}, [\href{http://arxiv.org/abs/hep-th/9711008}{{\tt
  hep-th/9711008}}].

\bibitem{Affleck98}
I.~{Affleck}, M.~{Oshikawa} and H.~{Saleur}, \emph{{Boundary critical phenomena
  in the three-state Potts model}}, {\emph{J. Phys. A: Math. Gen.} {\bf 31}
  (July, 1998) 5827}, [\href{http://arxiv.org/abs/cond-mat/9804117}{{\tt
  cond-mat/9804117}}].

\bibitem{FZZ00}
V.~Fateev, A.~B. Zamolodchikov and A.~B. Zamolodchikov, \emph{{Boundary
  Liouville field theory. 1. Boundary state and boundary two point function}},
  \href{http://arxiv.org/abs/hep-th/0001012}{{\tt hep-th/0001012}}.

\bibitem{Teschner01}
J.~Teschner, \emph{{Liouville theory revisited}}, {\emph{Class. Quant. Grav.}
  {\bf 18} (2001) R153}, [\href{http://arxiv.org/abs/hep-th/0104158}{{\tt
  hep-th/0104158}}].

\bibitem{ZZ01}
A.~B. Zamolodchikov and A.~B. Zamolodchikov, \emph{{Liouville field theory on a
  pseudosphere}},  \href{http://arxiv.org/abs/hep-th/0101152}{{\tt
  hep-th/0101152}}.

\bibitem{Seiberg03}
N.~Seiberg and D.~Shih, \emph{{Branes, rings and matrix models in minimal
  (super)string theory}}, {\emph{JHEP} {\bf 0402} (2004) 021},
  [\href{http://arxiv.org/abs/hep-th/0312170}{{\tt hep-th/0312170}}].

\bibitem{Seiberg04}
N.~Seiberg and D.~Shih, \emph{{Minimal string theory}}, {\emph{C. R. Phys.}
  {\bf 6} (2005) 165}, [\href{http://arxiv.org/abs/hep-th/0409306}{{\tt
  hep-th/0409306}}].

\bibitem{Atkin11}
M.~R. Atkin and J.~F. Wheater, \emph{{The Spectrum of FZZT Branes Beyond the
  Planar Limit}}, {\emph{JHEP} {\bf 1102} (2011) 084},
  [\href{http://arxiv.org/abs/1011.5989}{{\tt 1011.5989}}].

\bibitem{Atkin12}
M.~R. Atkin and S.~Zohren, \emph{{FZZT Brane Relations in the Presence of
  Boundary Magnetic Fields}}, {\emph{JHEP} {\bf 1211} (2012) 163},
  [\href{http://arxiv.org/abs/1204.4482}{{\tt 1204.4482}}].

\bibitem{Voiculescu92}
D.~Voiculescu, K.~Dykema and A.~Nica, \emph{Free Random Variables}.
\newblock CRM Monograph Series. American Mathematical Soc.

\bibitem{ksm13}
B.~Niedner, M.~R. Atkin and J.~F. Wheater, \emph{{Boundary States of the Potts
  Model on Random Planar Maps}},
  \href{http://dx.doi.org/10.1007/978-3-319-20046-0_47}{\emph{Springer Proc.
  Phys.} {\bf 170} (2016) 387}, [\href{http://arxiv.org/abs/1511.1525}{{\tt
  1511.1525}}].

\bibitem{Atkin15}
M.~R. Atkin, B.~Niedner and J.~F. Wheater, \emph{{Sums of Random Matrices and
  the Potts Model on Random Planar Maps}},
  \href{http://arxiv.org/abs/1511.3657}{{\tt 1511.3657}}.

\bibitem{Bouwknegt92}
P.~Bouwknegt, J.~G. McCarthy and K.~Pilch, \emph{{BRST analysis of physical
  states for 2-D gravity coupled to $c\leq 1$ matter}}, {\emph{Commun. Math.
  Phys.} {\bf 145} (1992) 541}.

\bibitem{Wronskians}
C.-T. Chan, H.~Irie, B.~Niedner and C.-H. Yeh, \emph{{Wronskians, dualities and
  FZZT-Cardy branes}},  \href{http://arxiv.org/abs/1601.4934}{{\tt 1601.4934}}.

\bibitem{Ginsparg93}
P.~H. Ginsparg and G.~W. Moore, \emph{{Lectures on 2-D gravity and 2-D string
  theory}},  in \emph{Boulder 1992, Proceedings, Recent directions in particle
  theory}, 1993.
\newblock \href{http://arxiv.org/abs/hep-th/9304011}{{\tt hep-th/9304011}}.

\bibitem{DiFrancesco93}
P.~Di~Francesco, P.~H. Ginsparg and J.~Zinn-Justin, \emph{{2-D Gravity and
  random matrices}}, {\emph{Phys. Rept.} {\bf 254} (1995) 1},
  [\href{http://arxiv.org/abs/hep-th/9306153}{{\tt hep-th/9306153}}].

\bibitem{Ambjorn97}
J.~Ambjorn, B.~Durhuus and T.~Jonsson, \emph{{Quantum geometry. A statistical
  field theory approach}}.
\newblock Cambridge Monographs on Mathematical Physics. Cambridge University
  Press, 1997.

\bibitem{Eynard11formal}
B.~Eynard, \emph{Formal matrix integrals and combinatorics of maps},  in
  \emph{Random Matrices, Random Processes and Integrable Systems} (J.~Harnad,
  ed.), CRM Series in Mathematical Physics, p.~415.
\newblock Springer New York, 2011.

\bibitem{Moore91}
G.~W. Moore, N.~Seiberg and M.~Staudacher, \emph{{From loops to states in 2-D
  quantum gravity}}, {\emph{Nucl. Phys.} {\bf B362} (1991) 665}.

\bibitem{legall13}
J.-F. Le~Gall, \emph{Uniqueness and universality of the brownian map},
  {\emph{Ann. Probab.} {\bf 41} (2013) 2880}.

\bibitem{DiFrancesco97}
P.~Di~Francesco, P.~Mathieu and D.~Senechal, \emph{{Conformal Field Theory}}.
\newblock Graduate Texts in Contemporary Physics. Springer-Verlag, New York,
  1997.

\bibitem{Polchinski94}
J.~Polchinski, \emph{{Combinatorics of boundaries in string theory}},
  {\emph{Phys. Rev.} {\bf D50} (1994) 6041},
  [\href{http://arxiv.org/abs/hep-th/9407031}{{\tt hep-th/9407031}}].

\bibitem{Bouwknegt92b}
P.~Bouwknegt and K.~Schoutens, \emph{{W symmetry in conformal field theory}},
  {\emph{Phys. Rept.} {\bf 223} (1993) 183},
  [\href{http://arxiv.org/abs/hep-th/9210010}{{\tt hep-th/9210010}}].

\bibitem{knizhnik88}
V.~G. Knizhnik, A.~M. Polyakov and A.~B. Zamolodchikov, \emph{{Fractal
  Structure of 2D Quantum Gravity}}, {\emph{Mod. Phys. Lett.} {\bf A3} (1988)
  819}.

\bibitem{Cappelli86}
A.~Cappelli, C.~Itzykson and J.~B. Zuber, \emph{{Modular Invariant Partition
  Functions in Two-Dimensions}}, {\emph{Nucl. Phys.} {\bf B280} (1987) 445}.

\bibitem{FateevZamolodchikov87}
V.~Fateev and A.~Zamolodchikov, \emph{{Conformal Quantum Field Theory Models in
  Two-Dimensions Having Z(3) Symmetry}}, {\emph{Nucl. Phys.} {\bf B280} (1987)
  644}.

\bibitem{Caldeira03}
A.~F. Caldeira, S.~Kawai and J.~F. Wheater, \emph{{Free boson formulation of
  boundary states in W(3) minimal models and the critical Potts model}},
  {\emph{JHEP} {\bf 0308} (2003) 041},
  [\href{http://arxiv.org/abs/hep-th/0306082}{{\tt hep-th/0306082}}].

\bibitem{Caldeira04}
A.~F. Caldeira and J.~F. Wheater, \emph{{Boundary states and broken bulk
  symmetries in W A(r) minimal models}},  in \emph{Shifman, M. (ed.) et al.:
  From fields to strings, vol. 2}, p.~1441, 2004.
\newblock \href{http://arxiv.org/abs/hep-th/0404052}{{\tt hep-th/0404052}}.

\bibitem{Bouwknegt91}
P.~Bouwknegt, J.~G. McCarthy and K.~Pilch, \emph{{Some aspects of free field
  resolutions in 2-D CFT with application to the quantum Drinfeld-Sokolov
  reduction}},  in \emph{{Strings and Symmetries 1991 Stony Brook, New York}},
  1991.
\newblock \href{http://arxiv.org/abs/hep-th/9110007}{{\tt hep-th/9110007}}.

\bibitem{Felder88}
G.~Felder, \emph{{BRST Approach to Minimal Models}}, {\emph{Nucl. Phys.} {\bf
  B317} (1989) 215}.

\bibitem{Mizoguchi91}
S.~Mizoguchi and T.~Nakatsu, \emph{{BRST structure of the W(3) minimal model}},
  {\emph{Prog. Theor. Phys.} {\bf 87} (1992) 727}.

\bibitem{Seiberg90}
N.~Seiberg, \emph{{Notes on quantum Liouville theory and quantum gravity}},
  {\emph{Prog. Theor. Phys. Suppl.} {\bf 102} (1990) 319}.

\bibitem{Teschner02}
B.~Ponsot and J.~Teschner, \emph{{Boundary Liouville field theory: Boundary
  three point function}}, {\emph{Nucl. Phys.} {\bf B622} (2002) 309},
  [\href{http://arxiv.org/abs/hep-th/0110244}{{\tt hep-th/0110244}}].

\bibitem{Cardy89}
J.~L. Cardy, \emph{Boundary conditions, fusion rules and the verlinde formula},
  {\emph{Nucl. Phys.} {\bf B324} (1989) 581}.

\bibitem{Polchinski98}
J.~Polchinski, \emph{String Theory: Volume 1, An Introduction to the Bosonic
  String}.
\newblock Cambridge Monographs on Mathematical Physics. Cambridge University
  Press, 1998.

\bibitem{JMU81a}
M.~Jimbo, T.~Miwa and K.~Ueno, \emph{Monodromy preserving deformation of linear
  ordinary differential equations with rational coefficients: I. general theory
  and $\tau$-function}, {\emph{Physica} {\bf D2} (1981) 306}.

\bibitem{JMU81b}
M.~Jimbo and T.~Miwa, \emph{Monodromy perserving deformation of linear ordinary
  differential equations with rational coefficients. ii}, {\emph{Physica} {\bf
  D2} (1981) 407}.

\bibitem{JMU81c}
M.~Jimbo and T.~Miwa, \emph{Monodromy preserving deformation of linear ordinary
  differential equations with rational coefficients. iii}, {\emph{Physica} {\bf
  D4} (1981) 26}.

\bibitem{Moore90a}
G.~W. Moore, \emph{{Matrix models of 2-D gravity and isomonodromic
  deformation}}, {\emph{Prog. Theor. Phys. Suppl.} {\bf 102} (1990) 255}.

\bibitem{Moore90b}
G.~W. Moore, \emph{{Geometry of the string equations}}, {\emph{Commun. Math.
  Phys.} {\bf 133} (1990) 261}.

\bibitem{Chan:2013wya}
C.-T. Chan, H.~Irie and C.-H. Yeh, \emph{{Duality Constraints on String Theory:
  Instantons and spectral networks}},
  \href{http://arxiv.org/abs/1308.6603}{{\tt 1308.6603}}.

\bibitem{Morozov94}
A.~Morozov, \emph{{Integrability and matrix models}}, {\emph{Phys. Usp.} {\bf
  37} (1994) 1}, [\href{http://arxiv.org/abs/hep-th/9303139}{{\tt
  hep-th/9303139}}].

\bibitem{Eynard07}
B.~Eynard and N.~Orantin, \emph{Invariants of algebraic curves and topological
  expansion}, {\emph{Commun. Number Theory Phys.} {\bf 1} (2007) 347}.

\bibitem{FKN92}
M.~Fukuma, H.~Kawai and R.~Nakayama, \emph{{Explicit solution for p - q duality
  in two-dimensional quantum gravity}}, {\emph{Commun. Math. Phys.} {\bf 148}
  (1992) 101}.

\bibitem{BEH01a}
M.~Bertola, B.~Eynard and J.~P. Harnad, \emph{{Duality, biorthogonal
  polynomials and multimatrix models}}, {\emph{Commun. Math. Phys.} {\bf 229}
  (2002) 73}, [\href{http://arxiv.org/abs/nlin/0108049}{{\tt nlin/0108049}}].

\bibitem{BEH01b}
M.~Bertola, B.~Eynard and J.~Harnad, \emph{{Duality of spectral curves arising
  in two matrix models}}, {\emph{Theor. Math. Phys.} {\bf 134} (2003) 27},
  [\href{http://arxiv.org/abs/nlin/0112006}{{\tt nlin/0112006}}].

\bibitem{BEH02}
M.~Bertola, B.~Eynard and J.~Harnad, \emph{{Differential systems for
  biorthogonal polynomials appearing in 2-matrix models and the associated
  Riemann-Hilbert problem}}, {\emph{Commun. Math. Phys.} {\bf 243} (2003) 193},
  [\href{http://arxiv.org/abs/nlin/0208002}{{\tt nlin/0208002}}].

\bibitem{BEH04}
M.~Bertola, B.~Eynard and J.~Harnad, \emph{{Semiclassical orthogonal
  polynomials, matrix models and isomonodromic tau functions}}, {\emph{Commun.
  Math. Phys.} {\bf 263} (2006) 401},
  [\href{http://arxiv.org/abs/nlin/0410043}{{\tt nlin/0410043}}].

\bibitem{Ambjorn05}
J.~Ambjorn and R.~A. Janik, \emph{{The Emergence of noncommutative target space
  in noncritical string theory}}, {\emph{JHEP} {\bf 08} (2005) 057},
  [\href{http://arxiv.org/abs/hep-th/0506197}{{\tt hep-th/0506197}}].

\bibitem{Gradshteyn07}
I.~S. Gradshteyn and I.~M. Ryzhik, \emph{Table of integrals, series, and
  products}.
\newblock Elsevier/Academic Press, Amsterdam, seventh~ed., 2007.

\bibitem{Behrend00}
R.~E. Behrend and P.~A. Pearce, \emph{{Integrable and conformal boundary
  conditions for sl(2) A-D-E lattice models and unitary minimal conformal field
  theories}}, {\emph{J. Statist. Phys.} {\bf 102} (2001) 577},
  [\href{http://arxiv.org/abs/hep-th/0006094}{{\tt hep-th/0006094}}].

\bibitem{ZJ99}
P.~Zinn-Justin, \emph{{The Dilute Potts model on random surfaces}}, {\emph{J.
  Statist. Phys.} {\bf 98} (2001) 245},
  [\href{http://arxiv.org/abs/cond-mat/9903385}{{\tt cond-mat/9903385}}].

\bibitem{Bonnet99}
G.~Bonnet, \emph{{Solution of Potts - 3 and Potts - infinity matrix models with
  the equations of motion method}}, {\emph{Phys. Lett.} {\bf B459} (1999) 575},
  [\href{http://arxiv.org/abs/hep-th/9904058}{{\tt hep-th/9904058}}].

\bibitem{Eynard99}
B.~Eynard and G.~Bonnet, \emph{{The Potts - q random matrix model: Loop
  equations, critical exponents, and rational case}}, {\emph{Phys. Lett.} {\bf
  B463} (1999) 273}, [\href{http://arxiv.org/abs/hep-th/9906130}{{\tt
  hep-th/9906130}}].

\bibitem{Guionnet10}
A.~Guionnet, V.~Jones, D.~Shlyakhtenko and P.~Zinn-Justin, \emph{{Loop models,
  random matrices and planar algebras}}, {\emph{Commun. Math. Phys.} {\bf 316}
  (2012) 45}, [\href{http://arxiv.org/abs/1012.0619}{{\tt 1012.0619}}].

\bibitem{Bernardi11}
O.~Bernardi and M.~Bousquet-M\'elou, \emph{Counting colored planar maps:
  Algebraicity results}, {\emph{J. Combin. Theory Ser.} {\bf B101} (2011) 315}.

\bibitem{Borot12}
G.~{Borot}, J.~{Bouttier} and E.~{Guitter}, \emph{{Loop models on random maps
  via nested loops: the case of domain symmetry breaking and application to the
  Potts model}}, {\emph{J. Phys. A: Math. Theor.} {\bf 45} (2012) 494017},
  [\href{http://arxiv.org/abs/1207.4878}{{\tt 1207.4878}}].

\bibitem{ZJ99-2}
P.~{Zinn-Justin}, \emph{{Adding and multiplying random matrices: A
  generalization of Voiculescu's formulas}}, {\emph{Phys. Rev. E} {\bf 59}
  (1999) 4884}, [\href{http://arxiv.org/abs/math-ph/9810010}{{\tt
  math-ph/9810010}}].

\bibitem{ZJ98}
P.~Zinn-Justin, \emph{Universality of correlation functions of hermitian random
  matrices in an external field}, {\emph{Commun. Math. Phys.} {\bf 194} (1998)
  631}.

\bibitem{Eynard03}
B.~Eynard, \emph{{Master loop equations, free energy and correlations for the
  chain of matrices}}, {\emph{JHEP} {\bf 0311} (2003) 018},
  [\href{http://arxiv.org/abs/hep-th/0309036}{{\tt hep-th/0309036}}].

\bibitem{Pasquier86}
V.~Pasquier, \emph{{Two-dimensional critical systems labelled by Dynkin
  diagrams}}, {\emph{Nucl. Phys.} {\bf B285} (1987) 162}.

\bibitem{Kostov95}
I.~K. Kostov, \emph{{Solvable statistical models on a random lattice}},
  {\emph{Nucl. Phys. Proc. Suppl.} {\bf 45A} (1996) 13},
  [\href{http://arxiv.org/abs/hep-th/9509124}{{\tt hep-th/9509124}}].

\bibitem{Matytsin93}
A.~Matytsin, \emph{{On the large N limit of the Itzykson-Zuber integral}},
  {\emph{Nucl. Phys.} {\bf B411} (1994) 805},
  [\href{http://arxiv.org/abs/hep-th/9306077}{{\tt hep-th/9306077}}].

\bibitem{Harishchandra56}
Harish-Chandra, \emph{A formula for semisimple lie groups}, {\emph{Nat. Acad.
  Sci.} {\bf 42} (1956) 538}.

\bibitem{Itzykson80}
C.~Itzykson and J.~Zuber, \emph{The planar approximation. ii}, {\emph{J. Math.
  Phys.} {\bf 21} (1980) 411}.

\bibitem{Gross91}
D.~J. Gross and M.~J. Newman, \emph{{Unitary and Hermitian matrices in an
  external field}}, {\emph{Phys. Lett.} {\bf B266} (1991) 291}.

\bibitem{Eynard95}
B.~Eynard and C.~Kristjansen, \emph{{More on the exact solution of the $O(n)$
  model on a random lattice and an investigation of the case $|n|>2$}},
  {\emph{Nucl. Phys.} {\bf B466} (1996) 463},
  [\href{http://arxiv.org/abs/hep-th/9512052}{{\tt hep-th/9512052}}].

\bibitem{Zee96}
A.~Zee, \emph{{Law of addition in random matrix theory}}, {\emph{Nucl. Phys.}
  {\bf B474} (1996) 726}, [\href{http://arxiv.org/abs/cond-mat/9602146}{{\tt
  cond-mat/9602146}}].

\bibitem{Staudacher89}
M.~Staudacher, \emph{{The Yang-lee Edge Singularity on a Dynamical Planar
  Random Surface}}, {\emph{Nucl. Phys.} {\bf B336} (1990) 349}.

\bibitem{Eynard02}
B.~Eynard, \emph{{Large N expansion of the 2 matrix model}}, {\emph{JHEP} {\bf
  0301} (2003) 051}, [\href{http://arxiv.org/abs/hep-th/0210047}{{\tt
  hep-th/0210047}}].

\bibitem{Eynard92}
B.~Eynard and J.~Zinn-Justin, \emph{{The O(n) model on a random surface:
  Critical points and large order behavior}}, {\emph{Nucl. Phys.} {\bf B386}
  (1992) 558}, [\href{http://arxiv.org/abs/hep-th/9204082}{{\tt
  hep-th/9204082}}].

\bibitem{Baxter71}
R.~Baxter, \emph{{Generalized ferroelectric model on a square lattice}},
  {\emph{Stud. Appl. Math.} {\bf 50} (1971) 51}.

\bibitem{Baxter73}
R.~J. Baxter, \emph{Potts model at the critical temperature}, {\emph{J. Phys.
  C} {\bf 6} (1973) L445}.

\bibitem{Durhuus96}
B.~Durhuus and C.~Kristjansen, \emph{{Phase structure of the O(n) model on a
  random lattice for $n>2$}}, {\emph{Nucl. Phys.} {\bf B483} (1997) 535},
  [\href{http://arxiv.org/abs/hep-th/9609008}{{\tt hep-th/9609008}}].

\bibitem{Kramers41}
H.~Kramers and G.~Wannier, \emph{{Statistics of the two-dimensional
  ferromagnet. Part 1.}}, {\emph{Phys. Rev.} {\bf 60} (1941) 252}.

\bibitem{eynard11b}
G.~{Borot} and B.~{Eynard}, \emph{{Enumeration of maps with self-avoiding loops
  and the $O(n)$ model on random lattices of all topologies}}, {\emph{J. Stat.
  Mech. Theor. Exp.} {\bf 1} (2011) 10},
  [\href{http://arxiv.org/abs/0910.5896}{{\tt 0910.5896}}].

\bibitem{Zamolodchikov86}
A.~Zamolodchikov, \emph{{Irreversibility of the Flux of the Renormalization
  Group in a 2D Field Theory}}, {\emph{JETP Lett.} {\bf 43} (1986) 730}.

\bibitem{Affleck91}
I.~Affleck and A.~W. Ludwig, \emph{{Universal noninteger 'ground state
  degeneracy' in critical quantum systems}}, {\emph{Phys. Rev. Lett.} {\bf 67}
  (1991) 161}.

\bibitem{Friedan03}
D.~Friedan and A.~Konechny, \emph{{On the boundary entropy of one-dimensional
  quantum systems at low temperature}}, {\emph{Phys. Rev. Lett.} {\bf 93}
  (2004) 030402}, [\href{http://arxiv.org/abs/hep-th/0312197}{{\tt
  hep-th/0312197}}].

\bibitem{Gervais84}
J.-L. Gervais and A.~Neveu, \emph{{Locality in Strong Coupling Liouville Field
  Theory and String Models for Seven-dimensions, Thirteen-dimensions and
  Nineteen-dimensions}}, {\emph{Phys. Lett.} {\bf B151} (1985) 271}.

\bibitem{Das89}
S.~R. Das, S.~Naik and S.~R. Wadia, \emph{{Quantization of the Liouville Mode
  and String Theory}}, {\emph{Mod. Phys. Lett.} {\bf A4} (1989) 1033}.

\bibitem{Bilal92}
A.~Bilal, \emph{{Remarks on the BRST cohomology for $c_M>1$ matter coupled to
  'Liouville gravity'}}, {\emph{Phys. Lett.} {\bf B282} (1992) 309},
  [\href{http://arxiv.org/abs/hep-th/9202035}{{\tt hep-th/9202035}}].

\bibitem{Gervais94}
J.-L. Gervais and A.~Neveu, \emph{{Locality in Strong Coupling Liouville Field
  Theory and String Models for Seven-dimensions, Thirteen-dimensions and
  Nineteen-dimensions}}, {\emph{Phys. Lett.} {\bf B151} (1985) 271}.

\bibitem{Gaberdiel13}
M.~R. Gaberdiel and R.~Gopakumar, \emph{{Minimal Model Holography}}, {\emph{J.
  Phys.} {\bf A46} (2013) 214002}, [\href{http://arxiv.org/abs/1207.6697}{{\tt
  1207.6697}}].

\bibitem{Compere08}
G.~Compere and D.~Marolf, \emph{{Setting the boundary free in AdS/CFT}},
  {\emph{Class. Quant. Grav.} {\bf 25} (2008) 195014},
  [\href{http://arxiv.org/abs/0805.1902}{{\tt 0805.1902}}].

\bibitem{LZ92}
B.~Lian and G.~Zuckerman, \emph{{Semiinfinite homology and 2-D gravity. 1}},
  {\emph{Commun. Math. Phys.} {\bf 145} (1992) 561}.

\bibitem{Bershadsky89}
M.~Bershadsky and H.~Ooguri, \emph{{Hidden SL(n) Symmetry in Conformal Field
  Theories}}, {\emph{Commun. Math. Phys.} {\bf 126} (1989) 49}.

\bibitem{Bouwknegt90}
P.~Bouwknegt, J.~G. McCarthy and K.~Pilch, \emph{{Free field approach to
  two-dimensional conformal field theories}}, {\emph{Prog. Theor. Phys. Suppl.}
  {\bf 102} (1990) 67}.

\bibitem{Martinec05}
A.~Basu and E.~J. Martinec, \emph{{Boundary ground ring in minimal string
  theory}}, {\emph{Phys. Rev.} {\bf D72} (2005) 106007},
  [\href{http://arxiv.org/abs/hep-th/0509142}{{\tt hep-th/0509142}}].

\bibitem{Takayanagi11}
T.~Takayanagi, \emph{{Holographic Dual of BCFT}}, {\emph{Phys. Rev. Lett.} {\bf
  107} (2011) 101602}, [\href{http://arxiv.org/abs/1105.5165}{{\tt
  1105.5165}}].

\bibitem{Verlinde15}
H.~Verlinde, \emph{{Poking Holes in AdS/CFT: Bulk Fields from Boundary
  States}},  \href{http://arxiv.org/abs/1505.5069}{{\tt 1505.5069}}.

\bibitem{kazakov89}
V.~Kazakov, \emph{{The Appearance of Matter Fields from Quantum Fluctuations of
  2D Gravity}}, {\emph{Mod. Phys. Lett.} {\bf A4} (1989) 2125}.

\bibitem{Gesser10}
J.~A. Gesser, \emph{{Non-Compact Geometries in 2D Euclidean Quantum Gravity}},
  \href{http://arxiv.org/abs/1010.5006}{{\tt 1010.5006}}.

\bibitem{IR10}
G.~Ishiki and C.~Rim, \emph{{Boundary correlation numbers in one matrix
  model}}, {\emph{Phys. Lett.} {\bf B694} (2010) 272},
  [\href{http://arxiv.org/abs/1006.3906}{{\tt 1006.3906}}].

\bibitem{BIR10}
J.-E. Bourgine, G.~Ishiki and C.~Rim, \emph{{Boundary operators in minimal
  Liouville gravity and matrix models}}, {\emph{JHEP} {\bf 1012} (2010) 046},
  [\href{http://arxiv.org/abs/1010.1363}{{\tt 1010.1363}}].

\bibitem{BIR12}
J.-E. Bourgine, G.~Ishiki and C.~Rim, \emph{{Bulk-boundary correlators in the
  hermitian matrix model and minimal Liouville gravity}}, {\emph{Nucl. Phys.}
  {\bf B854} (2012) 853}, [\href{http://arxiv.org/abs/1107.4186}{{\tt
  1107.4186}}].

\bibitem{Stokes47}
G.~Stokes, \emph{On the numerical calculation of a class of definite integrals
  and infinite series}, {\emph{Camb. Phil. Soc. Trans.} {\bf IX} (1847) 166}.

\bibitem{Stokes58}
G.~Stokes, \emph{On the discontinuity of arbitrary constants which appear in
  divergent developments}, {\emph{Camb. Phil. Soc. Trans.} {\bf X} (1858) 105}.

\bibitem{Chan10}
C.-T. Chan, H.~Irie and C.-H. Yeh, \emph{{Stokes Phenomena and Non-perturbative
  Completion in the Multi-cut Two-matrix Models}}, {\emph{Nucl. Phys.} {\bf
  B854} (2012) 67}, [\href{http://arxiv.org/abs/1011.5745}{{\tt 1011.5745}}].

\bibitem{Chan12a}
C.-T. Chan, H.~Irie and C.-H. Yeh, \emph{{Stokes Phenomena and Quantum
  Integrability in Non-critical String/M Theory}}, {\emph{Nucl. Phys.} {\bf
  B855} (2012) 46}, [\href{http://arxiv.org/abs/1109.2598}{{\tt 1109.2598}}].

\bibitem{Towse00}
C.~Towse, \emph{{Generalized Wronskians and Weierstrass Weights}},
  {\emph{Pacific. J. Math.} {\bf 193} (2000) 501}.

\bibitem{Anderson05}
G.~W. Anderson, \emph{{Lacunary Wronskians on genus one curves}}, {\emph{J.
  Number Theory} {\bf 115} (2005) 197}.

\bibitem{GattoScherbak12a}
L.~{Gatto} and I.~{Scherbak}, \emph{{Linear ODEs, Wronskians and Schubert
  Calculus}}, {\emph{Moscow Math. J.} {\bf 12} (2013) 275},
  [\href{http://arxiv.org/abs/1310.3345}{{\tt 1310.3345}}].

\bibitem{GattoScherbak12b}
L.~{Gatto} and I.~{Scherbak}, \emph{{On Generalized Wronskians}},
  {\emph{Impanga Lecture Notes, EMS Congress Series Report} (2013) 257},
  [\href{http://arxiv.org/abs/1310.4683}{{\tt 1310.4683}}].

\bibitem{Schmidt39}
F.~K. Schmidt, \emph{Die wronskische determinante in beliebigen
  differenzierbaren funktionenk\"orpern}, {\emph{Math. Z.} {\bf 45} (1939) 62}.

\bibitem{Fukuma99}
M.~Fukuma and S.~Yahikozawa, \emph{{Comments on D instantons in $c<1$
  strings}}, {\emph{Phys. Lett.} {\bf B460} (1999) 71},
  [\href{http://arxiv.org/abs/hep-th/9902169}{{\tt hep-th/9902169}}].

\bibitem{Maldacena04}
J.~M. Maldacena, G.~W. Moore, N.~Seiberg and D.~Shih, \emph{{Exact vs.
  semiclassical target space of the minimal string}}, {\emph{JHEP} {\bf 0410}
  (2004) 020}, [\href{http://arxiv.org/abs/hep-th/0408039}{{\tt
  hep-th/0408039}}].

\bibitem{Bergere13}
M.~Berg\`ere, G.~Borot and B.~Eynard, \emph{Rational differential systems, loop
  equations, and application to the qth reductions of kp}, {\emph{Annales Henri
  Poincar\'e} (2015) 1}, [\href{http://arxiv.org/abs/1312.4237}{{\tt
  1312.4237}}].

\bibitem{Wilshin08}
S.~Kawamoto, J.~F. Wheater and S.~Wilshin, \emph{{Charged boundary states in a
  Z(3) extended minimal string}}, {\emph{Int. J. Mod. Phys.} {\bf A23} (2008)
  2257}.

\bibitem{Fidkowski03}
L.~Fidkowski, V.~Hubeny, M.~Kleban and S.~Shenker, \emph{{The Black hole
  singularity in AdS / CFT}}, {\emph{JHEP} {\bf 02} (2004) 014},
  [\href{http://arxiv.org/abs/hep-th/0306170}{{\tt hep-th/0306170}}].

\bibitem{Witten91}
E.~Witten, \emph{{On string theory and black holes}}, {\emph{Phys. Rev.} {\bf
  D44} (1991) 314}.

\bibitem{Kazakov00}
V.~Kazakov, I.~K. Kostov and D.~Kutasov, \emph{{A Matrix model for the
  two-dimensional black hole}}, {\emph{Nucl. Phys.} {\bf B622} (2002) 141},
  [\href{http://arxiv.org/abs/hep-th/0101011}{{\tt hep-th/0101011}}].

\end{thebibliography}\endgroup
\thispagestyle{title}
\bibliographystyle{bib} 
\end{document}